\documentclass[twocolumn,showpacs,prd,nofootinbib,superscriptaddress,amsmath,amssymb,floatfix]{revtex4-1}

\usepackage{graphicx}  
\usepackage{dcolumn}   
\usepackage{bm}        
\usepackage{amssymb}   
\usepackage[usenames,dvipsnames]{xcolor}
\usepackage{lineno, soul, ulem}
\usepackage{rotating}

\newcommand{\BibitemShut}[1]{}

\def \aap  {A\&A}

\def \apj  {ApJ}
\def \apjs  {ApJS}
\def \apjl  {ApJL}

\def \araa  {ARA\&A}
\def \jcap  {JCAP}
\def \prd {Phy. Rev. D}

\def \mnras {MNRAS}

\def \prl {PRL}
  
\begin{document}


\title{Constraining the dark matter contribution of $\gamma$-rays in Cluster of galaxies using {\it Fermi}-LAT data}

\author{Mattia Di Mauro}\email{dimauro.mattia@gmail.com}
\affiliation{Dipartimento di Fisica, Universit\'a di Torino, Via P. Giuria 1, 10125 Torino, Italy}
\affiliation{Istituto Nazionale di Fisica Nucleare, Sezione di Torino, Via P. Giuria 1, 10125 Torino, Italy}
\author{Judit P\'erez-Romero}\email{judit.perez@uam.es}
\affiliation{Instituto de F\'isica Te\'orica, IFT UAM-CSIC, Departamento de F\'isica Te\'orica, Universidad Aut\'onoma de Madrid, ES-28049 Madrid (Spain)}
\author{Miguel A. S\'anchez-Conde}\email{miguel.sanchezconde@uam.es}
\affiliation{Instituto de F\'isica Te\'orica, IFT UAM-CSIC, Departamento de F\'isica Te\'orica, Universidad Aut\'onoma de Madrid, ES-28049 Madrid (Spain)}
\author{Nicolao Fornengo}\email{nicolao.fornengo@unito.it,nicolao.fornengo@to.infn.it}
\affiliation{Dipartimento di Fisica, Universit\'a di Torino, Via P. Giuria 1, 10125 Torino, Italy}
\affiliation{Istituto Nazionale di Fisica Nucleare, Sezione di Torino, Via P. Giuria 1, 10125 Torino, Italy}

\begin{abstract}
Clusters of galaxies are the largest gravitationally-bound systems in the Universe. Their dynamics are dominated by dark matter (DM), which makes them among the best targets for indirect DM searches. We analyze 12 years of data collected by the {\it Fermi} Large Area Telescope ({\it Fermi}-LAT) in the direction of 49 clusters of galaxies selected for their proximity to the Earth and their high X-ray flux, which makes them the most promising targets. We first create physically motivated models for the DM density around each cluster considering different assumptions for the substructure distribution. 
Then we perform a combined search for a $\gamma$-ray signal in the {\it Fermi}-LAT data between 500 MeV and 1 TeV.
We find a signal of $\gamma$ rays potentially associated with DM that is at a statistical significance of $2.5\sigma-3.0\sigma$ when considering a slope for the subhalo mass distribution $\alpha=1.9$ and minimum mass of $M_{\rm{min}}=10^{-6}$ $M_{\odot}$.
The best-fit DM mass and annihilation cross-sections for a $b\bar{b}$ annihilation channel are $m_{\chi}=40-60$ GeV and $\langle \sigma v \rangle = (2-4) \times 10^{-25}$ cm$^3$/s. When we consider $\alpha=2.0$ and $M_{\rm{min}}=10^{-9}$ $M_{\odot}$, the best-fit of the cross section reduces to $\langle \sigma v \rangle = (4-10) \times 10^{-26}$ cm$^3$/s. For both DM substructure models there is a tension between the values of $\langle \sigma v \rangle$ that we find and the upper limits obtained with the non-detection of a $\gamma$-ray flux from Milky Way dwarf spheroidal galaxies.
This signal is thus more likely associated with $\gamma$ rays produced in the intracluster region by cosmic rays colliding with gas and photon fields.

\end{abstract}

\maketitle

\section{Introduction}
\label{sec:intro}

In the standard cosmological model, clusters of galaxies are thought to form via a hierarchical sequence of mergers and accretion of smaller systems. This process is mainly driven by gravity and dissipationless dark matter (DM) that dominates the gravitational field (see, e.g., \citep{2012ARA&A..50..353K} for a review). During the cluster formation, most of the binding gravitational energy is dissipated into the hot, thermal and ionized gas phase.

Since clusters of galaxies are the largest gravitationally-bound systems in the Universe ($\sim$80\% of their mass being in the form of DM), they are also attractive astrophysical objects for indirect DM searches.
Revealing the nature of DM is one of the most important and challenging goals of modern physics. One of the possible strategies to solve this puzzle is through the detection of $\gamma$ rays, possibly produced from DM particles annihilating or decaying in astrophysical sources, where the DM density is predicted to be large~\citep{Fermi-LAT:2016afa}.
Previous studies searched for a signal of $\gamma$ rays from Milky Way dwarf spheroidal galaxies (dSphs) (see, e.g., \citep{Ackermann:2015zua,Fermi-LAT:2016uux,2019MNRAS.482.3480P,DiMauro:2021qcf}), nearby galaxies (see, e.g., \citep{DiMauro:2019frs,2021PhRvD.104h3026G}), the Milky Way halo \citep{2012ApJ...761...91A} and the Galactic center (see, e.g., \citep{TheFermi-LAT:2017vmf,DiMauro:2021qcf}).
One of the most interesting targets, among the above cited ones, is the Galactic center, for which several groups have detected an excess with morphological characteristics compatible with the expected DM particles annihilating in the central halo of the Milky Way (see, e.g., \citep{Goodenough:2009gk,Hooper:2010mq,Calore:2014nla,TheFermi-LAT:2017vmf,DiMauro:2021raz}). The flux of the signal is compatible with photons produced from DM particles with mass around $40-60$ GeV and annihilation cross-section close to the thermal one\footnote{The thermal cross-section is about $(2-3)\times 10^{-26}$ cm$^3$/s \citep{Steigman:2012nb} and it is the value that reproduces a relic abundance of DM compatible with the observed one in the thermal WIMP scenario.}. 
Yet, alternative interpretations, such as a $\gamma$-ray emission from a population of millisecond pulsars located in the Galactic halo (see, e.g., \citep{Macias:2016nev}), can equally well explain the properties of the excess.
Therefore, the origin of the Galactic center excess measured in the data collected by the Large Area Telescope (LAT) onboard the {\it Fermi} satellite ({\it Fermi}-LAT) remains a mystery and the search for a possible signal in other astrophysical targets is central to test the DM hypothesis.

The astrophysical signal of photons from clusters of galaxies is another very active matter of research, even if it is not the central subject of this paper.
Shock waves propagating in the intracluster medium (ICM) and turbulence are expected to accelerate high-energy electrons and protons, thus creating a non-thermal population of cosmic rays (CRs) that are confined within the cluster's magnetic field.
These CRs are predicted to generate photons across the entire electromagnetic spectrum via synchrotron radiation in the intracluster magnetic fields, bremsstrahlung and $\pi^0$ decay production\footnote{The $\pi^0$ decay production is due to CRs, mainly protons, interacting with the intracluster atoms, for the $90\%$ composed by Hydrogen, and producing $\pi^0$ mesons which subsequently decay into two photons.} through the interaction with intracluster gas and inverse Compton scattering on the photon fields. 
CRs may also be injected in the ICM from active galactic nuclei (AGN) outbursts (see, e.g., \citep{Bonafede:2014zfa}), or galactic winds associated with star formation activity in cluster member galaxies (see, e.g., \citep{2016PhRvD..93j1301R}). 

Extended regions of radio emission, also called halos and relics, have already been observed in many clusters \citep{Ferrari:2008jr,vanWeeren:2019vxy}, demonstrating  that electrons and positrons are accelerated in these sources.
Instead, searches for the non-thermal X-ray and $\gamma$-ray emission due to bremsstrahlung, $\pi^0$ decay and inverse Compton scattering have not yielded conclusive results (e.g., \citep{Rephaeli:2008jp,Fermi-LAT:2015rbk}).
There is, though, a growing evidence for a potential detection in the vicinity of the Coma cluster \citep{Fermi-LAT:2015rbk,XiEtAl2018,AdamEtAl2021,BaghmanyanEtAl2021}.
In particular, an analysis of 50 clusters using four years of data from {\it Fermi}-LAT resulted in upper limits on the CR-induced $\gamma$-ray emission \citep{Ackermann:2013iaq}. 
Instead, the authors of Ref.~\citep{Prokhorov:2013kca} performed a stacking analysis of 55 clusters using {\it Fermi}-LAT data above 10 GeV finding a signal coming from the central region of the sources ($\sim 0.25\;\mathrm{deg}$) at the $4.3\sigma$ significance that is probably due to the AGN activity. 
Recently, Refs.~\citep{Branchini:2016glc, Colavincenzo:2019jtj} reported a statistically-significant positive cross-correlation signal between the unresolved $\gamma$-ray emission measured by the {\it Fermi}-LAT and different galaxy cluster catalogs. The possible origin in terms of compact $\gamma$-ray emission from AGNs inside the clusters or diffuse emission from the ICM, still needs to be confirmed.
These results could be consistent with the ones published in Ref.~\citep{Reiss:2017cmj} where the authors have performed a stacking analysis of {\it Fermi}-LAT data for the 112 most massive, high latitude, extended clusters and they identified at the $5.8\sigma$ confidence level a bright, spectrally-flat $\gamma$-ray ring at the expected virial shock position around the sources. The ring signal implies that the shock deposits $0.6\%$ of the thermal energy in relativistic electrons over a Hubble time. 

Given their mass-to-light ratio of the order of 100 \citep{Popesso:2006uv}, clusters of galaxies represent interesting targets to search for a DM signal \citep{2011JCAP...12..011S}. Indeed, being the most massive structures in the Universe, some of the nearby galaxy clusters are not only ideal candidates for decaying DM \citep{Combet:2012tt}, for which the only relevant parameter is the mass, but also for annihilating DM, as the enhancement to the DM flux due to presence of halo substructures is expected to be maximal for these objects (see, e.g., \citep{Ando:2019xlm} for a review). 
Previous works have already performed this DM search by combining observations of samples of clusters \citep{2010JCAP...05..025A,Huang:2011xr, 2012MNRAS.427.1651H,Nezri:2012tu,Dugger:2010ys,Lisanti:2017qlb,Thorpe-Morgan:2020czg} or investigating the most promising objects individually \citep{2012JCAP...07..017A,2012ApJ...750..123A,Fermi-LAT:2015xij,MAGIC:2018tuz}. In the absence of a signal, these works resulted in constraints on the DM particle.

In this paper, we perform a combined search for a $\gamma$-ray signal in the direction of 49 clusters of galaxies selected in terms of their vicinity and brightness of their thermal emission in X-rays. We use 12 years of {\it Fermi}-LAT data with a state-of-the-art source catalog, which is the 4FGL-DR2 Fermi-LAT catalog \cite{Fermi-LAT:2019yla}. In our data analysis, we test directly the DM hypothesis by using physically-motivated templates of the DM density distribution in each object. 
In particular, for the latter we use three different models that assume different levels of the contribution of the halo substructures in these objects to the DM-related fluxes.
The robustness of the {\it Fermi}-LAT analysis is inspected by using different interstellar emission models, data selections and analysis setups.
We also apply the proper statistical framework for deriving the significance of the signal. In order to do so, we perform a search for a DM signal compatible with clusters in 3100 random sky directions, that allows us to properly calculate the statistics related to the null-signal hypothesis.
In addition to the $\gamma$-ray signal search, our combined analysis allows us to set stringent constraints on the DM particle properties.
The main novelties with respect to previous papers (e.g., \citep{2010JCAP...05..025A,Huang:2011xr, 2012MNRAS.427.1651H,Nezri:2012tu,Dugger:2010ys,Lisanti:2017qlb,Thorpe-Morgan:2020czg,2012JCAP...07..017A,2012ApJ...750..123A,Fermi-LAT:2015xij,MAGIC:2018tuz}) are the following: we use several more years of {\it Fermi}-LAT data, we include the spatial extension of the DM distribution taking into account the expected population of subhalos and we use  state-of-the-art interstellar emission models that are central for searching DM signals from extended sources.

The paper is organized as follow: in Sec.~\ref{sec:sample}, we build the sample of galaxy clusters that we use in the analysis, explain the criteria we followed to select the sample and provide details on some of the considered clusters. Sec.~\ref{sec:DMdistr} is dedicated to the modelling of the cluster DM content, paying particular attention to the subhalo population. In this same section, we also derive the expected DM fluxes for both DM annihilation and decay, and obtain 2D spatial templates of the expected emissions. In Sec.~\ref{sec:analysis} we report the selection we apply to {\it Fermi}-LAT data and the analysis method we use.
In Sec.~\ref{sec:results} we report our results for the combined analysis of a $\gamma$-ray signal coming from DM. Finally, in Sec.~\ref{sec: conclusions} we draw our conclusions.


\section{Selection of the sample}
\label{sec:sample}

One of the most popular and efficient methods to observe and derive properties of nearby galaxy clusters is by using X-ray observations. Indeed, data from different X-ray telescopes (e.g. ROSAT All-Sky Survey - RASS, Chandra, etc.) have been used to create galaxy cluster catalogs containing the most relevant source parameters, such as mass, distance, redshift, infrared flux, \citep{Popesso:2006uv, 2011A&A...534A.109P, Snowden:2007jg, Vikhlinin:2008cd}. Here, we will first identify those clusters that meet the best conditions for DM searches, e.g., for their large masses and small distances, and then we build our final target sample from this initial selection. 
As a starting point, we look back to previous \textit{Fermi}-LAT Collaboration works that analyzed galaxy cluster data searching for $\gamma$-ray signals (either originated from DM or not), e.g., \citep{2010JCAP...05..025A, 2012JCAP...07..017A, Ackermann:2013iaq}. A careful look into these papers reveals that most clusters were extracted from the well-known HIFLUGCS catalog \citep{2002ApJ...567..716R}, an X-ray, flux-limited catalog containing the 63 brightest clusters in the X-ray band.\footnote{We note that authors in Ref.~\citep{Kafer:2019kuq} point out a bias towards including mostly cool-cored clusters in the HIFLUGCS catalog, which would imply an implicit bias in our sample as well.} All these clusters were scrutinized for DM searches in the past \citep{2011JCAP...12..011S, JeltemaEtAl2009}, sometimes also including Perseus, Ophiuchus, M49, Virgo and the galaxy groups NGC~5044 and NGC~5846 (some of them included in the so-called extended HIFLUGCS \citep{2002ApJ...567..716R}) given their optimal properties for DM searches. Thus, as our initial sample, we select the clusters in the HIFLUGCS catalog plus the individual ones mentioned above. 

\begin{figure*}
\includegraphics[width=\textwidth]{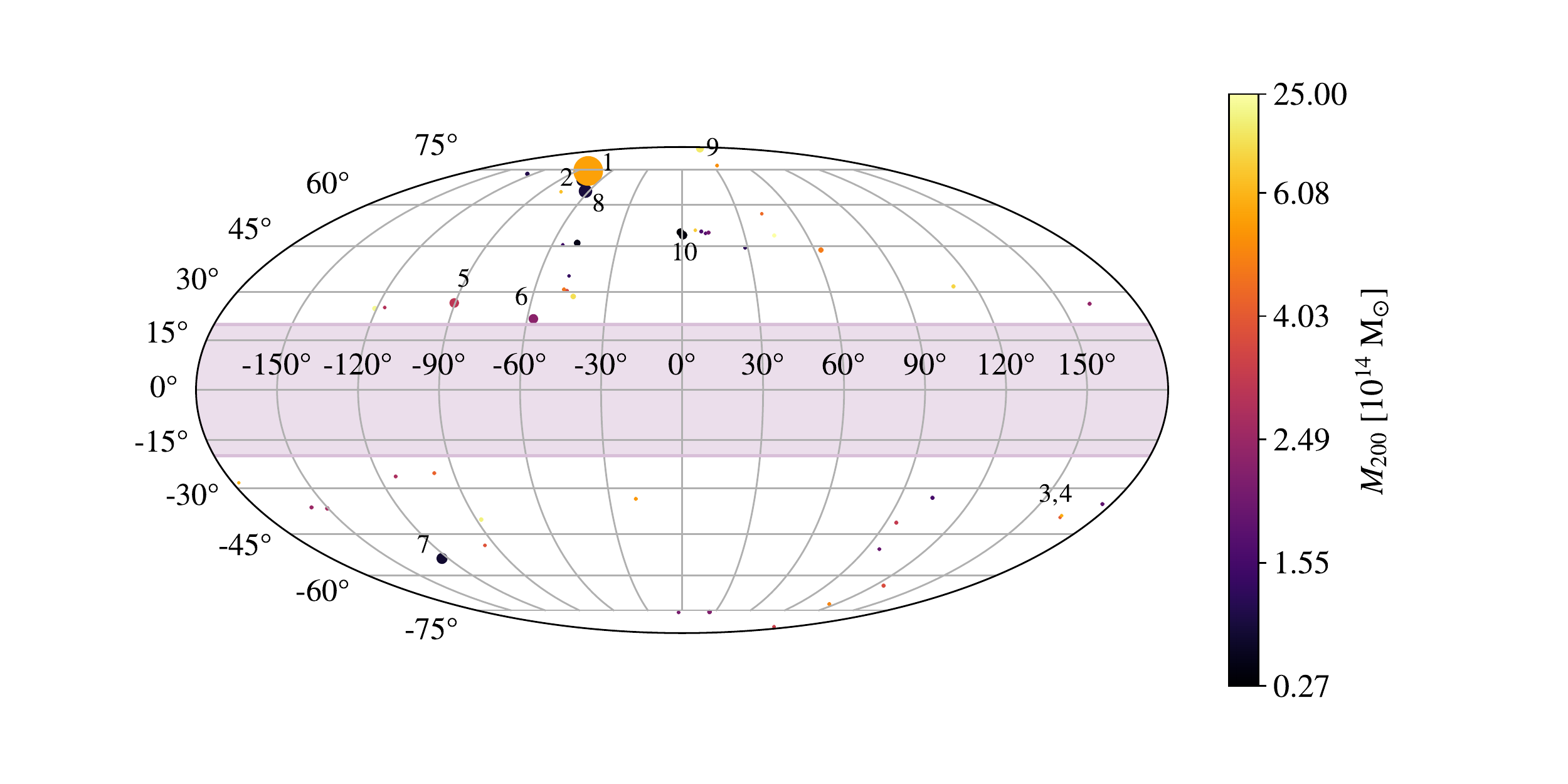}
\caption{Sky map in Mollweide projection showing the position of the 49 galaxy clusters composing our final target sample, summarized in Tabs.~\ref{tab:clusters_all_1}, \ref{tab:clusters_all_2}, \ref{tab:clusters_all_3} and \ref{tab:clusters_all_4}. The marker size represents the angle subtended by the virial radius in each case, while the vertical bar on the right denotes virial masses. Our mask excluding clusters with Galactic latitude $|b|<20$ deg is shown in pink. All clusters are at redshifts $z<0.1$. The numbers in the plot correspond to the following relevant clusters : 1 - Virgo, 2 - M49, 3 - A0399, 4 - A0401, 5 - A1060-Hydra, 6 - A3526-Centaurus, 7 - NGC~1399-Fornax, 8 - NGC~4636, 9 - A1656-Coma, 10 - NGC~5813.}
\label{fig:moll_proj_sample}
\end{figure*}

Yet, some of the clusters in this initial sample present major observation drawbacks for our purposes. First, it would be desirable to avoid those clusters lying across the Galactic plane, as the level of the Galactic diffuse $\gamma$-ray emission in this area would make it extremely challenging to disentangle between such Galactic signal and one originated in the galaxy cluster itself. Therefore, we decide to apply a mask in Galactic latitude of $|b|<20$ deg.  This removes Ophiuchus and Perseus from our initial sample above. Second, we take into account that the DM flux is proportional to $1/d_L^2$, where $d_L$ is the luminosity distance. Thus we also apply a cut in distance and remove those clusters beyond $z>0.1$, that are expected to be attenuated already by a factor $\sim 60$ 
relative to clusters in our most immediate vicinity (see, e.g., \cite{Fermi-LAT:2018lqt}). Our final galaxy cluster sample consists of 49 galaxy clusters, all of them having an X-ray flux $f_{X} \geq 1.7\times10^{-11} $erg s$^{-1}$ cm$^{-2}$. To derive the DM density profile of the clusters, we use the results from \citep{Schellenberger:2017wdw}, where the authors perform a new analysis of the clusters of interest using new data from Chandra X-ray Observatory. 
Clusters' masses obtained under the hydrostatic assumption are all given in Tabs.~\ref{tab:clusters_all_1}, \ref{tab:clusters_all_2}, \ref{tab:clusters_all_3} and \ref{tab:clusters_all_4}. Since M49, Virgo, NGC~5044 and NGC~5846 are not included in the original HIFLUGCS catalog, we adopt the masses derived in Ref.~\citep{2002ApJ...567..716R} for M49, NGC~5044 and NGC~5846, and the mass value quoted in Ref.~\citep{2011JCAP...12..011S} for Virgo. Sky positions, angular sizes and virial masses of our sample are shown in Fig.~\ref{fig:moll_proj_sample}. In this same figure it can be seen that there are two pairs of overlapping clusters: M49 \& Virgo, and A0399 \& A0401. We analyze these clusters separately since the shared $\gamma$-ray flux of these adjacent objects contributes less than $10\%$ of the total.
For example, M49 and Virgo are about 8 degrees apart. In the middle between them, the geometrical factor for the two clusters decreases at the few percent level with respect to the value close to their center. From the figure, we also conclude that the most massive and closest clusters are the ones exhibiting the highest DM fluxes and will thus dominate the analysis, e.g., Virgo (the largest and most massive one), A1060-Hydra, A3526-Centaurus and NGC~1399-Fornax, among others.

\section{Clusters modelling and DM-induced fluxes}
\label{sec:DMdistr} 

In this section, we perform the DM modelling of all clusters in our sample and compute their expected DM annihilation and decay fluxes.

In this work, we assume that all the DM is composed by Weakly Interacting Massive Particles (WIMPs \citep{Bertone:2004pz, Hooper:2009zm}), which can very weakly interact with the known particles of the Standard Model (SM). The expected $\gamma$-ray flux from either the annihilation of two WIMPs or its decay can be computed as \citep{Bergstrom:1997fj}:
\begin{equation}
\frac{d\Phi_{\gamma}}{dE}(E, \Delta\Omega, l.o.s)= \frac{d\phi_{\gamma}}{dE}(E)\times \begin{cases}
 J(\Delta\Omega, l.o.s) \\
 D(\Delta\Omega, l.o.s),\\
\end{cases}
\label{eq:gamma-flux}
\end{equation}
where $\frac{d\Phi_{\gamma}}{dE}$ is the DM-induced $\gamma$-ray flux. The $\frac{d\phi_{\gamma}}{dE}$ term is the so-called ``particle physics term'', which encodes the spectrum features of the WIMPs. The terms $D(\Delta\Omega, l.o.s)$ and $J(\Delta\Omega, l.o.s)$ refer to the so-called astrophysical $D$-factor, for decay, and $J$-factor, for annihilation, computed along the line of sight ($l.o.s.$) and within a given solid angle $\Delta\Omega$. The particle physics term is then computed as:
\begin{equation}
\frac{d\phi_{\gamma}}{dE}(E)=\frac{1}{4\pi m_{\chi}}\frac{dN_{\gamma}}{dE}(E)\times \begin{cases}
\frac{\langle\sigma v\rangle}{2 m_{\chi}}\\
\frac{1}{\tau},\\
\end{cases}
\label{eq:gamma-spectrum}
\end{equation}
where $m_{\chi}$ is the DM mass, $\frac{dN_{\gamma}}{dE}$ is the WIMP photon spectrum\footnote{We recall that for decay, the energy budget is half of the annihilation case and the end of the spectrum happens at $m_{\chi}/2$.}, $\langle\sigma v\rangle$ is the thermally-averaged annihilation cross-section and $\tau$ is the DM particle lifetime. In the above expression, we assume that the DM particles are their own antiparticles, i.e., Majorana particles.

As it can be seen in Eq.~(\ref{eq:gamma-flux}), the computation of the flux can be factorized in two terms: the first term in Eq.~(\ref{eq:gamma-flux}) encapsulates the spectral information of the expected signal (the DM mass and annihilation/decay channels, annihilation cross-section or decay lifetime), 
while the $J$- and $D$-factors carry the information about the morphology of the DM signal. Indeed, for our purposes, we can safely assume that the spatial distribution of the DM signal is independent of energy. This is due to the fact that we are assuming the so-called prompt emission for which $\gamma$ rays are produced after hadronization or electromagnetic cascade from particles produced after DM annihilation or decay. These processes are basically point like considering the dimension of the Galaxy. Therefore, the spatial morphology of the signal is due to the density distribution, that is energy independent, and not by the particle physics process that produce photons.

We follow Ref.~\citep{Cirelli:2010xx} to calculate $\frac{dN_{\gamma}}{dE}$ including electro-weak corrections. As for the calculation of the $J$- and $D$-factors, we define them as:
\begin{equation}\label{eq:j-factor}
J(\Delta\Omega, l.o.s) = \int_0^{\Delta\Omega}d\Omega\int_{l.o.s}\rho_{\rm{tot}}^2(r)dl,
\end{equation}
\begin{equation}\label{eq:d-factor}
D(\Delta\Omega, l.o.s) = \int_0^{\Delta\Omega}d\Omega\int_{l.o.s}\rho_{\rm{tot}}(r)dl,
\end{equation}
where $\Delta\Omega=2\pi(1-\cos\alpha_{\rm int})$, where $\alpha_{\rm{int}}$ is the integration angle, i.e., the angle between the $l.o.s$ and the direction that points toward the center of the cluster, and $\rho_{\rm{tot}}(r)$ the DM density profile. $\rho_{\rm{tot}}(r)$ describes the DM distribution inside the object and its modelling is key to obtaining realistic $D$ and $J$-factor values.
For each galaxy cluster, we model this DM distribution as follows:
\begin{equation}
\label{eq:rho_decomposition}
\rho_{\rm{tot}}(r) = \rho_{\rm{main}}(r) + \langle \rho_{\rm{subs}}\rangle(r),
\end{equation}
where $\rho_{\rm{main}}(r)$ is the smooth DM distribution in the main halo where the cluster resides, and $\langle \rho_{subs} \rangle(r)$ refers to the population of subhalos expected to exist according to $\Lambda$CDM (e.g., \citep{Kuhlen:2012ft, Zavala:2019gpq}). 
Because of their masses and distances, subhalos will not be individually resolved by the LAT, which has an instrumental angular resolution of order of a few tenth of degree. This fact, together with their large number, also allows us to avoid generating each subhalo individually, but rather to use an average description of the whole subhalo population in our work (as we describe below, subhalos are drawn from distribution functions obtained from cosmological simulations). 
In the next subsections we describe in detail the modelling that we performed for each of these components.

\subsection{Main halo modeling}\label{subsec:DM_main_halo}

Following results from DM-only $\Lambda$CDM cosmological simulations, we model the cluster's main halos with the Navarro-Frenk-White (NFW; \citep{Navarro:1995iw, NavarroEtAl1997}) DM density profile. Although deviations from this profile may exist for individual clusters, given the LAT angular resolution this description is expected to provide a realistic portrait for our purposes. The NFW profile reads as: 
\begin{equation}\label{eqn:NFW}
 \rho (r)=\frac{\rho_{0}}{\left(\frac{r}{r_{s}}\right)\left(1+\frac{r}{r_{s}}\right)^{2}}\,,
\end{equation}
where $r_{s}$ is the scale radius and $\rho_{0}$ the characteristic DM density.

We build a comprehensive DM density profile for each cluster starting from its measured mass. For nearby galaxy clusters as the ones in our sample, the mass, defined as $M_{200}$ (see Eq.~\ref{eq:M200}), is well constrained by their X-ray surface brightness profiles. In this work, we adopt the X-ray mass estimates presented in Ref.~\citep{Schellenberger:2017wdw} (except for M49, NGC~5044, NGC~5846 and Virgo, see Sec.~\ref{sec:sample} for details) and from them we derive both $r_{s}$ and $\rho_{0}$ for each cluster.

We first compute the virial radius, $R_{200}$, assuming a spherical overdensity with $\Delta =200$ (called $\Delta_{200}$) times the critical density of the Universe:
\begin{equation}
R_{200}=\left(\frac{3 M_{200}}{4\pi \Delta_{200}\rho_{\rm crit}}\right)^{1/3},
\label{eq:R200}
\end{equation}
with the critical density $\rho_{\rm crit}\,=\,137\,$M$_{\odot}\,$kpc$^{-3}$, computed assuming $H_0 = 70$ km$\;$s$^{-1}$Mpc$^{-1}$. 
We can now compute the NFW scale radius $r_{s}$ as
\begin{equation}\label{eqn:scale radius}
r_s \equiv R_{200}/c_{200},
\end{equation}
where $c_{200}$ is the so-called halo concentration. For this parameter, in our work we adopt the concentration-mass $(c-M)$ relation proposed in \citep{Sanchez-CondeEtAl2014} for main halos:
\begin{equation}
    c_{200}(M_{200},z=0)=\sum_{i=0}^5 c_i \times \left[\ln\left(\frac{M_{200}}{h^{-1}M_\odot}\right)\right]^i,
    \label{eq:c_M}
\end{equation}
which has proven to work well for objects in the mass range between dSphs and galaxy clusters. 

As for the scale density $\rho_{0}$, it can be computed by imposing
\begin{equation}
\label{eq:M200}
M_{200}=\int_0^{R_{200}}\rho_{\rm NFW}(r)r^2drd\Omega,
\end{equation}
and then we get 
\begin{equation} 
\rho_0 = \frac{2~\Delta_{200}~\rho_{\rm crit}~c_{200}}{3~f(c_{200})},
\end{equation}
where $f(c_{200})=\frac{2}{c_{200}^2}\left(\ln{(1+c_{200})}-\frac{c_{200}}{1+c_{200}}\right)$.

Finally, the variable that describes the extension\footnote{The full angular extension will correspond to the angle subtended by $2\times R_{200}$.} of the clusters in the sky is $\theta_{200}$, i.e. the angle subtended by $R_{200}$:
\begin{equation}\label{eqn:theta}
    \theta_{200} = \arctan\left(\frac{R_{200}}{d_L}\right).
\end{equation}

We apply the formalism described in this section to all clusters in our sample. The resulting NFW parameter values are included in Tabs.~\ref{tab:clusters_all_1}, \ref{tab:clusters_all_2}, \ref{tab:clusters_all_3} and \ref{tab:clusters_all_4}. 

\subsection{Modeling of the subhalo population}
\label{subsec:subhalo_modelling}
Since galaxy clusters are the largest gravitational bound objects in the Universe we expect them to host a large number of subhalos. The subhalo population can be parametrized as:
\begin{equation}\label{eq:subhalos-distribution}
\frac{d^3N}{dVdMdc}= N_{\rm{tot}}\frac{dP_V}{dV}(R)\frac{dP_M}{dM}(M)\frac{dP_c}{dc}(M,c),
\end{equation}
where $N_{\rm{tot}}$ is the total number of subhalos; and $P_i$ with $i = V, M, c$ is the probability distribution in each of the domains normalized to 1; $V$ referring to main halo volume, $M$ to the distribution of the subhalo masses and $c$ to subhalo concentration. Note that with this parametrization we are able to model the population of subhalos independently for each of the mentioned variables. This parametrization allows us to directly implement analytical models, the result of N-body cosmological simulations, for each distribution.

Unfortunately, there are still significant uncertainties pertaining to the properties of the subhalo population. Numerical cosmological simulations have been instrumental to shed light on halo substructures in the past years (for a review, see, e.g., \citep{Zavala:2019gpq} and references therein); yet many questions remain and are still matter of debate, e.g., minimum mass to form clumps \citep{2006PhRvL..97c1301P}, impact of tidal stripping on subhalo survival \citep{Aguirre-Santaella:2022kkm}, precise shape of subhalo DM density profiles \citep{2020MNRAS.491.4591E}, etc. All these uncertainties translate into uncertainties in the computation of the DM-induced $\gamma$-ray flux. In the following, we describe in detail how we plan to tackle this important issue through our work:
\begin{itemize}
    \item $\frac{dP_V}{dV}$: Since we are assuming spherical symmetry for the main halo, the only dependence regarding the distribution of subhalos within its volume is the relative distance of the subhalos to the center of the host. Because of this, in the following we will refer to this distribution as the subhalo radial distribution (SHRD). We adopt the SHRD results from high-resolution Milky-Way-size numerical simulations, namely \citep{Springel:2008cc} (Aquarius simulation) and \citep{Diemand:2008in} (Via Lactea II - VL-II simulation), which are some of the most used in the community. 
    We use both the Aquarius and VL-II SHRDs to encapsulate the current uncertainty on this parameter.
    
    \item $\frac{dP_M}{dM}$: The mass distribution of subhalos is known as the subhalo mass function (SHMF). Different studies based on N-body DM-only simulations agree that the SHMF can be parametrized as follows:
    \begin{equation}\label{eq:subhalo-mass-function}
    \frac{dN}{dM}\propto M^{-\alpha}.
    \end{equation}
    Typical values are $\alpha = 1.9$ \citep{Springel:2008cc} and $\alpha = 2.0$ \citep{Diemand:2008in}, the former being more conservative, as it implies a smaller number of subhalos, and being also more in line with other recent results \citep{Zavala:2019gpq}. The total mass in the form of substructures is typically expressed as a fraction of the total mass of the system, $f_{\rm sub}$, and depends on the minimum and maximum values adopted for the subhalo masses. The lower the minimum subhalo mass considered, the more mass would be bound in the form of substructures. 
    Note that different values of $f_{\rm sub}$ are needed for different values of $\alpha$ in order to conserve the total mass. If we adopt the ratio of the maximum subhalo mass to the host mass to be $M_{\rm max}^{\%}=0.01$ \citep{2011MNRAS.413.1373W, 2010MNRAS.401.1796A}, we obtain $f_{\rm sub}=0.18$ for $\alpha=1.9, M_{\rm min}=10^{-6}$M$_{\odot}$ and $f_{\rm sub}=0.34$ for $\alpha=2.0, M_{\rm min}=10^{-9}$M$_{\odot}$.
    
    \item $\frac{dP_c}{dc}$: Subhalos are subject to tidal forces, which produce an important mass loss in most cases, especially in the outskirts, e.g. \citep{Springel:2008cc,Diemand:2008in,Kazantzidis:2003im, Pe_arrubia_2010, errani2020asymptotic, Aguirre-Santaella:2022kkm}. Because of this, subhalos are known to be  more concentrated than field halos of the same mass \citep{Moline:2016pbm, Moline:2021rza, Newton:2021cjn}
    . In our work, we adopt the $(c-M)$ relation in \citep{Moline:2016pbm}, that was derived from VL-II data and includes a radial dependence of the concentration to account for the location of the subhalos within the main halo (with subhalos closer to the host halo center being more concentrated than those at outer radii). 
\end{itemize}

We now proceed by defining three benchmark models that will bracket the mentioned uncertainties on the properties of the subhalo population. Note though that this will be relevant only for annihilation. Indeed, the role of the substructures in the case of decay fluxes is negligible, as the dependence of the $D$-factor is simply linear with the mass of the system. The three benchmark models are:

\begin{itemize}
    \item MIN: does not include subhalos, thus the whole DM in the cluster is supposed to be smoothly distributed following an NFW profile, with the parameters derived as in Sec.~\ref{subsec:DM_main_halo}. We recall that this is the only benchmark model that will be considered for decay DM.
    \item MED: this model represents -- according to current knowledge -- the most realistic contribution of the subhalo population to the $\gamma$-ray flux due to DM annihilation. We adopt the VL-II SHRD \citep{Diemand:2008in}, $\alpha = 1.9$ for the SHMF, and $M_{\rm min}=10^{-6}$M$_{\odot}$.
    \item MAX: this model is defined so as to provide an upper bound to the contribution of the subhalo population to the annihilation flux. We adopt the Aquarius SHRD \citep{Springel:2008cc}, $\alpha = 2.0$, and $M_{\rm min}=10^{-9}$M$_{\odot}$.
\end{itemize}
 
A summary of these benchmark models is given in Tab.~\ref{tab:dm_models}.
 
\begin{table}
\begin{tabular}{|c|c|c|c|c|c|}
\hline
Model & SHRD & $\alpha$ & $c(M)$ & $M_{min}$ & $f_{sub}$ \\ 
\hline
\hline
MIN & - & - & - & - & - \\
\hline
MED & VL-II~\citep{Diemand:2008in} & 1.9 & \citep{Moline:2016pbm} & $10^{-6}~\mathrm{M}_{\odot}$ & 0.18 \\
\hline
MAX & Aquarius~\citep{Springel:2008cc} & 2.0 & \citep{Moline:2016pbm} & $10^{-9}~\mathrm{M}_{\odot}$ & 0.34\\
\hline 
\end{tabular} 
\caption{\label{tab:dm_models} Summary of the three benchmark models that we consider to quantify the contribution of the subhalo population to the $\gamma$-ray flux from DM. 
See Sec.~\ref{subsec:subhalo_modelling} for full details of each of the parameters.}
\end{table}

The effect of taking into account the halo substructures in our calculations is an enhancement of the annihilation flux, usually quantified in terms of the so-called substructure boost factor, $B$. This boost can range from $B=0$, where the contribution of the substructure is absent, up to almost $\sim 2$ orders of magnitude, depending on the adopted description of the subhalo population and of the host halo mass (for a review, see \citep{Ando:2019xlm}). In our work, we note that we consider the number of substructure levels to be $N_{\rm lvl}=2$ (subhalos inside subhalos, \citep{Sanchez-CondeEtAl2014, Bonnivard:2015pia}). 

\subsection{Annihilation and decay fluxes}

\begin{figure*}
\includegraphics[width=0.49\textwidth]{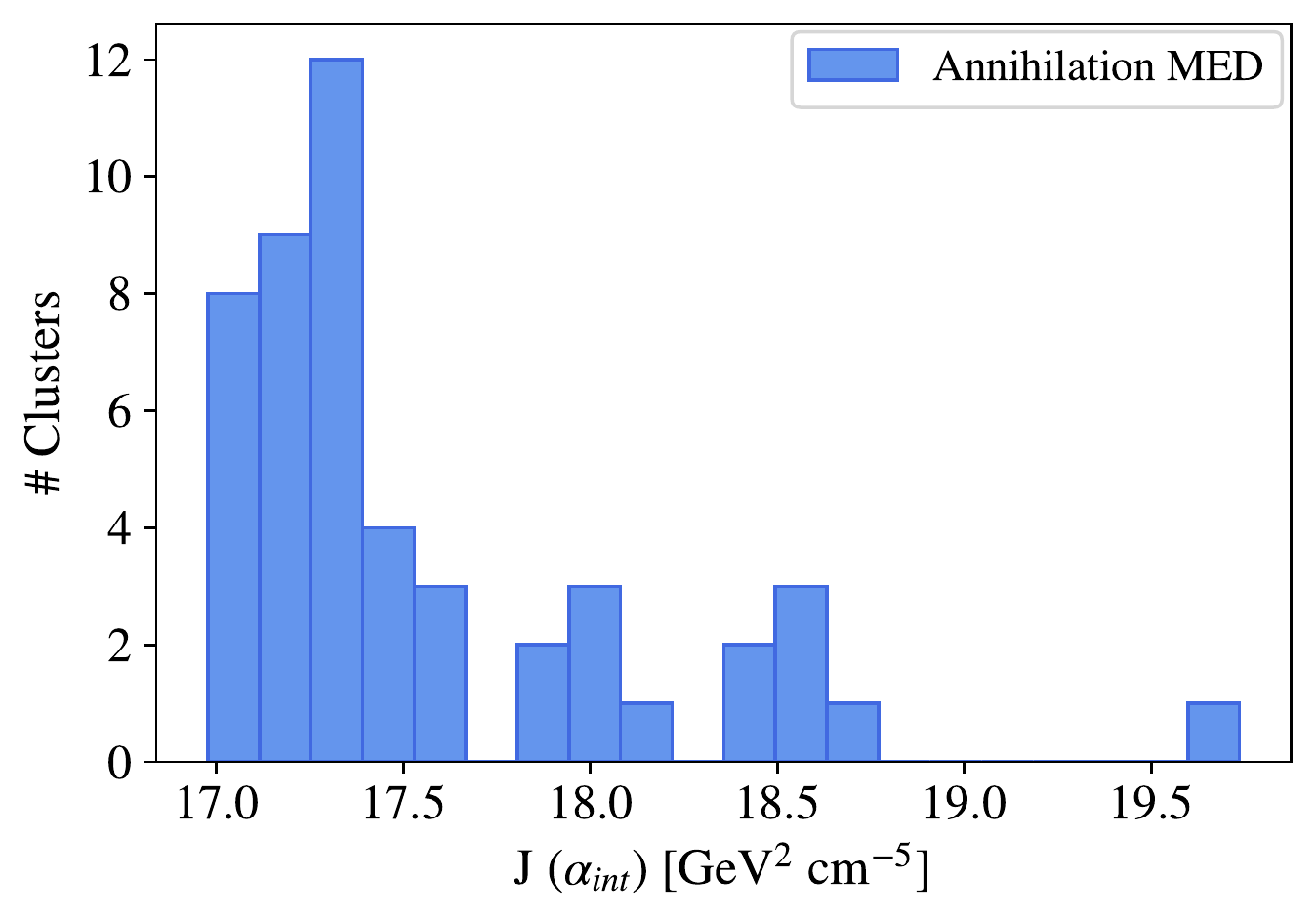}
\includegraphics[width=0.49\textwidth]{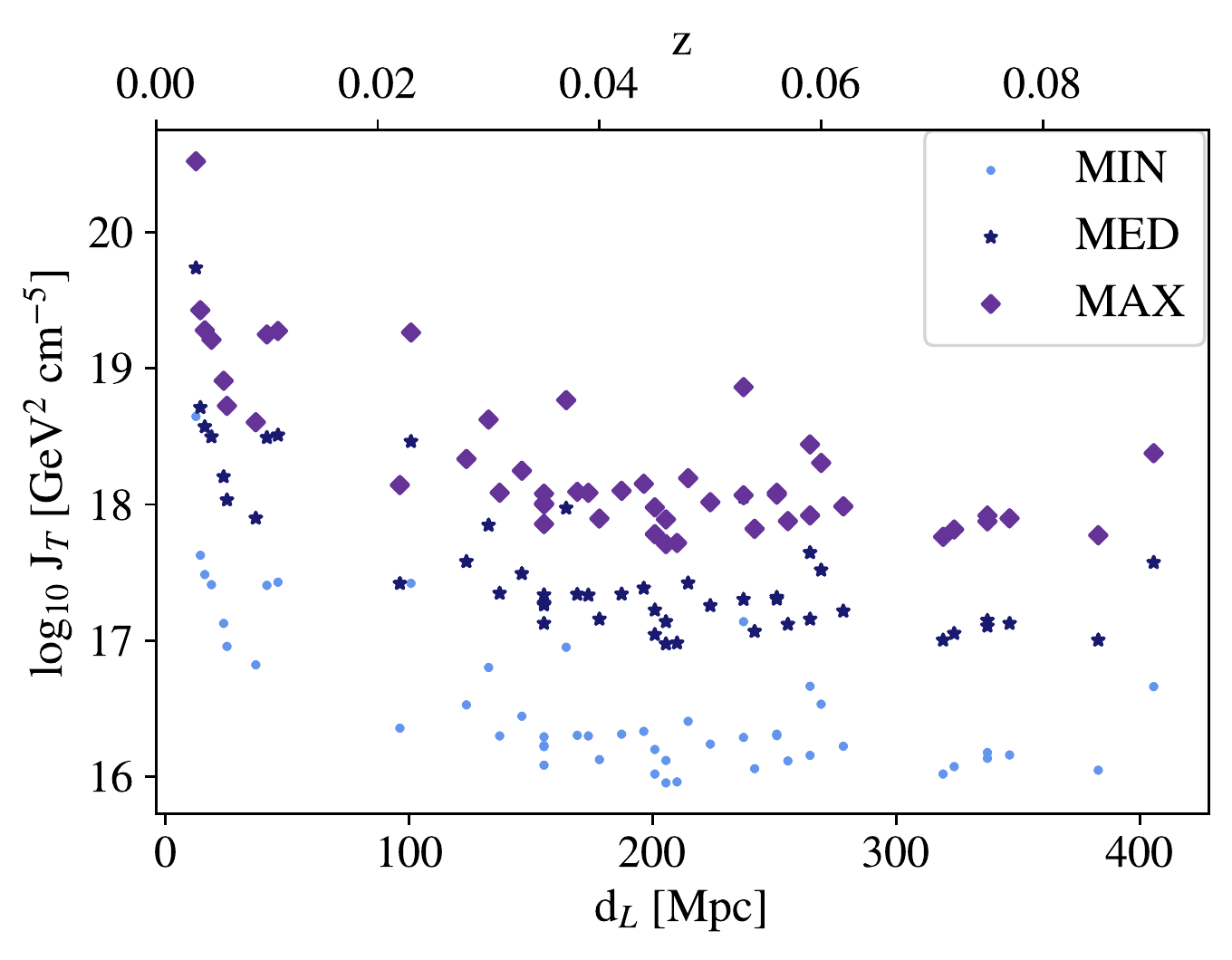}
\vspace{2cm}
\includegraphics[width=0.49\textwidth]{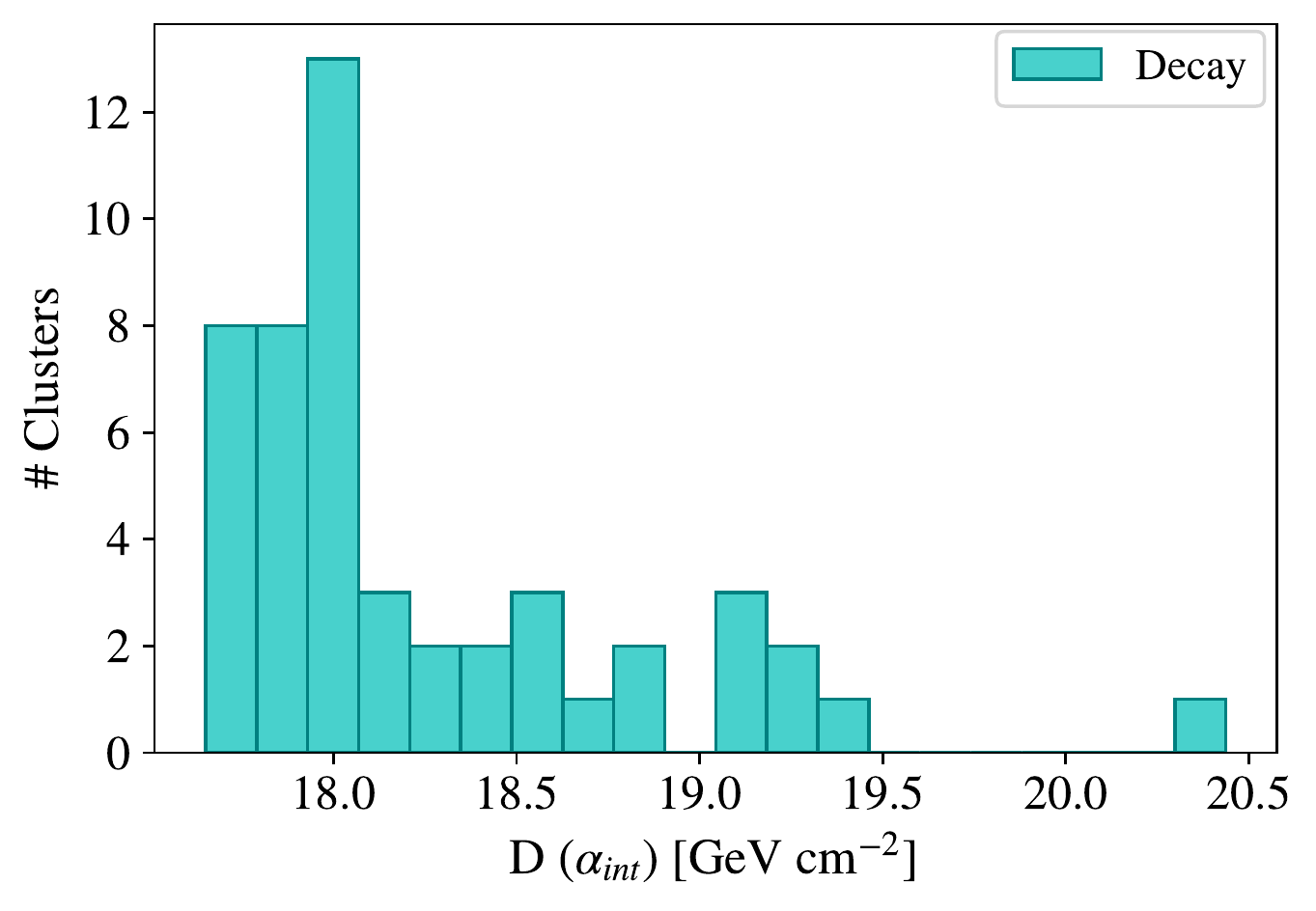}
\includegraphics[width=0.49\textwidth]{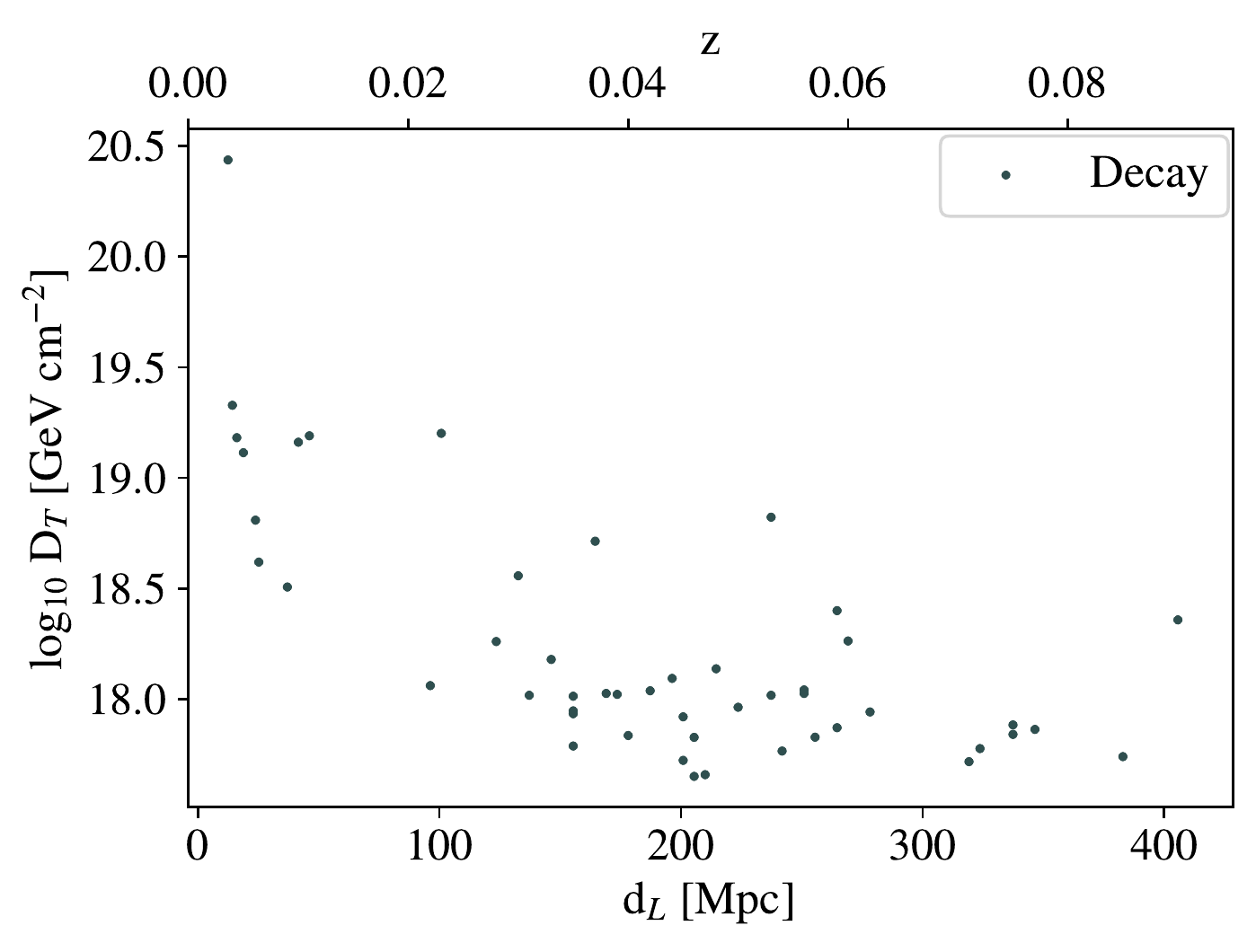}
\caption{\textbf{Left panels:} Distribution of the integrated $J$-factors (top) and $D$-factors (bottom) for all clusters in our sample assuming the MED model. \textbf{Right panels:} Integrated $J$ and $D$-factors versus their luminosity distance ($d_{L}$) and redshift ($z$). A detailed analysis on the dependence of the $J$- and $D$-factors with the distance is included in App.~\ref{app:JD_parametrization}.
}
\label{fig:JD-factors_sample}
\end{figure*}

We compute the $J$- and $D$-factors for all clusters in our sample and for the different benchmark models in Tab.~\ref{tab:dm_models} using the \texttt{CLUMPY} software \citep{Charbonnier:2012gf, Bonnivard:2015pia, Hutten:2018aix}. We summarize the results we obtain for both the integrated $J$- and $D$-factors ($J_{T}, D_{T}$) and subhalo boosts in Tabs.~\ref{tab:clusters_all_1}, \ref{tab:clusters_all_2}, \ref{tab:clusters_all_3} and \ref{tab:clusters_all_4}. Part of this information is depicted in Fig.~\ref{fig:JD-factors_sample} as well: the left panels show histograms of the $J$- and $D$-factors while the right panels also show their dependence with the distance to Earth. A detailed analysis on the latter is included in App.~\ref{app:JD_parametrization}, where we also provide useful parametrizations to compute $J$- and $D$-factors from the distance alone, that rely on our clusters sample.
We note that we do not use these parametrizations in this analysis because we have calculated the exact geometrical factor for each object instead. However, the relations between distance and the $J$ (or $D$) in App.~\ref{app:JD_parametrization} can be used elsewhere to estimate $J$ or $D$ by knowing only the cluster distance.

From the left panel in Fig.~\ref{fig:JD-factors_sample} we can already anticipate that the results of our combined data analysis in the next section are going to be dominated by a few clusters ($\sim 8$), i.e., those exhibiting the highest $J$- and $D$-factors, indeed far from the typical values of the rest of the sample. We can identify these clusters as Virgo (which gives the highest DM flux for any benchmark model and any scenario), NGC~4636, M49, A1060-Hydra, A1656-Coma, A3526-Centaurus, and NGC~1399-Fornax. According to Tabs.~\ref{tab:clusters_all_1}, \ref{tab:clusters_all_2}, \ref{tab:clusters_all_3} and \ref{tab:clusters_all_4}, these are also among the closest and largest in angular size ($d_L\lesssim$ 100 Mpc; see right panels of Fig.~\ref{fig:JD-factors_sample}). The only exception is Coma, which is slightly further but is one of the most massive clusters in the sample. 

Another information we can extract from the left panels of Fig.~\ref{fig:JD-factors_sample} is the enhancement of the $J$-factors due to the inclusion of halo substructure in the calculations. We can appreciate an enhancement of approximately one order of magnitude from MIN to MED, and almost another order of magnitude increase from MED to MAX. A quantitative description of the corresponding boost values is provided in Tabs.~\ref{tab:clusters_all_1}, \ref{tab:clusters_all_2}, \ref{tab:clusters_all_3} and \ref{tab:clusters_all_4}. The cluster with the highest boost is Virgo, reaching $B_{\rm MED}=12.30$ and $B_{\rm MAX}=74.80$. For the whole sample, we find mean boost values of $B_{\rm MED}=10.61$ and $B_{\rm MAX}=60.18$ for the MED and MAX scenarios, respectively. This can be compared with the expected boosts for objects in the same mass range according to \citep{Moline:2016pbm}: $B\approx9.0$ for $\alpha=1.9$ and $B\approx65.0$ for $\alpha=2.0$, both for $M_{200}=10^{14}$M$_{\odot}$. 
Despite the fact that their description of the subhalo population is not exactly the same as the one we adopt in this work, we can notice a clear correspondence between our MED model and the case of $\alpha=1.9$ in \citep{Moline:2016pbm}, as well as between the MAX model and their $\alpha=2.0$ case. We conclude that the obtained boosts are thus compatible with expectations.

We discuss other sources of uncertainties regarding the DM modelling of the clusters in our sample. The first one comes from the estimate of cluster masses as derived from X-rays surface brightness data, also known as hydrostatic masses. It is well known that different observational methods can yield different mass estimates. The deviation from the X-ray mass estimates is parametrized through the so-called hydrostatic bias. However, the clusters community has not yet reached an agreement on how to measure or precisely quantify the latter \citep{Piffaretti:2008ah, Salvati:2019zdp}. According to our main reference for the hydrostatic masses \citep{Schellenberger:2017wdw}, for low mass clusters ($M_{200}\lesssim 10^{14}$M$_{\odot}$), our hydrostatic masses may be underestimated by up to $\sim$20\% with respect to Sunyaev-Zeldovich (SZ) mass estimates \citep{Planck:2015koh}. For more massive clusters ($M_{200}\gtrsim 10^{14}$M$_{\odot}$), our hydrostatic masses can be overestimated by up to $\sim$50\% with respect to SZ mass estimates \citep{Planck:2015koh}. However we recall that an SZ mass estimate for all the clusters in our sample is not available. In comparison with dynamical mass estimates \citep{Zhang:2016rtu}, hydrostatic masses provide $\sim$97\% of agreement. The second major source of uncertainty at play comes from the intrinsic scatter of the concentration-mass relation that is adopted for the host halo, typically assumed to be of $\sim 0.14$ dex (e.g. \citep{Sanchez-CondeEtAl2014}). These uncertainties combined translate into $J$-factor values as $\sigma_J \approx 0.2$ dex\footnote{For the $D$-factors their impact is quantified as $\sigma_D\sim 10^{-3}$ dex.}. Yet, we note that, by considering three largely different benchmark models for the description of the subhalo population, we bracket a wider range of possible values for the annihilation fluxes, as can be seen in Tabs.~\ref{tab:clusters_all_1}, \ref{tab:clusters_all_2}, \ref{tab:clusters_all_3} and \ref{tab:clusters_all_4}. 

\begin{figure*}
\includegraphics[width=0.8\textwidth]{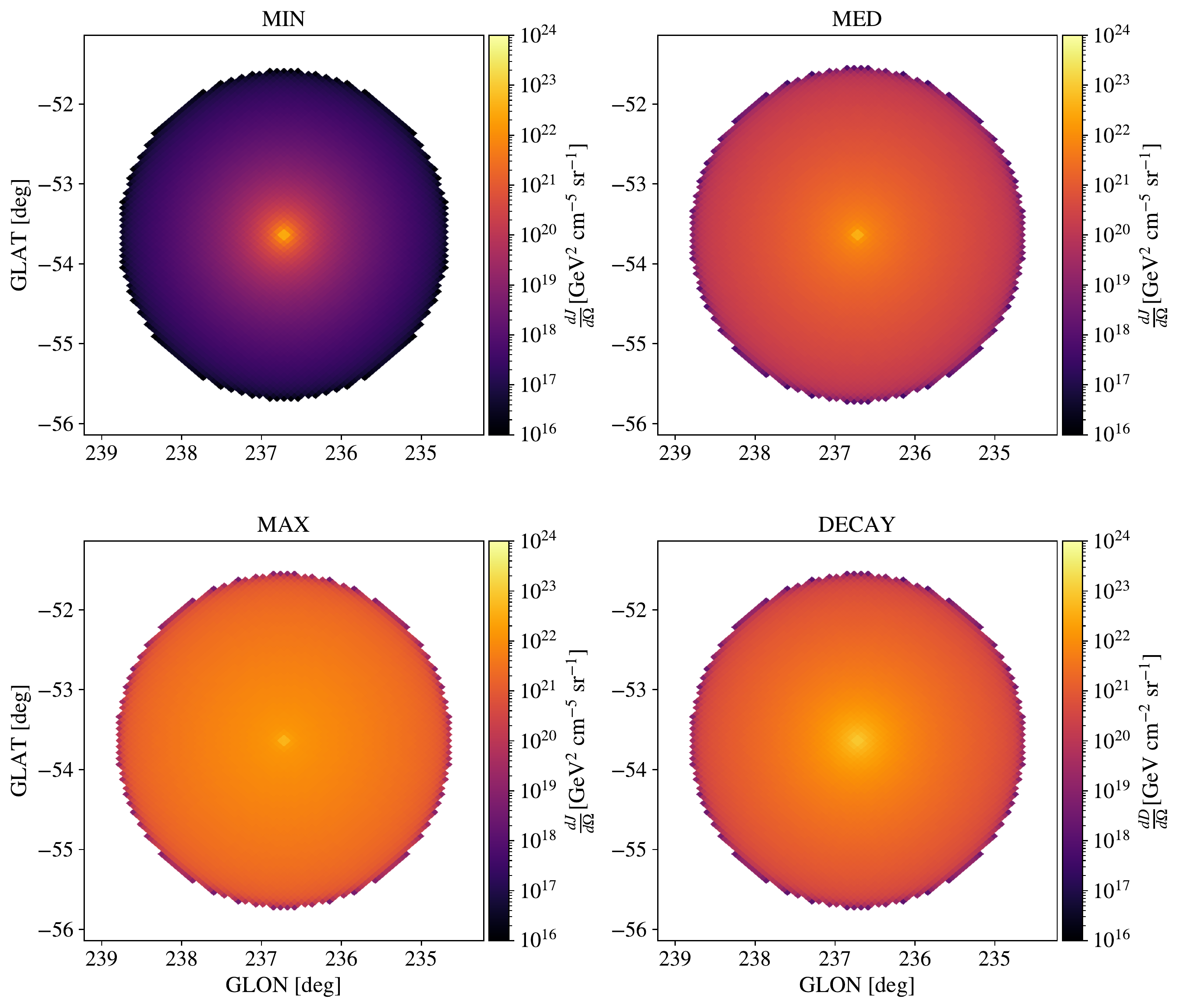}
\caption{Example of a two-dimensional spatial template of the expected DM emission in the Fornax cluster, for the cases of annihilation (top panels and left bottom panel, corresponding respectively to the MIN, MED, MAX substructure benchmark models; see Tab.~\ref{tab:dm_models}), and  decay (bottom right panel).}
\label{fig:clumpy_map}
\end{figure*}

The final output of our cluster DM modelling is a two-dimensional template containing both the level and spatial morphology of the expected DM signal in the cluster under consideration. Templates for all clusters in our sample are again obtained using the \texttt{CLUMPY} software. In total, we obtain $49\times4 = 196$ templates (49 clusters in the sample, 3 models for annihilation and one for decay). As an example, we show in Fig.~\ref{fig:clumpy_map} the four maps obtained for Fornax. As mentioned previously, in the case of DM annihilation the role of subhalos becomes more important in the outskirts of the cluster, while the central part is always dominated by the ``cusp'' of the main halo NFW profile. These maps constitute the input models for our LAT analysis presented in the next section, and will be used as reference models to be fitted to the data. 

\section{Analysis Framework}
\label{sec:analysis} 
In this section we provide the details of our selection of {\it Fermi}-LAT data, background models and analysis pipeline.

\subsection{Data selection}
\label{sec:analysisdata} 

We use 12 years of {\it Fermi}-LAT data from 2008 August 4 to 2020 August 2 passing standard data quality selection criteria\footnote{We select mission elapsed time (MET) starting at 239557417 and ending at 618050000. See the following webpage for the quality selection criteria \url{https://fermi.gsfc.nasa.gov/ssc/data/analysis/documentation/Cicerone/Cicerone_Data_Exploration/Data_preparation.html}}.
We choose an energy range from 500 MeV to 1 TeV with 8 energy bins for decade. 
The choice of the lower-end of the energy range is dictated by the extension in the sky of the clusters in our sample that can be at the degree level for several sources (see Tabs.~\ref{tab:clusters_all_1}, \ref{tab:clusters_all_2}, \ref{tab:clusters_all_3} and \ref{tab:clusters_all_4}). For such extended objects, the mismodeling of background components can affect the results of the analysis. In order to minimize this systematic, we decided to remove the very-low-energy {\it Fermi}-LAT data that have both a poor angular and energy resolution, and that are dominated by astrophysical background emission. Nevertheless, we also run the analysis with an energy range $0.1-1000$ GeV, finding similar results (see Sec.~\ref{sec:sys}).

We consider a region of interest (ROI) of $20\;\mathrm{deg} \times 20\;\mathrm{deg}$ centered at the source of interest for most of the clusters in our sample. We enlarge the ROI to $26\;\mathrm{deg} \times 26\;\mathrm{deg}$ for Virgo, that has a DM distribution spanning several degrees across the sky (see Tabs.~\ref{tab:clusters_all_1}, \ref{tab:clusters_all_2}, \ref{tab:clusters_all_3} and \ref{tab:clusters_all_4}).  
We select photon data belonging to the Pass~8 {\tt SOURCEVETO} event class and we employ the corresponding instrument response functions {\tt P8R3\_SOURCEVETO\_V2}.
The choice of {\tt SOURCEVETO} data is motivated by the fact that it has the same background rate as of the {\tt SOURCE} class, typically used for point-source analysis, up to 10 GeV and the same as the {\tt ULTRACLEANVETO}, usually used for diffuse emission analysis, above 50 GeV. However, {\tt SOURCEVETO} has 15\% more acceptance than {\tt ULTRACLEANVETO}\footnote{\url{https://fermi.gsfc.nasa.gov/ssc/data/analysis/documentation/Cicerone/Cicerone_Data/LAT_DP.html}.}.
The above mentioned characteristics make {\tt SOURCEVETO} data ideal to analyze extended sources such as clusters of galaxies.
In Sec.~\ref{sec:sys} we report the results obtained with {\tt SOURCE} and {\tt ULTRACLEANVETO} data and instrument response functions to demonstrate that our results are not affected by different data selections.
We also tested larger ROIs of $26\;\mathrm{deg} \times 26\;\mathrm{deg}$.
The data are binned using a pixel size of $0.08\;\mathrm{deg}$.

We apply the energy dispersion to all the components of our model using the method implemented in the {\tt Fermitools}\footnote{For a complete description see \url{https://fermi.gsfc.nasa.gov/ssc/data/analysis/documentation/Pass8_edisp_usage.html}.}.

\subsection{Background components}
\label{sec:background} 

{\it Fermi}-LAT data are fitted in our analysis using the following components: fluxes from individual sources, isotropic emission and Galactic interstellar emission (IEM) and flux from DM. The latter is modelled using the templates generated following Sec.~\ref{sec:DMdistr}.
Point-like and extended sources are taken from the 4FGL-DR2 {\it Fermi}-LAT catalog \citep{Fermi-LAT:2019yla}, i.e., the list of sources detected in 10 years of mission. 
We select from the catalog the sources that are in a region $24\;\mathrm{deg}\times24\;\mathrm{deg}$ centered at the position of the source of interest: we include also sources that are $2\;\mathrm{deg}$ outside our ROI since they can still contribute to the data selected in our analysis.

The choice of the IEM is central in modeling {\it Fermi}-LAT data when searching for very extended sources.
Therefore, we decide to use a IEM divided into different components to leave each of them more freedom in fitting the data.
In particular we use one template for the bremsstrahlung and one for the $\pi^0$ decay production. The inverse Compton scattering contribution is divided into the three interstellar radiation field components: cosmic microwave background (CMB), starlight and infrared. 
The IEM is thus divided into 5 templates each with the normalization and spectral index free to vary in the fit.

We adopt the templates used in Ref.~\citep{TheFermi-LAT:2017vmf}, which have been optimized to fit the data on the Galactic plane and can be used for the full sky but can be used to analyze all the directions of the sky.
We refer to Ref.~\citep{TheFermi-LAT:2017vmf} for all the details of these models and we summarize below the main characteristics.
The templates have been created with the {\tt Galprop} code\footnote{\url{http://galprop.stanford.edu}} \citep{1998ApJ...493..694M,2000ApJ...537..763S,2004ApJ...613..962S}, which calculates the propagation and interactions of CRs in the Galaxy by numerically solving the transport equations given a model for the CR source distribution, injection spectrum, and interaction targets.
We consider the baseline model that assumes a CR source distribution traced by the distribution of pulsars reported in Ref.~\citep{Lorimer:2006qs}.
The CR confinement volume has a height of 10 kpc and a radius of 20 kpc. This model assumes HI column densities derived from the 21-cm line intensities for a spin temperature of 150 K. 
The dust reddening map of Ref.~\citep{1998ApJ...500..525S} is used to correct the HI maps to account for the presence of dark neutral gas not traced by the combination of HI and CO surveys \citep{2012ApJ...750....3A}. 
Moreover, it includes the inverse Compton model reported in Ref.~\citep{Porter_2008} and divided into the starlight, infrared and CMB components. 

Finally, the model contains the Loop I, Sun, and Moon emissions merged into a unique template, and the {\it Fermi} bubbles template, which includes both low-latitude and high latitude components. Summarizing, our model counts a total of 8 components: 5 IEM components, the isotropic emission component, the Fermi Bubbles template, and Loop I+Sun+Moon component.
In addition to this model we also derive the results carried out considering the IEM model labeled as {\tt Yusifov}, which is generated using the pulsar distribution reported in Ref.~\citep{2004A&A...422..545Y}.

We are not using the official IEM model and isotropic template released together with the 4FGL-DR2 catalog, called {\tt gll\_iem\_v07.fits}\footnote{\url{https://fermi.gsfc.nasa.gov/ssc/data/access/lat/BackgroundModels.html}}, because this model has patches added to absorb residuals present at several positions in the sky.
These patches can absorb the large scale emission possibly present in and around galaxy clusters.

\subsection{Analysis technique}
\label{sec:analysistec} 

Our analysis pipeline is entirely based on {\tt FermiPy}, which is widely used in the scientific community to perform analysis of {\it Fermi}-LAT data. 
{\tt FermiPy} is a Python package that automates high-level analyses with the {\tt Fermitools} \citep{2017ICRC...35..824W}\footnote{See \url{http://fermipy.readthedocs.io/en/latest/}.}.
We use versions {\tt 1.0.1} of {\tt Fermipy} and {\tt 1.2.3} of the {\tt Fermitools}.

The analysis applied in this paper is very similar to the ones recently used for the search for a DM signal from other astrophysical targets such as the Galactic center \citep{DiMauro:2021raz}, M31 and M33 \citep{DiMauro:2019frs}, dSphs \citep{DiMauro:2021qcf} and dwarf irregular galaxies \citep{2021PhRvD.104h3026G}.
We describe below the main steps of the analysis and we refer to the previously cited papers for further details.

\begin{itemize}
    \item We first perform a fit to the ROI using the background components reported in Sec.~\ref{sec:background} and the DM template generated using the model explained in Sec.~\ref{sec:DMdistr}. The spectral energy distribution (SED) parameters of all the sources in the ROI, the normalization and spectral index of each IEM components and the normalization of the isotropic template are left free in the fit.
    \item The sources that are detected with a Test Statistic ($TS$) lower than 25 are removed from the model of the ROI\footnote{The Test Statistic ($TS$) is defined as twice the difference in maximum log-likelihood between the null hypothesis (i.e., no source present) and the test hypothesis: $TS = 2 ( \log\mathcal{L}_{\rm test} -\log\mathcal{L}_{\rm null} )$ \citep{1996ApJ...461..396M}.}. A $TS$ of 25 corresponds roughly to a detection at $5\sigma$ significance for a source modeled with two free parameters. We choose to remove faint sources with significance $<5\sigma$ from the model because this is the usual cut in significance above which sources are included in {\it Fermi}-LAT catalogs (see, e.g., \cite{Fermi-LAT:2019yla}). 
    \item We perform the search for new sources with a $TS>25$ since we are using 12 years of data but we select sources from the 4FGL-DR2 catalog, which was obtained with 10 years of data. For this scope we use the tool {\tt find\_sources} implemented in Fermipy. The tool  generates a $TS$ map in each pixel of the ROI and searches for values larger than 25. If any are found, {\tt FermiPy} adds a source at the pixel of the $TS$ peak with a power-law SED shape and then fits the source free parameter to the data. Then, a new fit to the entire ROI is performed with all the new sources included in the model. The new sources found with the analysis have typically $TS$ values between 25 and 40. No new sources have been added within a few degrees of angular distance from the cluster position so this step of the analysis does not compromise the search of a possible DM signal.
    \item We compute the SED for the DM template by providing for each energy bin the likelihood as a function of the energy flux. In each energy bin, the only free parameter is the normalization, which is computed independently for each bin. This approach permits to test a variety of DM channels with theoretically different $\gamma$-ray spectral shapes.
    \item We compute the logarithm of the likelihood as a function of DM mass and annihilation cross-section (or decay time) $\mathrm{log}(\mathcal{L}_{i,j}\left(\mu,\theta_{i,j}|\mathcal{D}_{i,j}\right))$, where $i$ runs over the targets list and $j$ is the index of each energy bin of the \textit{Fermi}-LAT data ($\mathcal{D}$), $\mu$ are the DM parameters ($\langle\sigma v\rangle$ (or decay time $\tau$) and $m_\chi$), and $\theta$ are the parameters in the background model, i.e., the nuisance parameters. In this last part of the anaylysis we assume a specific annihilation or decay channel.
    \item We combine the results for the individual clusters by summing together the likelihood profiles independently for each energy bin:
    \begin{equation}
    \mathrm{log}(\mathcal{L}_j\left(\mu,\theta_j|\mathcal{D}_j\right))=\sum_i \mathrm{log}(\mathcal{L}_{i,j}\left(\mu,\theta_{i,j}|\mathcal{D}_{i,j}\right)), 
    \end{equation}
    where $\mathrm{log}(\mathcal{L}_j\left(\mu,\theta_j|\mathcal{D}_j\right))$ represents the likelihood profile for a specific DM annihilation (decay) channel as a function of the DM mass and annihilation cross-section (decay time) obtained for the combined analysis of all clusters.
    
\end{itemize}

Most of the results are shown for the two annihilation or decay channels: $b\bar{b}$ and $\tau^+\tau^-$. However, for some cases, we report the results obtained with other channels. 
We select WIMP masses ranging from 5 GeV up to 10 TeV.
We include the statistical uncertainty on the $J$-factor by adding an additional likelihood term to the binned Poisson likelihood for the LAT data:
\begin{eqnarray}
    \mathcal{L}_i&&\left(J_i|J_{\rm{obs},i},\sigma_i\right)= \frac{1}{\mathrm{log}(10)J_{\rm{obs},i} \sqrt{2\pi}\sigma_i} \times \nonumber \\ 
    &\times& \exp{\left[- \left( \frac{\mathrm{log}_{10}(J_i) - \mathrm{log}_{10}(J_{\rm{obs},i})}{\sqrt{2}\sigma_i} \right)^2\right]},
\end{eqnarray}
where $J_{\rm{obs},i}$ is the best fit for the observed $J/D$-factor for the $i$-th cluster while $\sigma_i$ is the error in $\mathrm{log}_{10}(J_{\rm{obs},i})$ space; $J_i$ is the value of the $J/D$-factor for which the likelihood is calculated. This term of $\mathcal{L}$ disfavors values of $J_i$ very different from values in the range $[J_{\rm{obs},i} - \sigma_i,J_{\rm{obs},i} + \sigma_i]$ weighting the difference with the error $\sigma_i$. 
For all the clusters in our sample, we decide to set $\sigma_i$ to $\sigma_J=0.2$ dex namely the error on $\mathrm{log}_{10}(J)$ according to the estimation discussed in the previous section.
However, we also test the case with an error equal to 0 or 0.4. As we will show in Sec.~\ref{sec:sys}, the larger the value of $\sigma_J$ is the larger is the $TS$ of the signal. However, signal significance as well as the best-fit values for the DM parameters (mass and annihilation cross-section) are not affected by the value of $\sigma_J$.
We also stress that the uncertainty on individual $J$-factor values is subdominant compared to the one coming from the use of different DM substructure models for each cluster. 

The significance of the DM hypothesis can be evaluated comparing the likelihood obtained when the DM template is added into the model, i.e., test hypothesis, and when it is not, i.e., null hypothesis.
The likelihood for the test hypothesis is calculated as:
\begin{equation}
    \mathrm{log}(\mathcal{L}\left(\mu \right))=\sum_j \mathrm{log}(\mathcal{L}_{j}\left(\mu,\theta_{j}|\mathcal{D}_{j}\right)).
\end{equation}
As a result the $TS$ value for the DM emission is calculated as:
\begin{equation}
\label{eq:TS}
TS=2~\Delta\mathrm{log}(\mathcal{L})=2~\mathrm{log}\left[\frac{\mathcal{L}\left(\mu\right)}{\mathcal{L}_{\rm{null}}}\right],
\end{equation}
where $\mathcal{L}_{\rm{null}}$ is the likelihood in the case of null hypothesis, i.e., no DM, and $\mathcal{L}$ is the likelihood for the DM hypothesis.\\
Assuming that the position of the cluster is fixed in the analysis, the DM template has two free parameters: the DM mass $m_{\chi}$ and the annihilation cross-section $\langle \sigma v \rangle$ or the mean particle lifetime $\tau_{\chi}$, for DM decay.
Based on the asymptotic theorem of Chernoff \citep{10.1214/aoms/1177728725}, the $TS$ can be converted to a significance of the signal based on a mixture of $\chi^2$ distributions.
In particular we can assume that the $TS$ distribution follows the $\chi^2$ distribution for two degrees of freedom divided by 2. This brings to a relation between the $TS$ and the significance of $TS\sim\sigma^2$. 
Therefore, a discovery ($5\sigma$) would be given by $TS\sim25$. 
However, as we will demonstrate in Sec.~\ref{sec:null}, the $TS$ distribution in our analysis deviates significantly from the asymptotic expectation, i.e., from the $\chi_2^2/2$ distribution. 

This is due to the fact that clusters can be
very extended, typically a few degrees across the sky (see Tab.~\ref{tab:clusters_all_1}, \ref{tab:clusters_all_2}, \ref{tab:clusters_all_3} and \ref{tab:clusters_all_4}), and that at such large scales there are a lot of unmodelled components, i.e., residuals in the data that do not follow the Poissonian statistic. These residuals yield $TS$ values for the detection of extended sources in random directions that are much larger than the expected one from a $\chi_2^2/2$ distribution. Therefore, the null hypothesis will have a much broader distribution, with large tails and, as a result, the $5\sigma$ significance will be found for a $TS$ value much larger than 25.
 

\section{Results}
\label{sec:results}

\subsection{Combined search}
\label{sec:TS}

The first result we obtain with our analysis is the individual $TS$ for a DM signal in each cluster of our sample.
We focus our results on the $b\bar{b}$ and $\tau^+\tau-$ annihilation and decay channels.

\begin{figure*}[t!]
\includegraphics[width=0.49\textwidth]{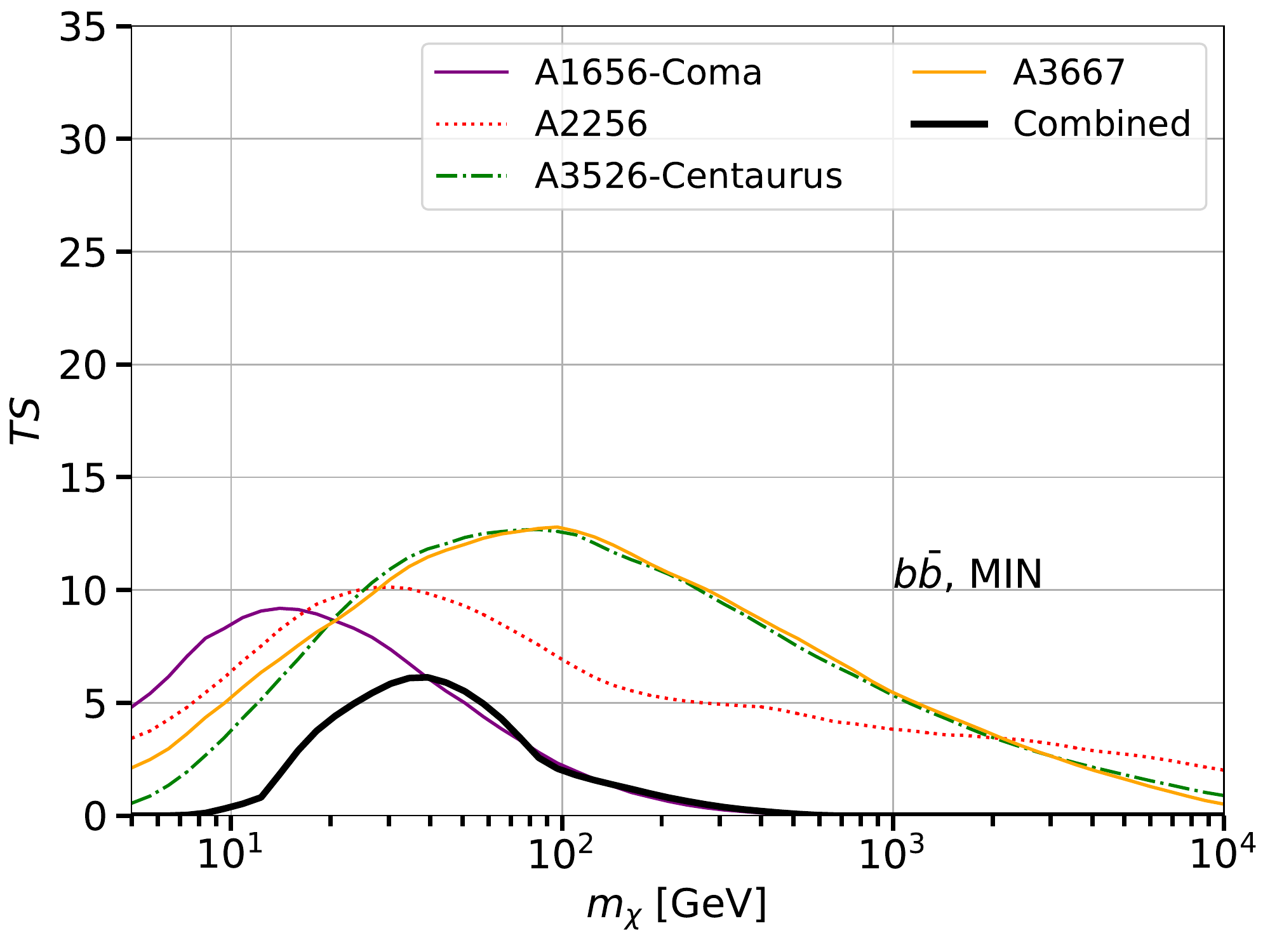}
\includegraphics[width=0.49\textwidth]{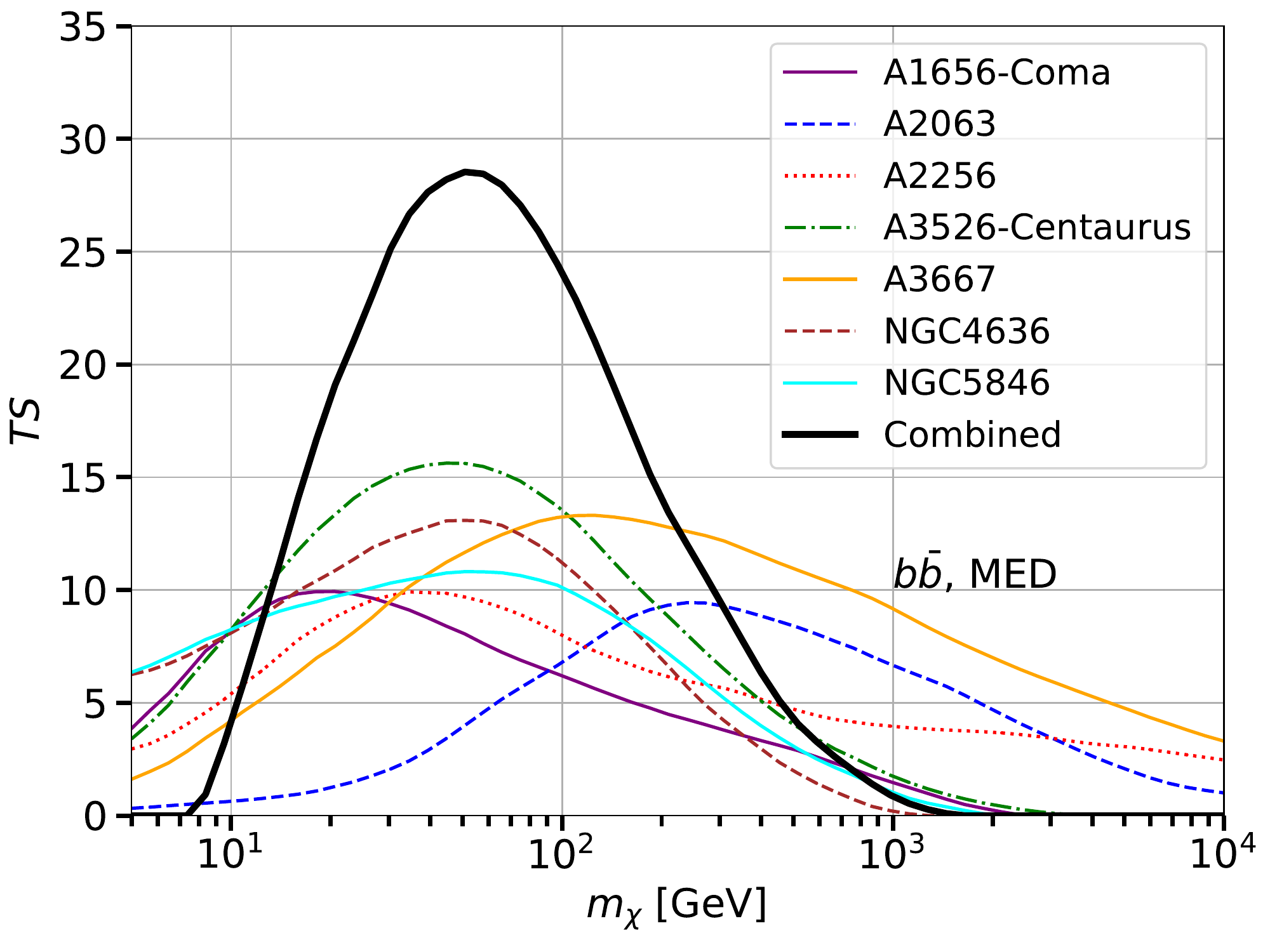}
\includegraphics[width=0.49\textwidth]{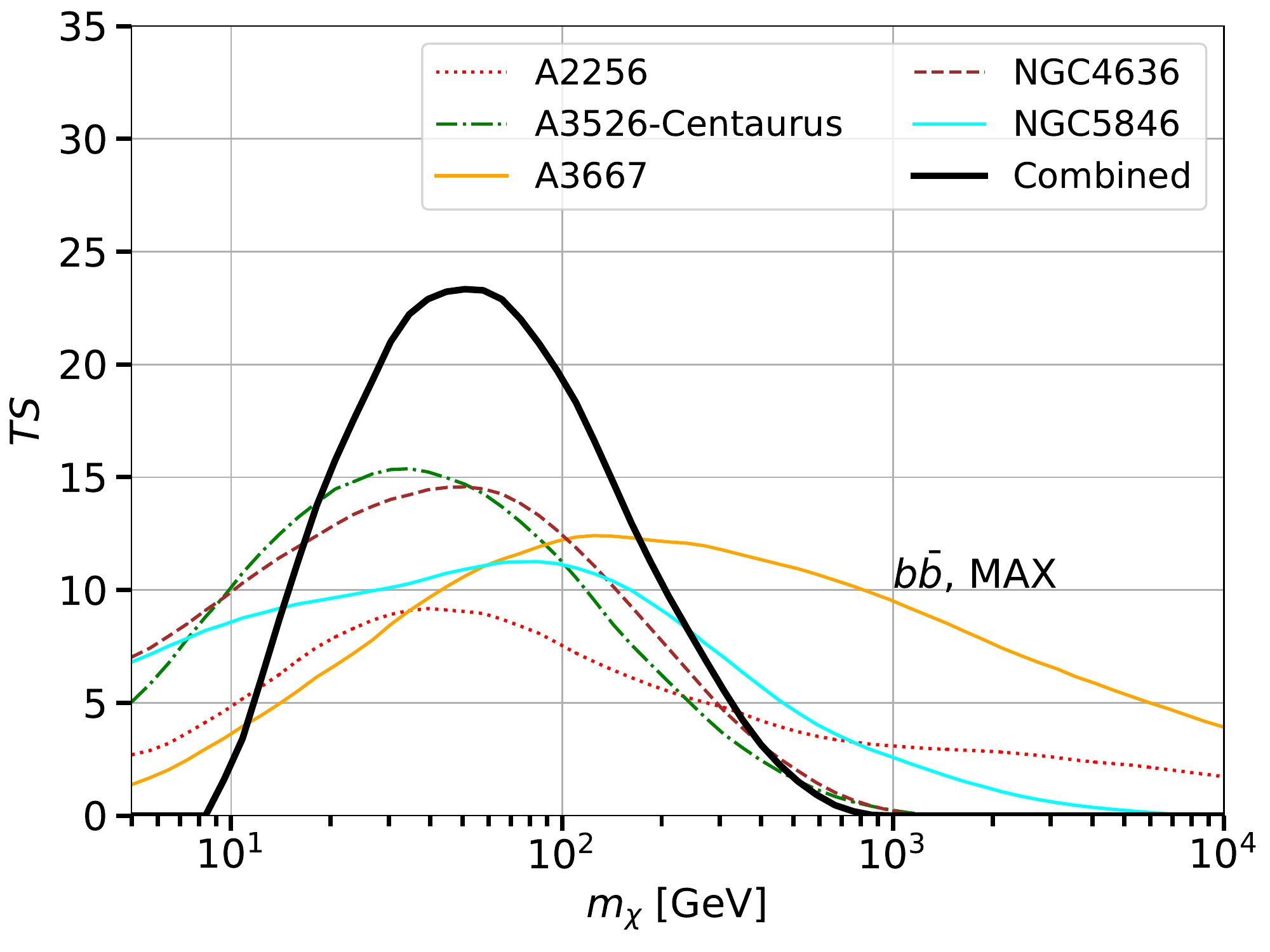}
\includegraphics[width=0.49\textwidth]{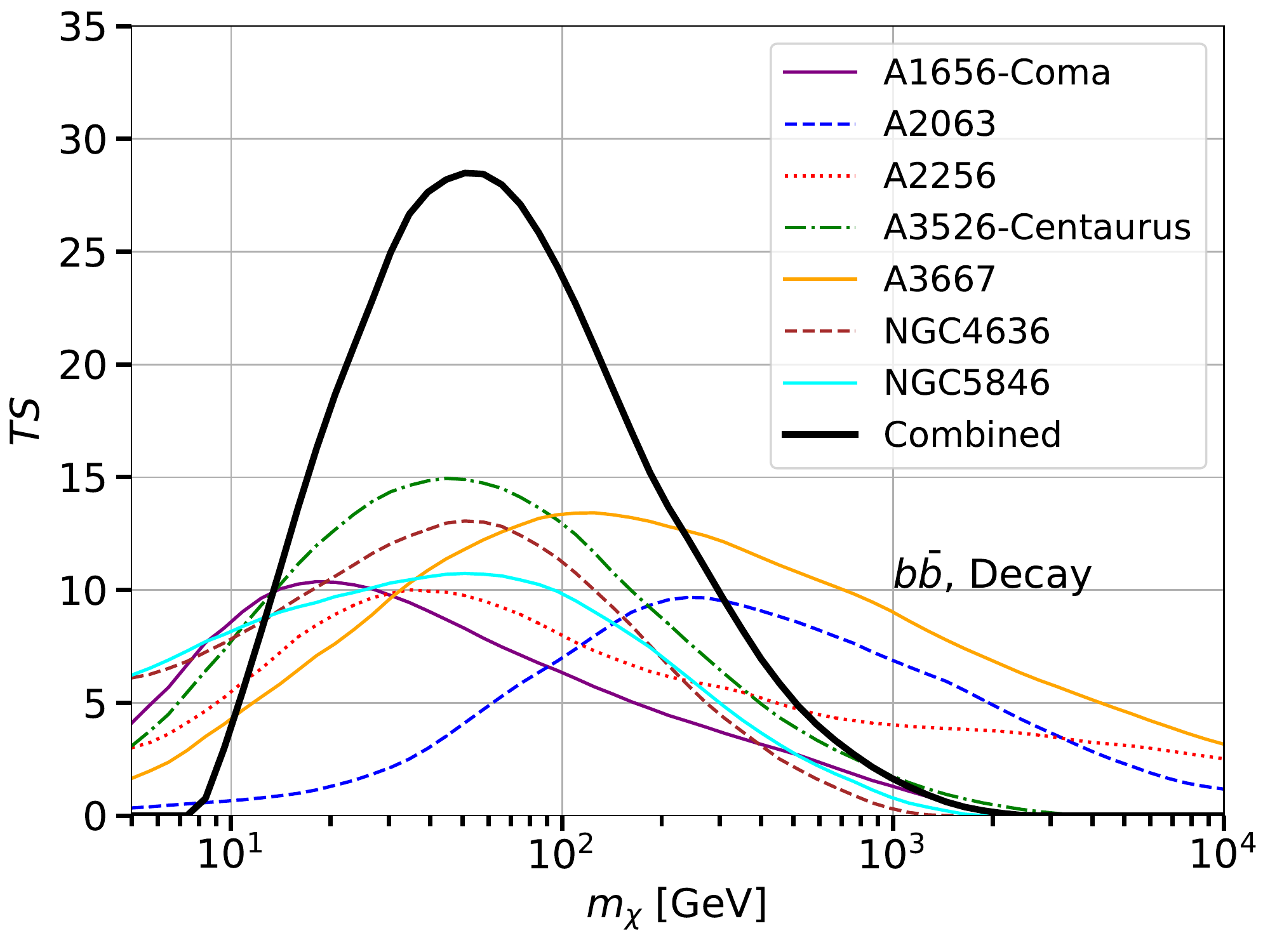}
\caption{$TS$ as a function of the DM mass ($m_{\chi}$) obtained from the analysis of individual clusters. We show only the objects for which we obtain $TS>9$. We show, from top left to bottom right, the case for the MIN, MED, MAX models and annihilating DM and for decaying DM with the $b\bar{b}$ channel. We also report the result we obtain for the combined analysis of all the sources.
}
\label{fig:tscluster}
\end{figure*}

\begin{figure*}[t!]
\includegraphics[width=0.49\textwidth]{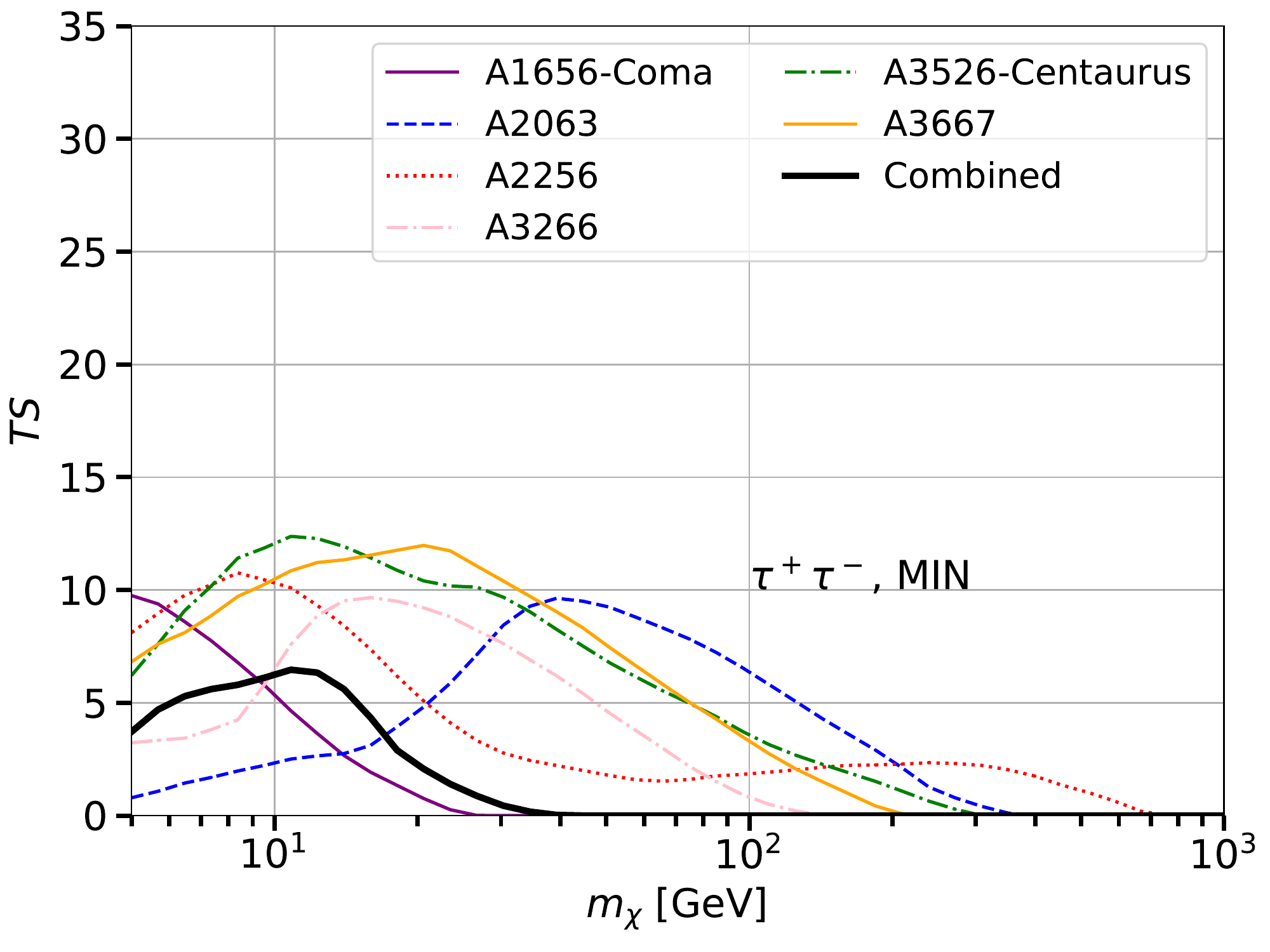}
\includegraphics[width=0.49\textwidth]{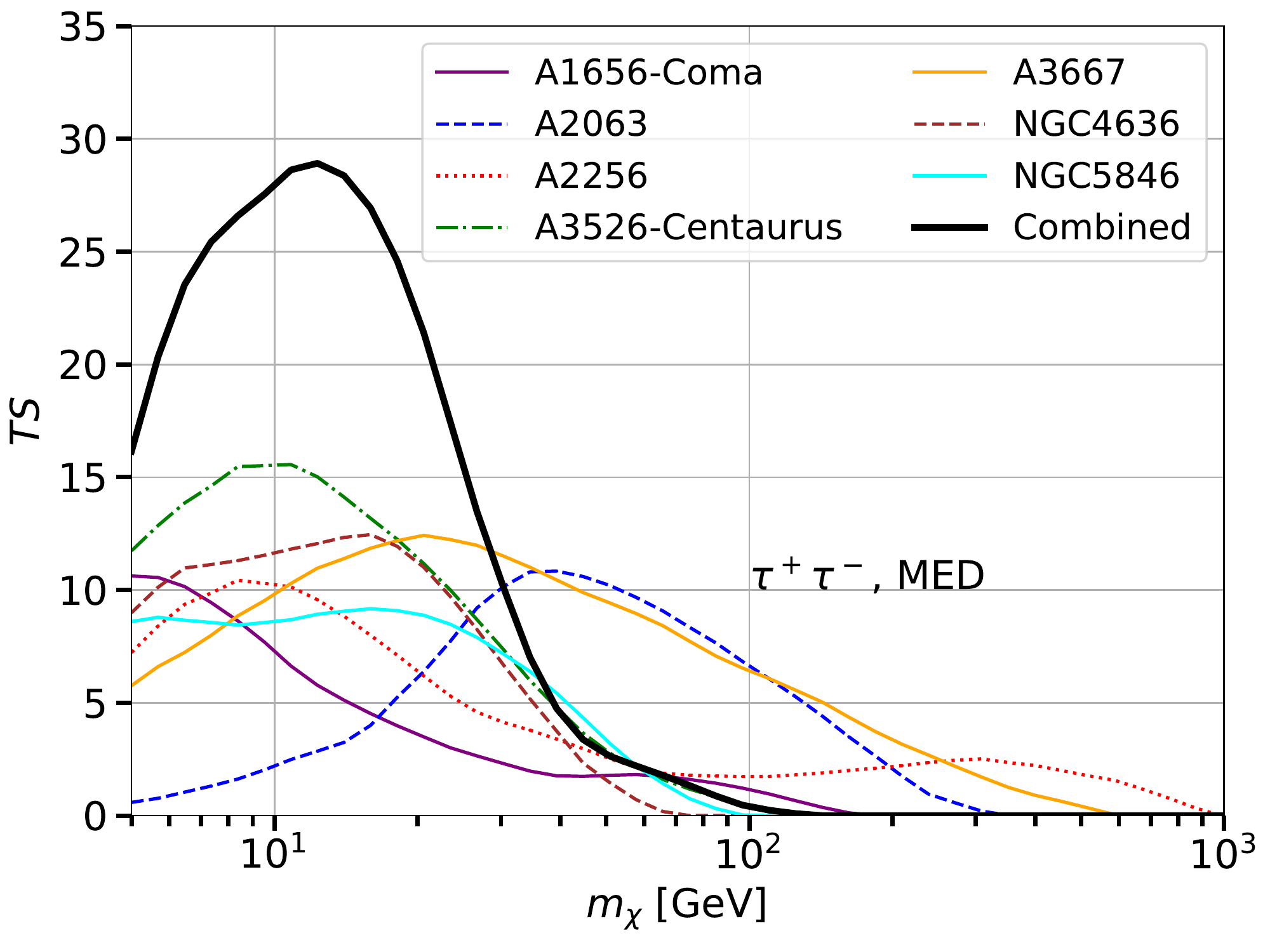}
\includegraphics[width=0.49\textwidth]{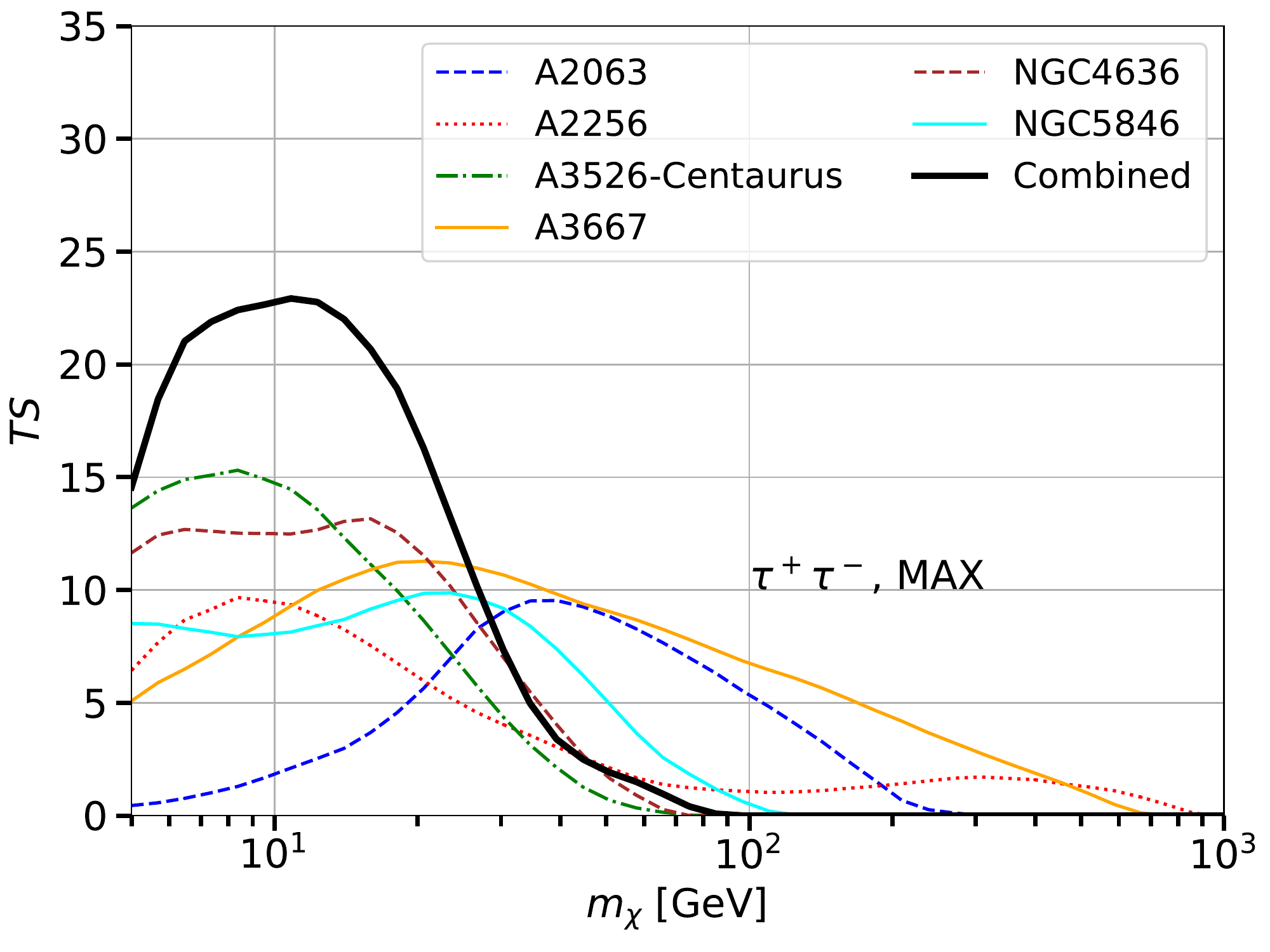}
\includegraphics[width=0.49\textwidth]{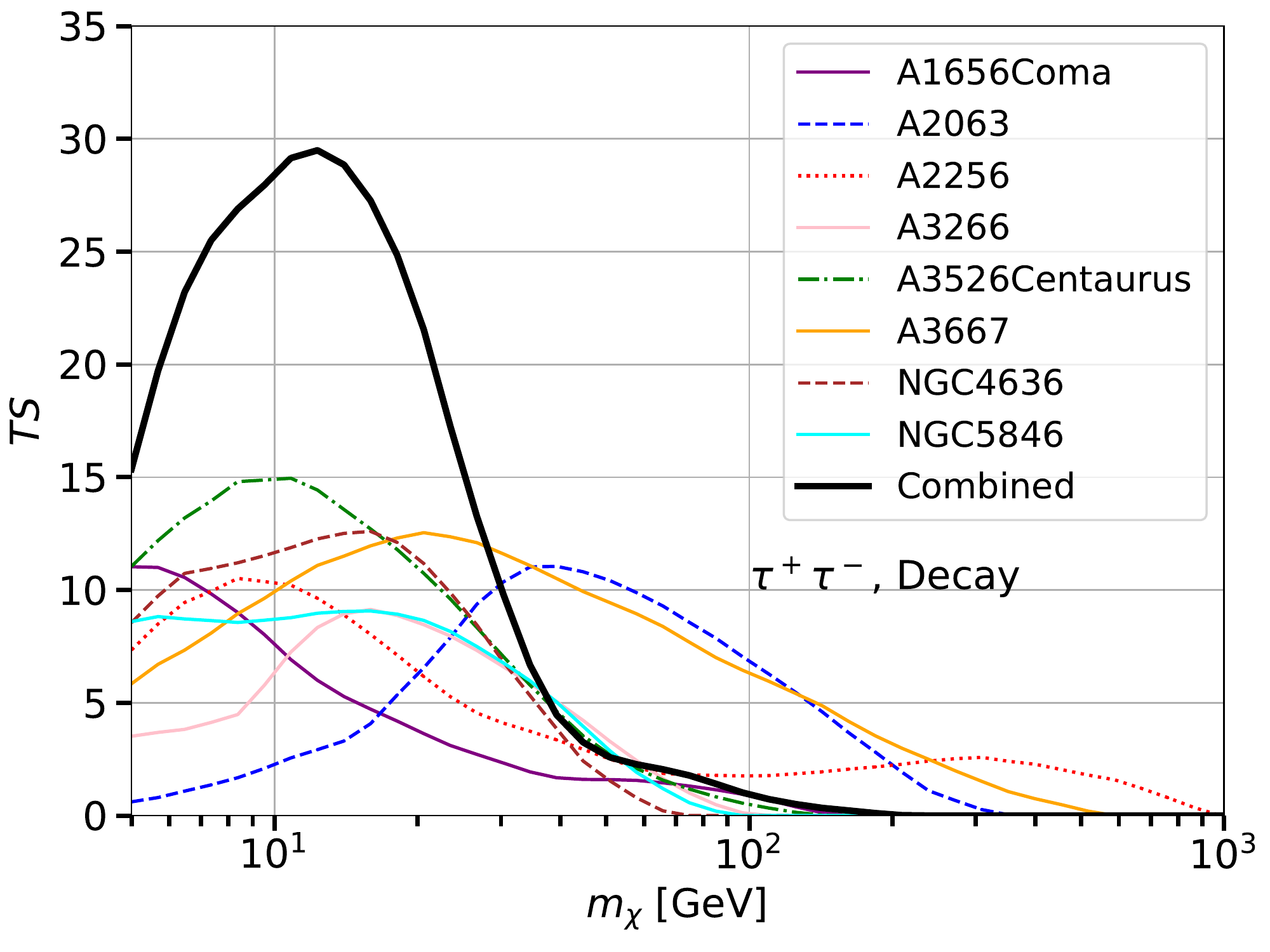}
\caption{Same as figure Fig.~\ref{fig:tscluster} for the $\tau^+\tau^-$ annihilation and decay channel.}
\label{fig:tscluster_tau}
\end{figure*}

We report in Tabs.~\ref{tab:clusters_all_1}, \ref{tab:clusters_all_2}, \ref{tab:clusters_all_3} and \ref{tab:clusters_all_4}, the $TS$ we obtain for each source assuming the $b\bar{b}$ annihilation channel and the MED model. In Fig.~\ref{fig:tscluster} and \ref{fig:tscluster_tau} we show the $TS$ as a function of the DM mass ($m_{\chi}$) for the clusters detected with the highest significance.
In the two figures we show the cases with the MIN, MED and MAX DM models and for the decaying DM hypothesis.
The objects for which we find the highest $TS$ values are: A3526-Centaurus, A1656-Coma, NGC~5846, NGC~4636, A2256, A3667.
Yet, the highest $TS$ is 15 (obtained for A3526-Centaurus in the MED model), which is much smaller than the value of 25 typically used to include a source in {\it Fermi}-LAT catalogs.
Therefore, we do not detect individual clusters in {\it Fermi}-LAT data.
If we reduce the lower-end of the energy range considered in our analysis to 100 MeV, the $TS$ of A3526-Centaurus increases to 34, with a best fit for the DM mass between 10-30 GeV and for the annihilation cross-section of $2\times 10^{-26}$ cm$^3$/s. However, even this value is well below the $5\sigma$ significance once the actual $TS$ distribution is considered, as we show in Sec.~\ref{sec:null}.

Refs.~\citep{Xi:2017uzz,AdamEtAl2021,Baghmanyan:2021jwg} reported a detection of an extended emission from the Coma cluster at the level of $TS\sim 20-50$.
These papers have assumed a specific model for the hadronic emission of photons or simple geometrical extended templates such as a uniform disk.
Instead, in Ref.~\citep{Fermi-LAT:2015rbk} the {\it Fermi}-LAT Collaboration has used a cored profile, which is motivated by observations of the radio halo in Coma, as well as a point source and a disk template.
The maximum $TS$ they obtain is around 13 which corresponds to a global significance of about $1.8\sigma$, after correcting for trial factors.
Our results for this cluster are compatible with the one of Ref.~\citep{Fermi-LAT:2015rbk}. Indeed, we find a maximum $TS\sim 13$ when we analyze the data above 100 MeV, and 10 when we analyze the data above 500 MeV, both for the MED model. 

\begin{figure}
\includegraphics[width=0.49\textwidth]{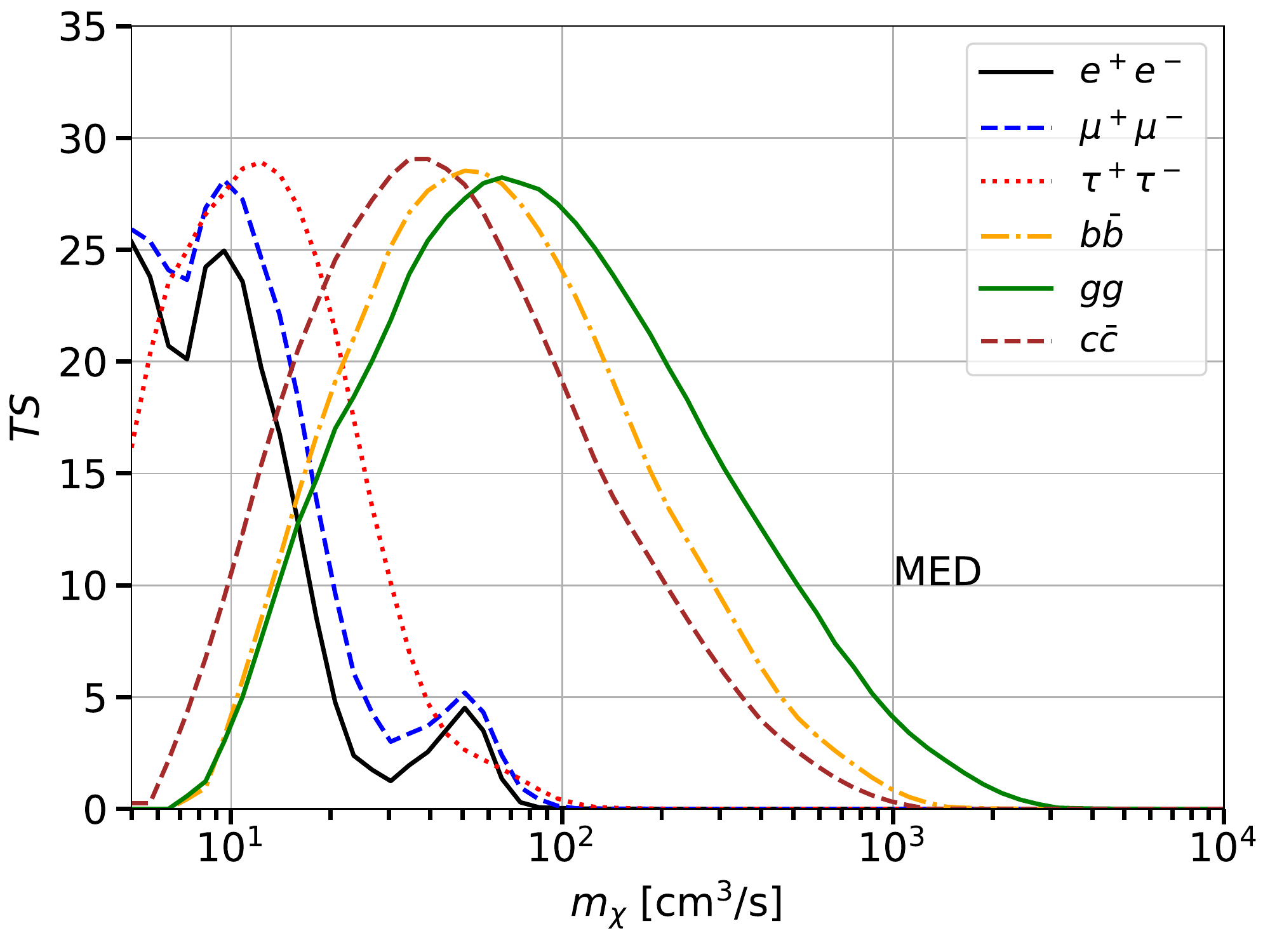}
\caption{$TS$ as a function of $m_{\chi}$ obtained in the combined analysis for different annihilation channels.}
\label{fig:tschannels}
\end{figure}

We also report in Figs.~\ref{fig:tscluster} and \ref{fig:tscluster_tau} the result obtained for the combined analysis of the entire sample. The $TS$ is at most of the order of 5 for the MIN model, 27 for the MED and 23 for the MAX DM models. For decaying DM, the TS reaches a value of about 28.
The best fit masses found for $b\bar{b}$ are between 40-70 GeV while for the $\tau^+\tau^-$ channel are between 8-20 GeV.
Interestingly, the value of the $TS$ at the peak of the distribution is quite different using the MIN and MED/MAX models. This points in favor of a signal that is more compatible with a DM distribution that includes substructures and that, as a result, is more extended than the one of the MIN case.
In Fig.~\ref{fig:tschannels} we report the $TS$ as a function of $m_{\chi}$ for other leptonic and hadronic channels.
We do not consider the $W^{\pm}$, $Z$ gauge bosons, the Higgs scalar boson and the $t$ quark because for these channels the DM mass is forced to be larger than the bosons or the $t$ quark mass. For this range of DM masses the $TS$ of the signal is much smaller than the results we obtain for lower DM mass values.
In addition to this, the source spectra of $\gamma$ rays produced from these channels are similar to the ones obtained with the $b$ quark (see, e.g., \cite{Cirelli:2010xx}).
The best-fit DM masses for the $e^+e^-$ and $\mu^+\mu^-$ annihilation channels are at about 10 GeV, i.e., lower than what obtained for $\tau^+\tau^-$. Yet, we note that we are not including the secondary production of $\gamma$ rays from $e^{\pm}$ produced by DM that subsequently inverse Compton scatter against the intracluster low-energy photon fields or photons produced by bremsstrahlung on intracluster atoms. These additional secondary contributions increase the overall flux and as a consequence reduce the best-fit values obtained for $\langle \sigma v \rangle$.

\begin{figure*}[t!]
\includegraphics[width=0.49\textwidth]{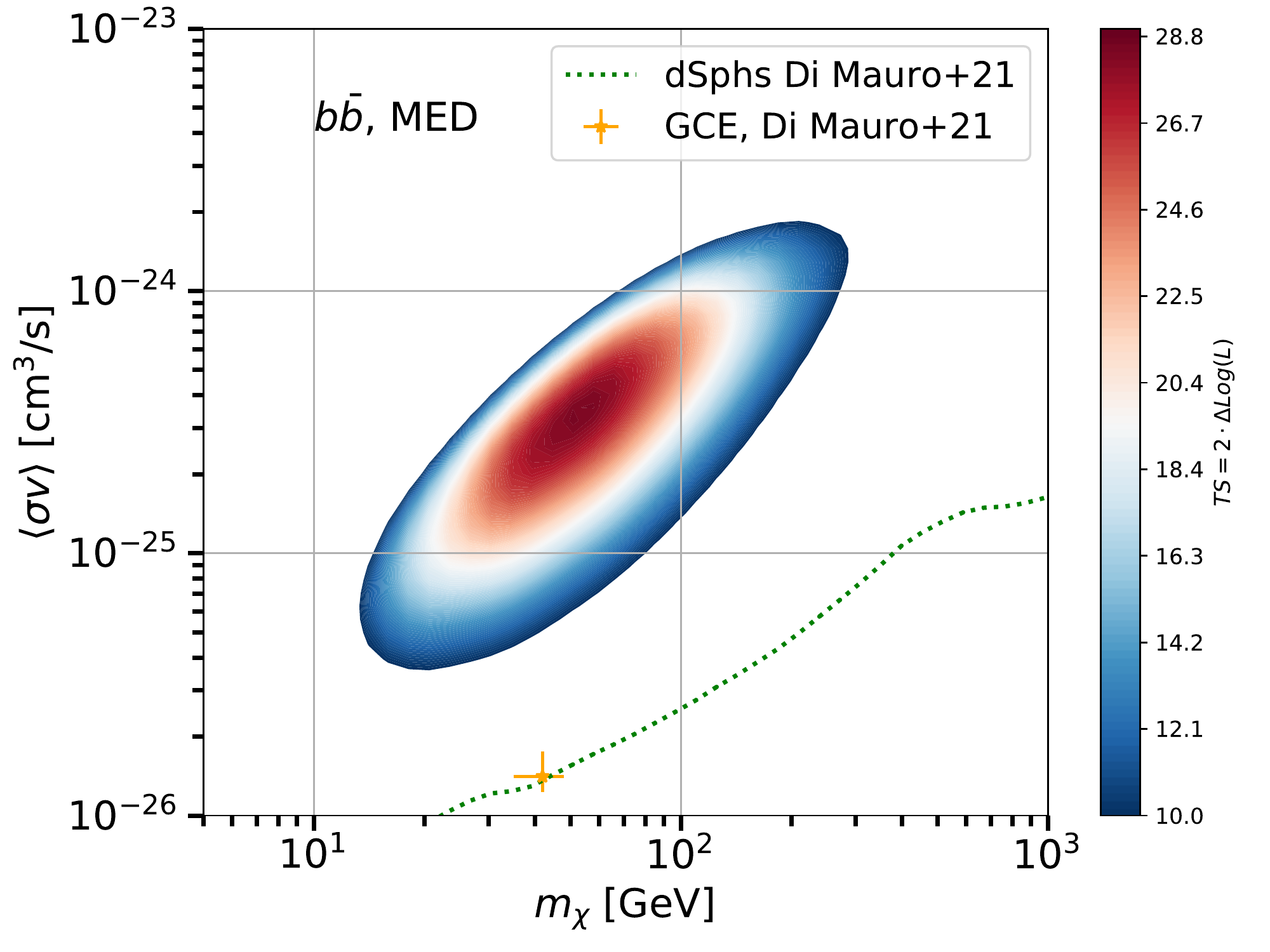}
\includegraphics[width=0.49\textwidth]{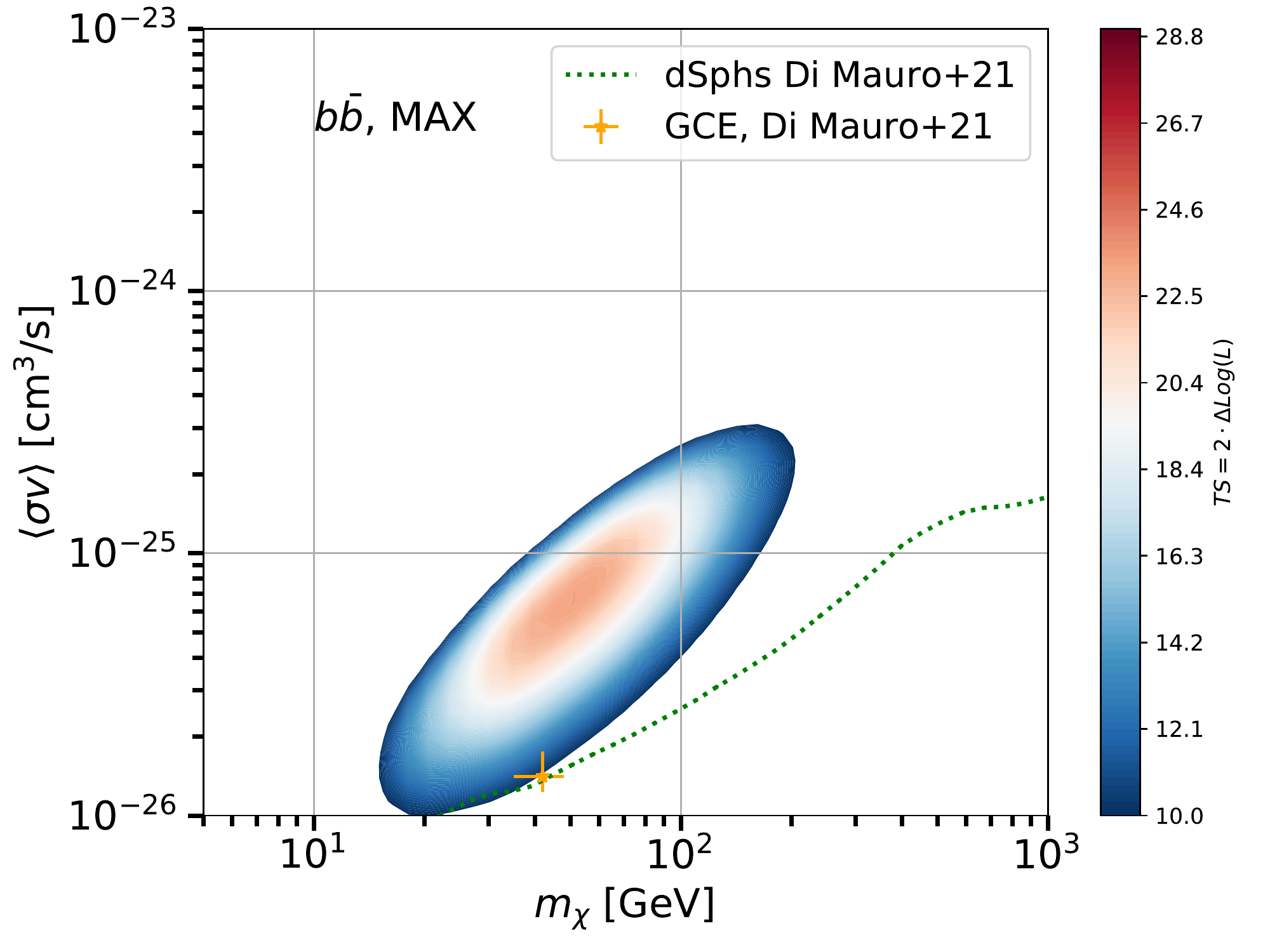}
\includegraphics[width=0.49\textwidth]{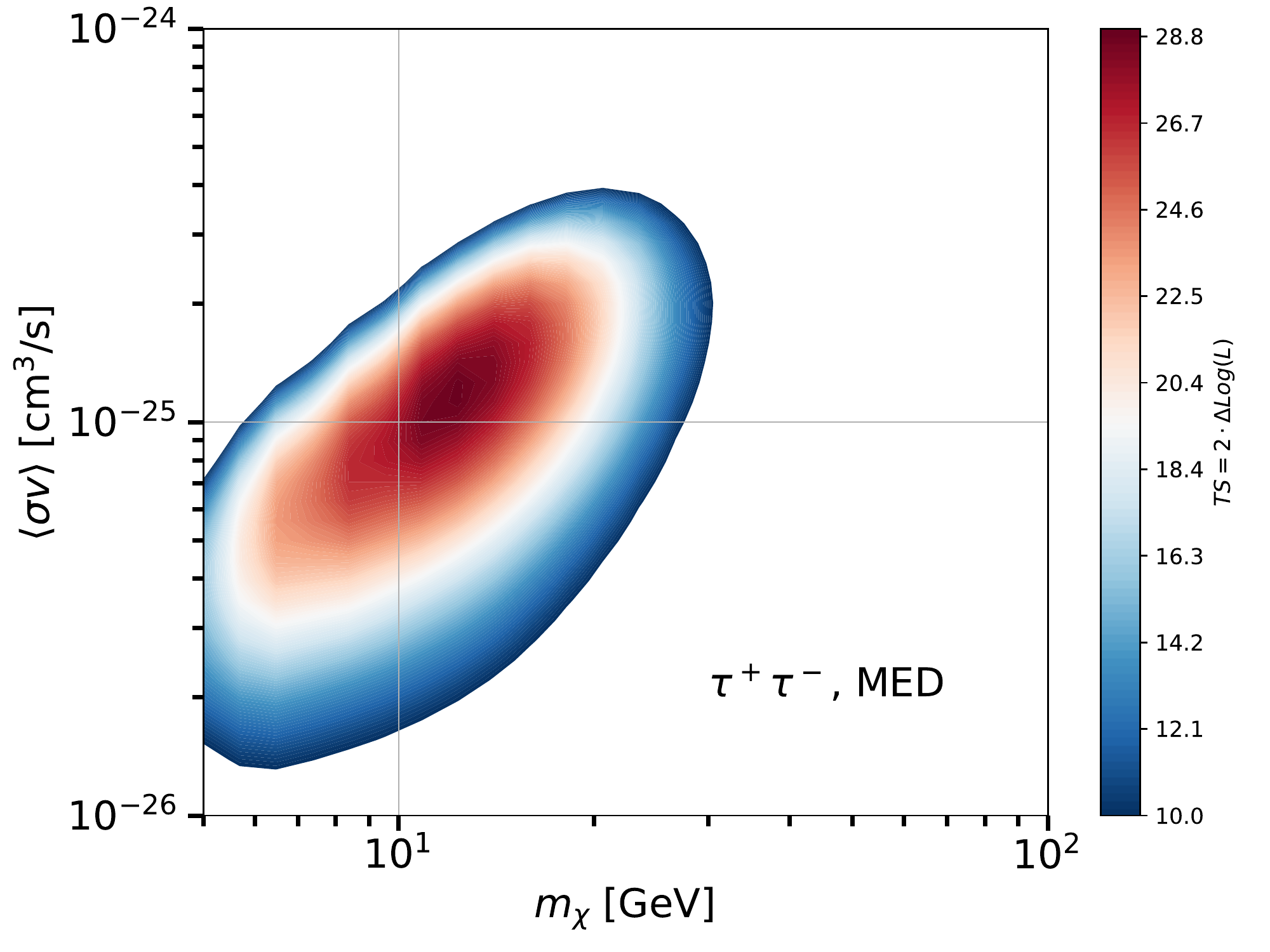}
\includegraphics[width=0.49\textwidth]{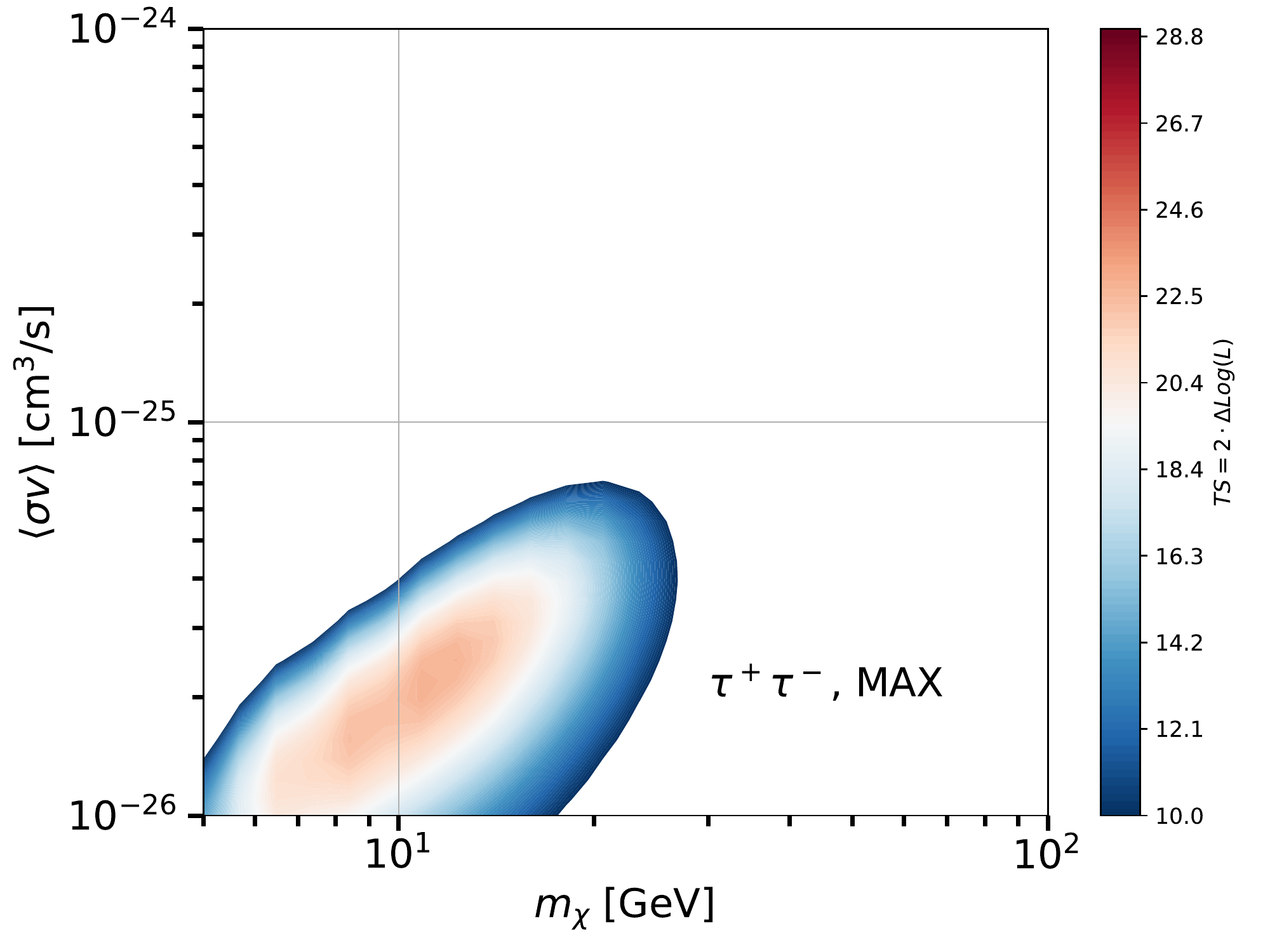}
\caption{Contour plot for the $TS$ as a function of the DM mass and annihilation cross-section obtained with the combined analysis and the $b\bar{b}$ (top panels) and $\tau^+\tau^-$ (bottom panels) annihilation channels. We display the result obtained for the MED (left panels) and MAX (right panels) DM models. In case of the $b\bar{b}$ annihilation channel we also report the best-fit obtained in \citep{DiMauro:2021raz} for the fit to the Galactic center excess and the upper limits from a sample of dSphs.}
\label{fig:contourcluster}
\end{figure*}

\begin{figure*}[t!]
\includegraphics[width=0.49\textwidth]{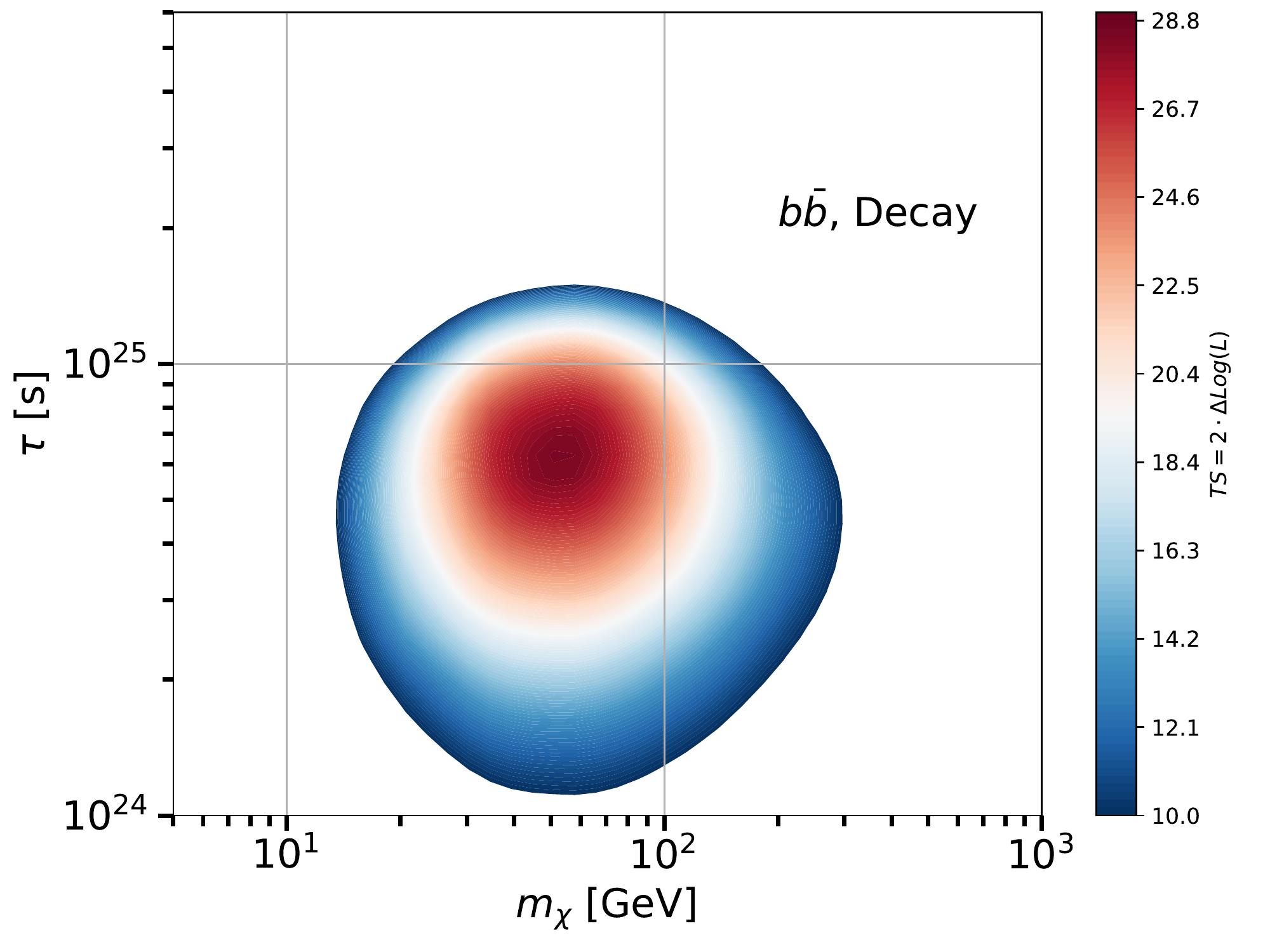}
\includegraphics[width=0.49\textwidth]{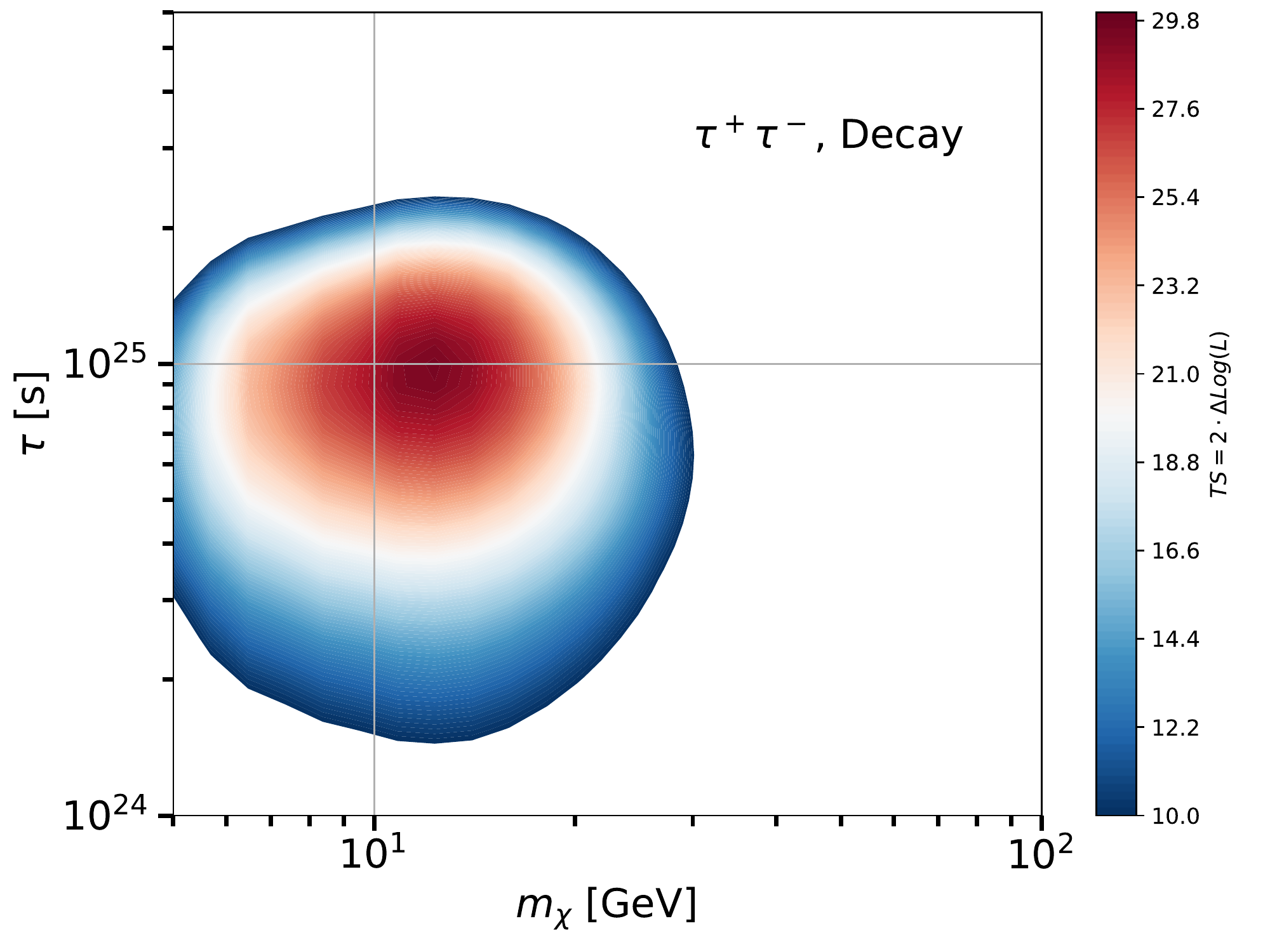}
\caption{Contour plot for the $TS$ as a function of the DM mass and decay time $\tau$ obtained with the combined analysis and the $b\bar{b}$ (left panels) and $\tau^+\tau^-$ (right panels) annihilation channels.}
\label{fig:contourclusterdecay}
\end{figure*}

In Fig.~\ref{fig:contourcluster} we show the contour plot of the $TS$ as a function of the DM mass and annihilation cross section for the $b\bar{b}$ and $\tau^+\tau^-$ annihilation channels.
The peak of the $TS$ is around 27 for the MED model and 23 for the MAX model. For the $b\bar{b}$ ($\tau^+\tau^-$) channel the best fit mass is about 40-60 GeV (8-20 GeV) and the annihilation cross section is $2-4\times 10^{-25}$ ($8-20\times 10^{-26}$) cm$^3$/s for the MED and $4-9\times 10^{-26}$ ($1-3\times 10^{-26}$) cm$^3$/s for the MAX model.
When we perform the analysis starting from 100 MeV we obtain a value of 33 for the peak of the $TS$ and similar best-fit values for the DM mass and annihilation cross section. 
The values of the annihilation cross sections that we obtain for the MED model are ruled out by the constraints obtained with Milky Way dSphs (see, e.g., \citep{DiMauro:2021qcf}). In fact, the upper limits for $\langle \sigma v \rangle$ obtained from a combined analysis of dSphs is about a factor of 20 stronger than the best-fit region shown in Fig.~\ref{fig:contourcases} for the MED case.
The annihilation cross sections obtained for the MAX model could be marginally compatible with the upper limits obtained by {\it Fermi}-LAT for other astrophysical targets and consistent with the best-fit found for the DM interpretation of the Galactic center excess \citep{DiMauro:2021qcf}.
Yet, as we will see in Sec.~\ref{sec:null}, the significance obtained with our combined analysis of clusters is much below $5 \sigma$ in all cases once the actual $TS$ distribution is properly computed and adopted for the null hypothesis.

In Fig.~\ref{fig:contourclusterdecay} we show the contour plot of the $TS$ as a function of the DM mass and the decay time for the $b\bar{b}$ and $\tau^+\tau^-$ annihilation channels.
The best-fit for the DM mass is similar to the one obtained for the annihilation case while the decay time is $5-8 \times 10^{24}$ s and $8-12 \times 10^{24}$ s for the two channels.
The peak of the $TS$ is about 29 for both decay channels.
These values of $\tau$ are ruled out by the lower limits obtained with the Isotropic diffuse $\gamma$-ray background \cite{Blanco:2018esa,Ando:2015qda} that are at the level of $10^{27}-10^{28}$ s.
Instead, the lower limits obtained with the CMB are at the level of the best-fit values so they do not rule out the DM interpretation (see, e.g., \cite{Slatyer:2016qyl}).

Our analysis contains a few important improvements with respect to previous ones (see, e.g., \citep{2010JCAP...05..025A,Huang:2011xr, 2012MNRAS.427.1651H,Nezri:2012tu,Dugger:2010ys,Lisanti:2017qlb,Thorpe-Morgan:2020czg,2012JCAP...07..017A,2012ApJ...750..123A,Fermi-LAT:2015xij,MAGIC:2018tuz}). First of all, we use an extended template for the DM density distribution for each of the clusters in our sample. By using a point source model we would find a much smaller significance for the combined signal that approaches zero. This is due to the fact that the possible $\gamma$-ray signal, regardless of whether it originates from DM annihilating particles or from CR interactions in the ICM, is expected to be very extended in the sky. Moreover, we use more years of data with respect to previous papers. For example, with respect to Ref.~\cite{Fermi-LAT:2015rbk} we used four times more data, and to Ref.~\cite{Lisanti:2017qlb} $50\%$ more data. Analyzing more years of data has the consequence of increasing {\it linearly} the $TS$ for the detection of the signal that could be present in the data.

\subsection{Null hypothesis $TS$ distribution}
\label{sec:null}

In the case where we have a perfect knowledge of the background components, the $TS$ values found with an analysis of $\gamma$-ray data should follow a Poissonian distribution.
As a result, the asymptotic theorem of Chernoff \citep{10.1214/aoms/1177728725} is satisfied and $TS$ can be converted to a significance based on combinations of $\chi^2$ distributions.
In our case, with two DM parameters $m_{\chi}$ and $\langle \sigma v \rangle$ the $TS$ histogram of the residuals should be compatible with the $\chi^2$ distribution for two degrees of freedom divided by two ($\chi_2^2/2$).
However, the analysis of real data at relatively low energies and for extended sources such as clusters of galaxies probably deviates significantly from the asymptotic case.

In order to properly convert the $TS$ for a DM signal into a significance, we thus have to build the $TS$ distribution using random blank sky directions.
We perform, for each cluster in our sample, the analysis in 3100 random directions in the sky using real data. The random directions have been chosen to remove directions that point towards the disk of the Milky Way since all the objects in our sample are located away from the Galactic plane. In particular, we select random directions that satisfy the condition $|b|>20\;\mathrm{deg}$. 
We have selected directions of the sky farther at least $2\;\mathrm{deg}$ from known sources.
Given the number of sources and random directions the ROIs chosen for this analysis are not fully independent, i.e., they have an overlap. However, the distance between the different ROI centers are farther than the typical extension of the DM templates for the clusters in our sample. Therefore, the fact that the ROIs are not fully independent does not affect significantly the results of our analysis.
The number of random directions we have chosen is limited due to the fact that with a larger number of random directions the ROIs would significantly overlap and thus the analysis of the different regions of the sky would not be truly independent.
We use the DM distribution associated with the MED model.
For each ROI we run the same analysis explained in Sec.~\ref{sec:analysistec}. We decide to fix the annihilation channel to $b\bar{b}$ and the mass to 50 GeV because, as we have seen in the previous section, this is the best-fit value we will obtain from the combined analysis of the clusters in our sample assuming the $b\bar{b}$ annihilation channel. 
This implies that the significance we derive in this section is local, i.e., it does not take into account look elsewhere effect and differs thus from the global significance.
For each random direction we find the combined $TS$ for the list of clusters in our sample. This gives us a list of 3100 values for the $TS$ of DM associated to the null hypothesis.

\begin{figure}[h!]
\includegraphics[width=0.49\textwidth]{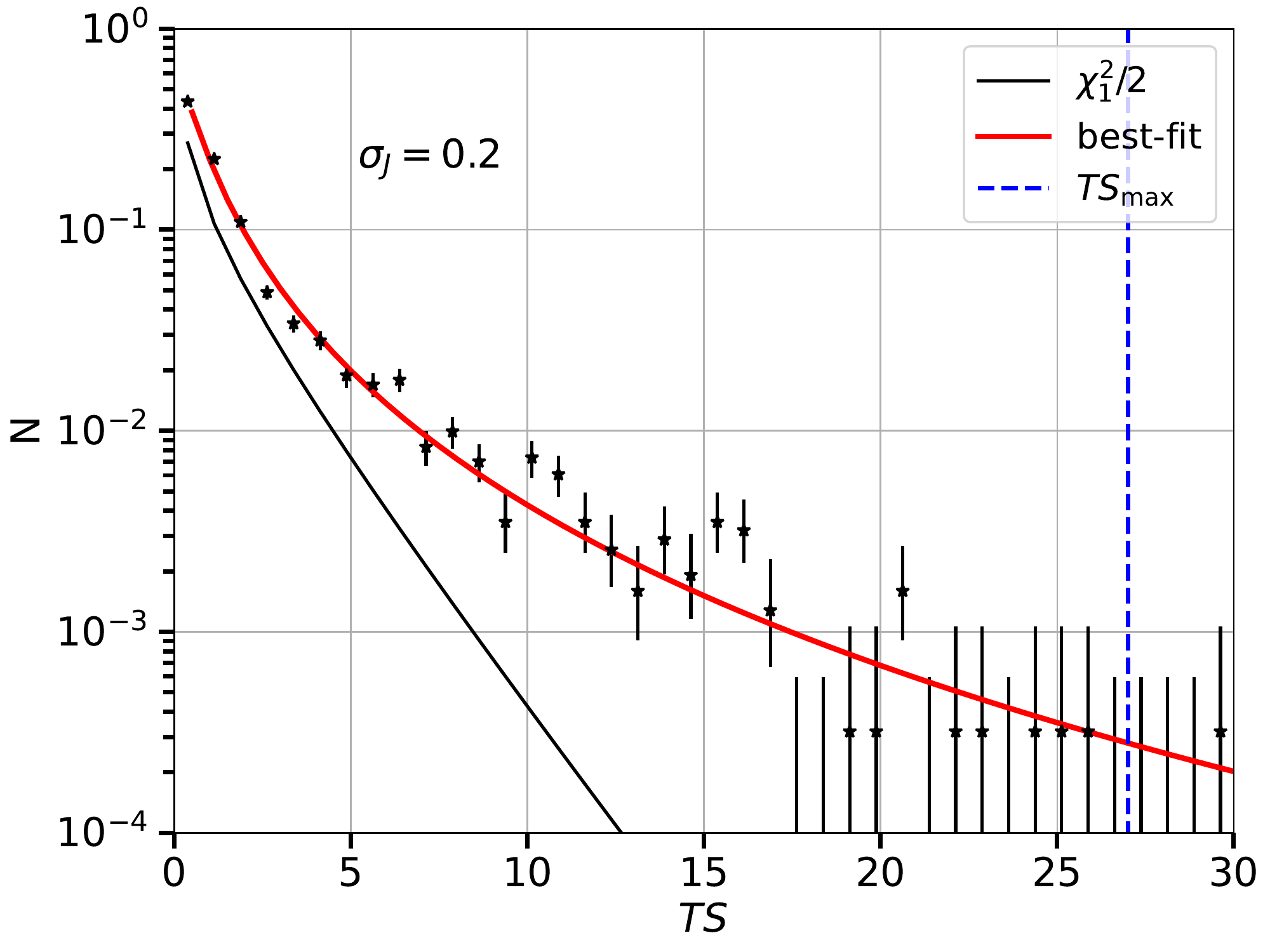}
\includegraphics[width=0.49\textwidth]{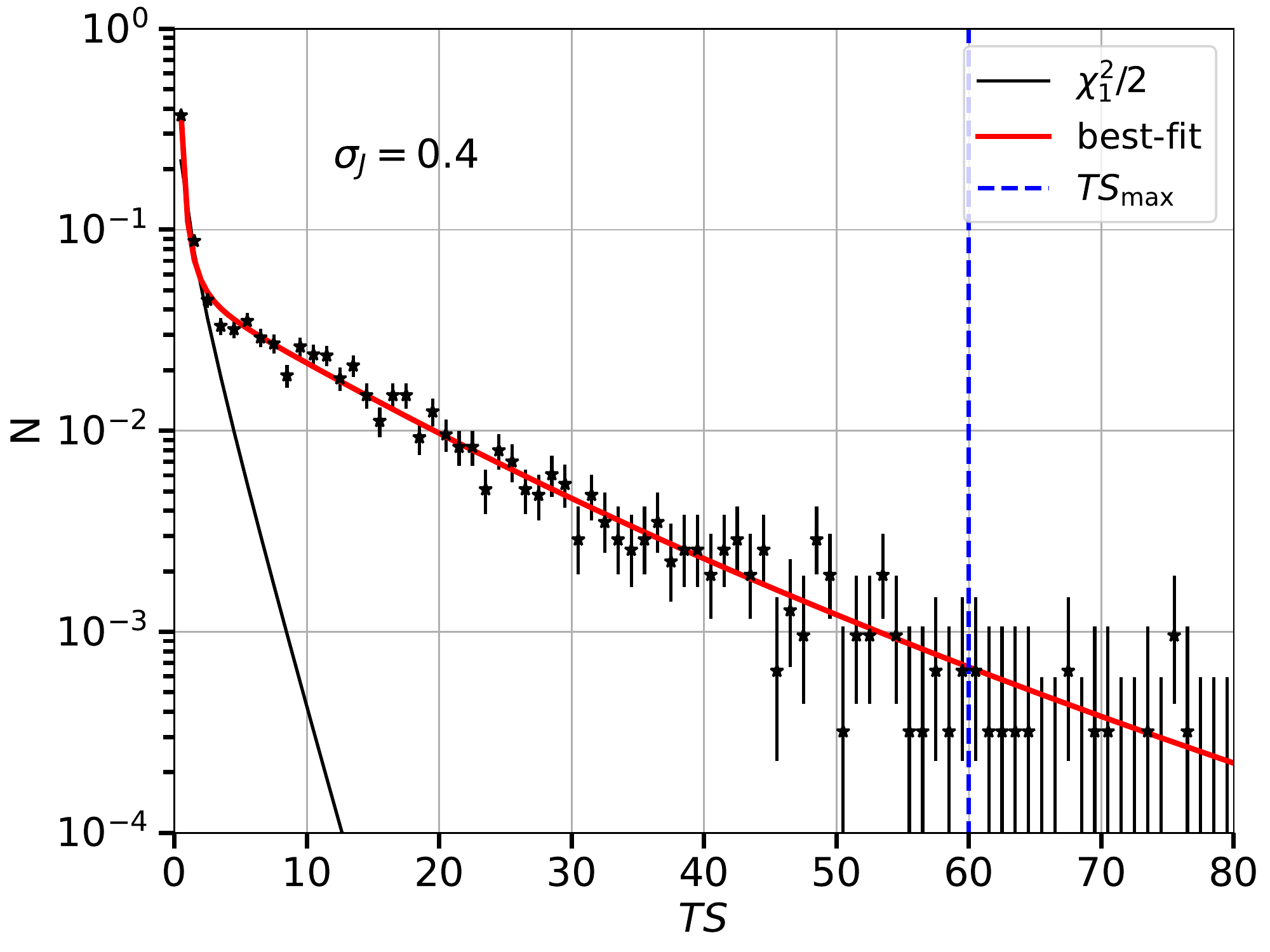}
\caption{Normalized histogram of the $TS$ distribution for the analysis we have performed in random directions (black data points). Data error bars represent the Poissonian error. We report also the $\chi^2$ distribution for 1 degree of freedom (the annihilation cross section) and the best fit we find with the analytic function in Eq.~(\ref{eq:null}) (red line). The vertical blue dashed line represents the peak of the $TS$ obtained in the combined analysis of our cluster sample (see Fig.~\ref{fig:tscluster} and \ref{fig:TScases}). We show the results for the case with $\sigma_J=0.2$ (top panel) and $\sigma_J=0.4$ (bottom panel).
}
\label{fig:null}
\end{figure}

The distribution of the $TS$ we derive with this type of analysis is reported in the top panel of Fig.~\ref{fig:null} together with the distribution of the $\chi^2$ distribution for 1 degree of freedom ($\chi_1^2/2$). We consider one degree of freedom because the DM mass is fixed. 
As anticipated, the $TS$ distribution is very different from the $\chi_1^2/2$. In particular, there is a prominent tail at larger $TS$ values compared to that of the TS distribution following $\chi_1^2/2$.
This conclusion brings us to find an alternative function that fits well the observed distribution:
\begin{equation}
    N_{\rm{norm}} (TS) = 0.22 \times (TS)^{-1.29-0.31\log{(TS/2.55)}}~,
\label{eq:null}
\end{equation}
also shown in Fig.~\ref{fig:null}. The $TS$ for the detection of a DM signal from our cluster sample is 27 for the MED model and this is the highest value we obtain by changing the DM distribution model.
This $TS$ value is associated to a $p$-value of $3.1\times 10^{-3}$ and a significance of 2.7$\sigma$.
By using different bins for the $TS$ histogram and different analytic shape for fitting it, we find the significance for the MED model to be always between 2.5-3.0$\sigma$.

We also run this analysis for the case where we use an uncertainty for $\log_{10}{J}$ equal to $\sigma_J=0.4$.
This case is reported in the bottom panel of Fig.~\ref{fig:null}. The $TS$ distribution for the random direction has, in this case, a much larger tail, as expected since we increase $\sigma_J$ with respect to the previous case.
We fit the distribution with a function similar to Eq.~(\ref{eq:null}) and find that the significance of the signal corresponding to $TS=60$ (see next section) is at the level of 2.5$\sigma$.

We can draw similar conclusions also for all the other cases tested in Sec.~\ref{sec:sys}.

\subsection{Systematics on the significance of the signal due to analysis setup}
\label{sec:sys}

In this section we perform the data analysis by changing some of the assumptions we have made in Sec.~\ref{sec:results}. We will label the model used up to now as {\tt Baseline}.
Below we report the list of these checks.
\begin{itemize}
    \item {\tt Source} and {\tt Ultracleanveto}: We change the data selection using the Pass~8 {\tt SOURCE} and {\tt ULTRACLEANVETO} event class and we employ the corresponding instrument response functions.
    \item {\tt IEM}: We change the IEM by using the model called {\tt Yusifov} in \citep{DiMauro:2021qcf}. This model assumes that the Galactic source distribution is taken from the pulsar one \citep{Yusifov:2004fr}. The $\pi^0$, bremsstrahlung and inverse Compton scattering components of the IEM are re-evaluated with this model according to the different source distribution with respect to the {\tt Baseline} model.
    \item {\tt 100 MeV}: We reduce the low-energy end of the analysis to 100 MeV. For this scope we reduce the value of the max zenith angle ($z_{\rm max}$) to $90\;\mathrm{deg}$ as suggested in the LAT data selection recommendations\footnote{\url{https://fermi.gsfc.nasa.gov/ssc/data/analysis/documentation/Cicerone/Cicerone_Data_Exploration/Data_preparation.html}}.
    \item {\tt ROI}: We run the analysis with a larger ROI of $26\;\mathrm{deg}\times 26\;\mathrm{deg}$ and include the sources in the model if they are inside a cube of size $30\;\mathrm{deg}$. This check permits to verify that the signal does not suffer from leakage outside of the ROI and that there are no edge effects impacting the results.
    \item {\tt ROI 100 MeV}: We run the analysis with a larger ROI of $26\;\mathrm{deg}\times 26\;\mathrm{deg}$ and include the sources in the model if they are inside a cube of size $30\;\mathrm{deg}$ and select data above 100 MeV.
    \item {\tt $\sigma_J=0$} and {\tt $\sigma_J=0.4$}. We vary the error on the logarithm of the $J$-factor to 0 or 0.4. This check has been performed to test whether the significance of the signal changes assuming a perfectly known or more uncertain knowledge of the DM density.
\end{itemize}

\begin{figure}[h!]
\includegraphics[width=0.49\textwidth]{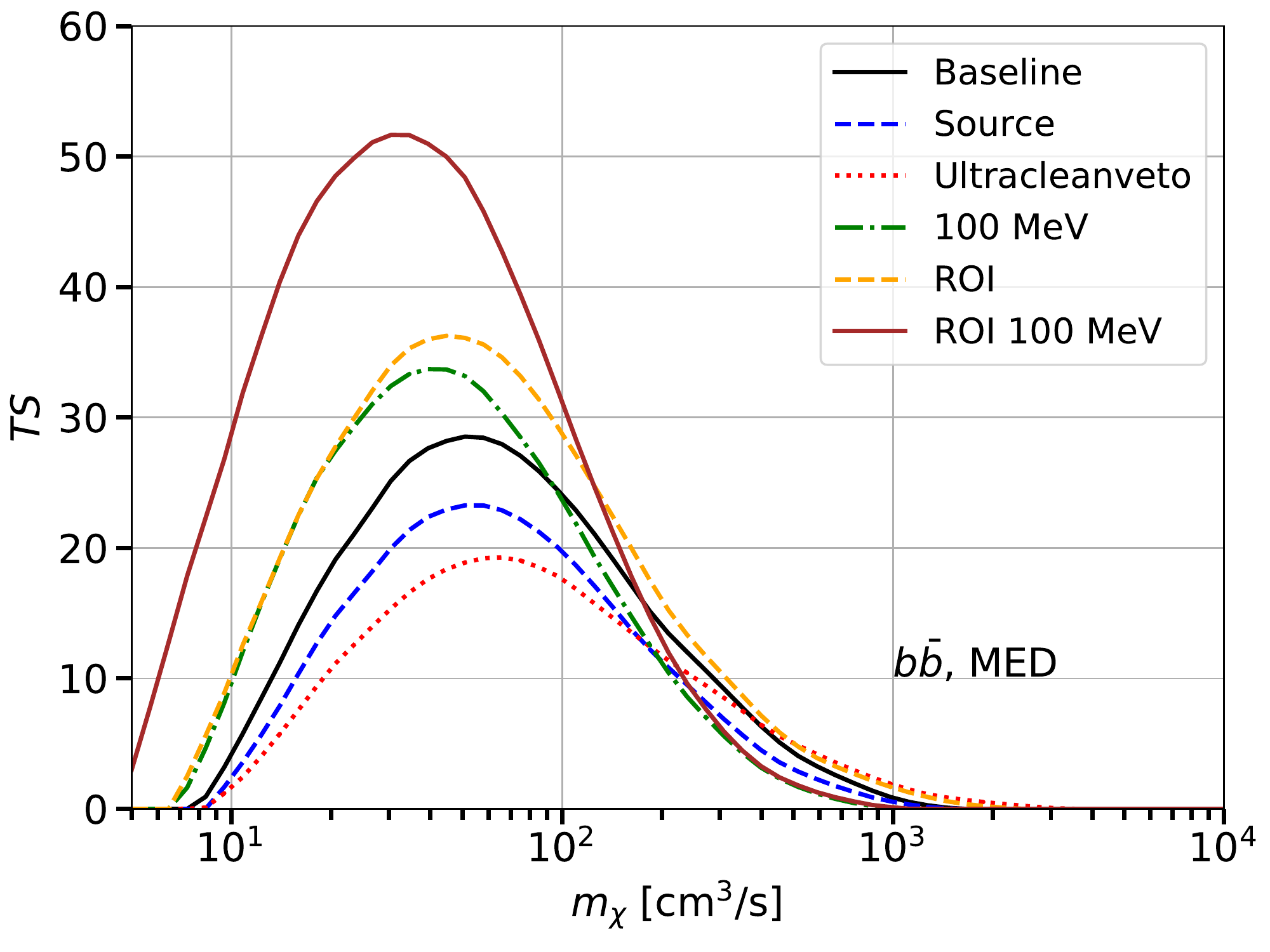}
\includegraphics[width=0.49\textwidth]{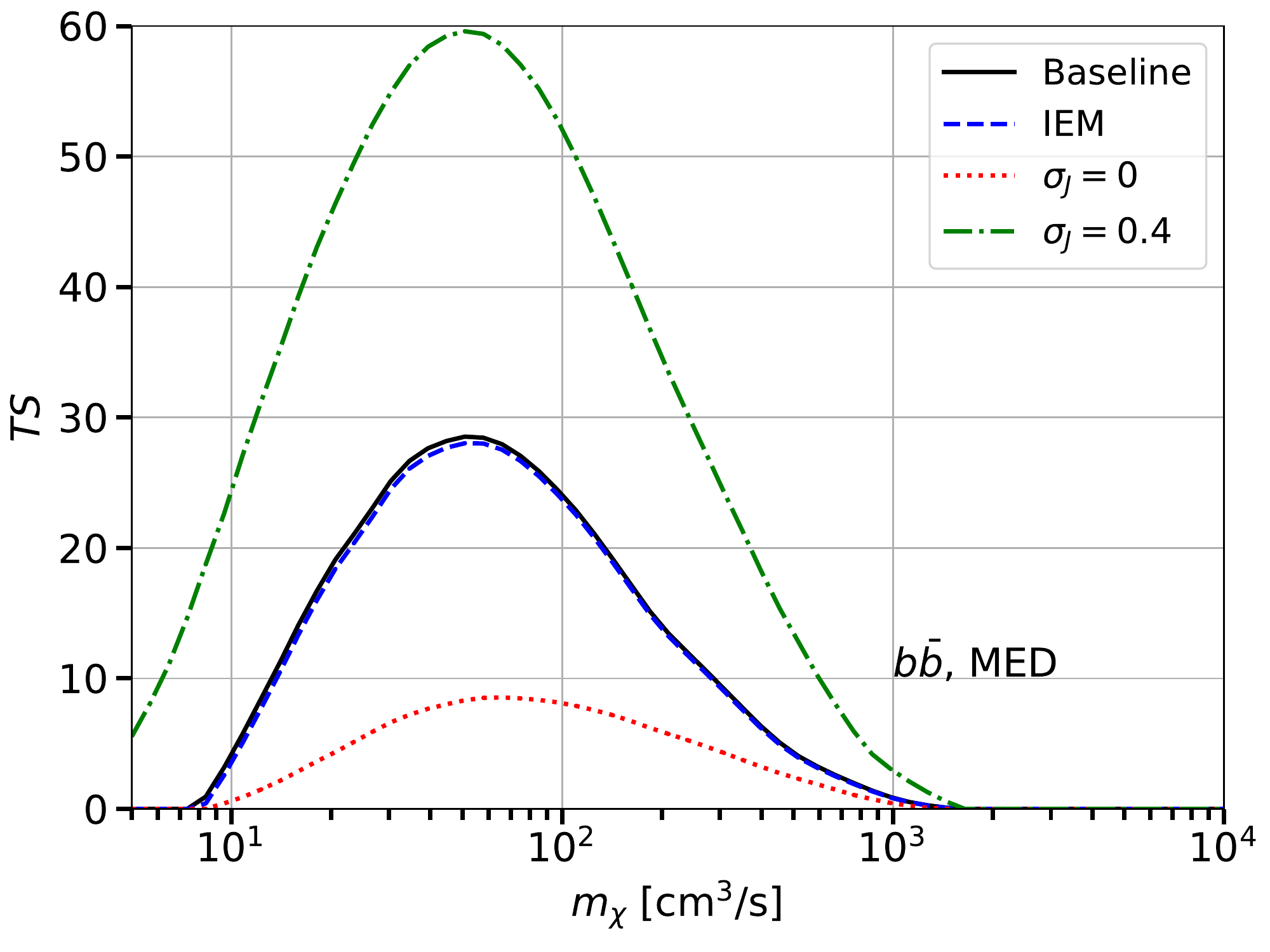}
\caption{Same as Fig.~\ref{fig:tscluster} for the {\tt Baseline} setup (black solid) and for all the other cases tested in the analysis to check the systematics on the results due to data selection, energy range, ROI size, different choice of the IEM and error on the $J$-factor. See the text for the explanation of the tested cases.}
\label{fig:TScases}
\end{figure}

\begin{figure*}
\includegraphics[width=0.49\textwidth]{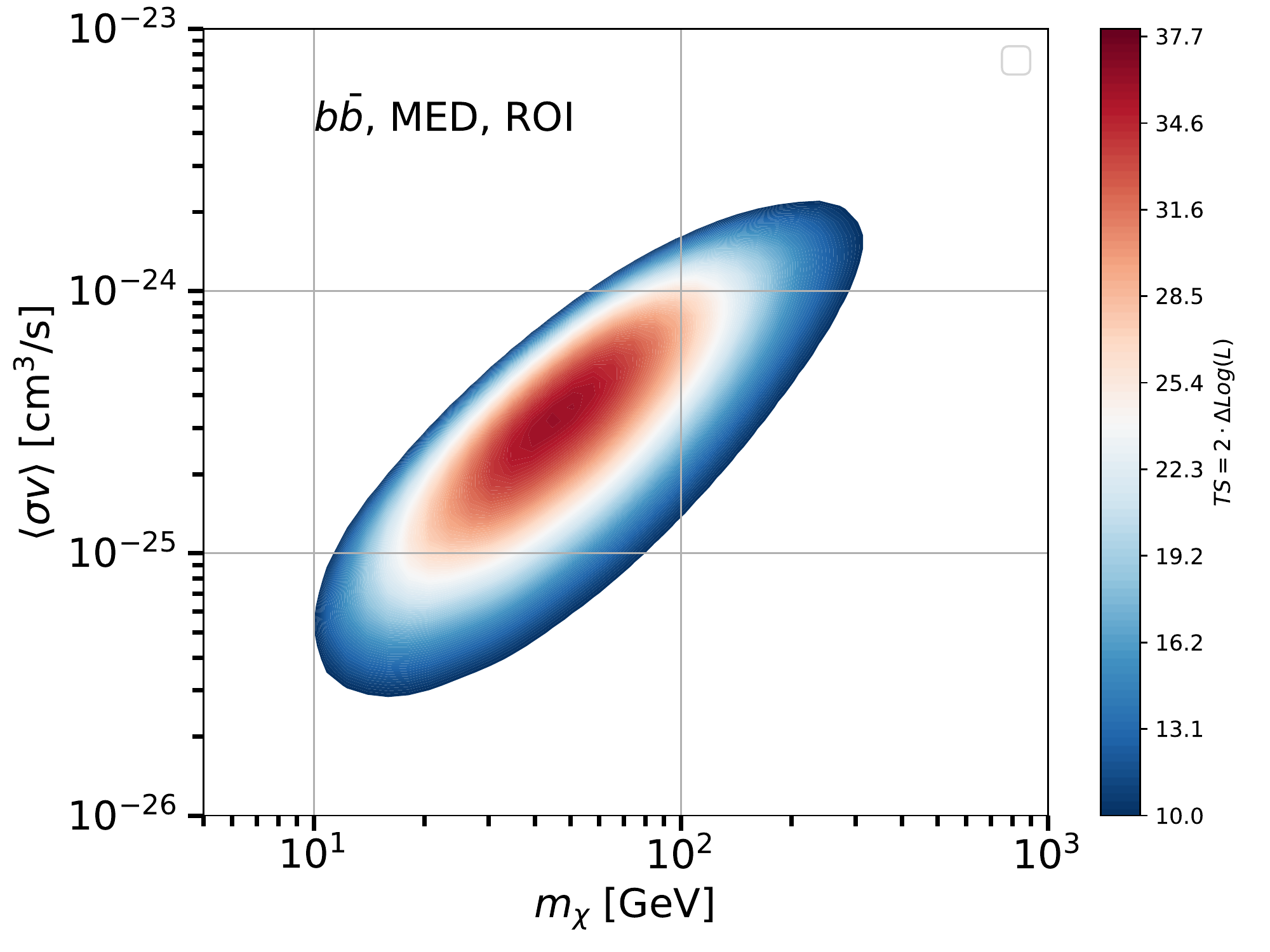}
\includegraphics[width=0.49\textwidth]{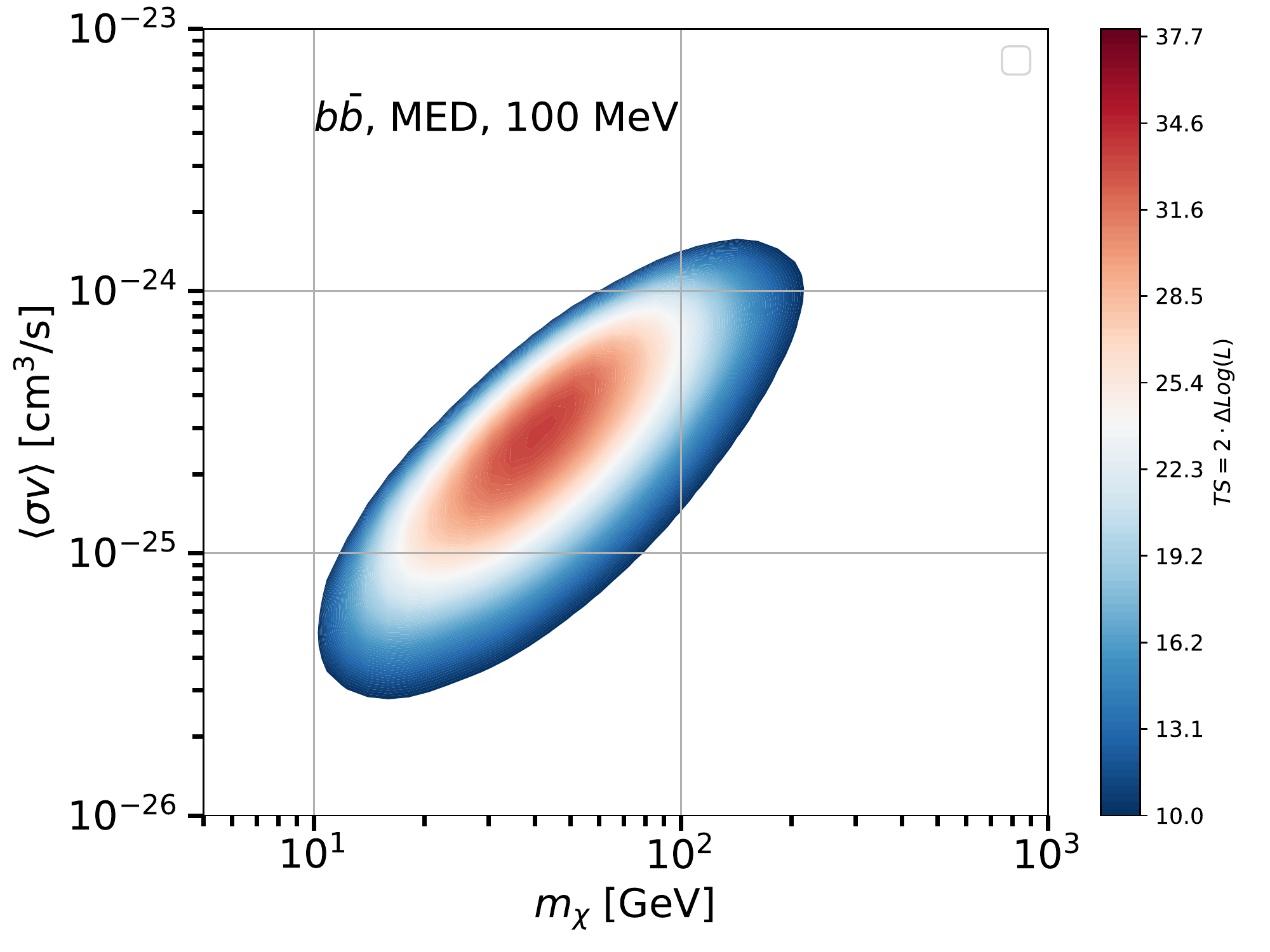}
\includegraphics[width=0.49\textwidth]{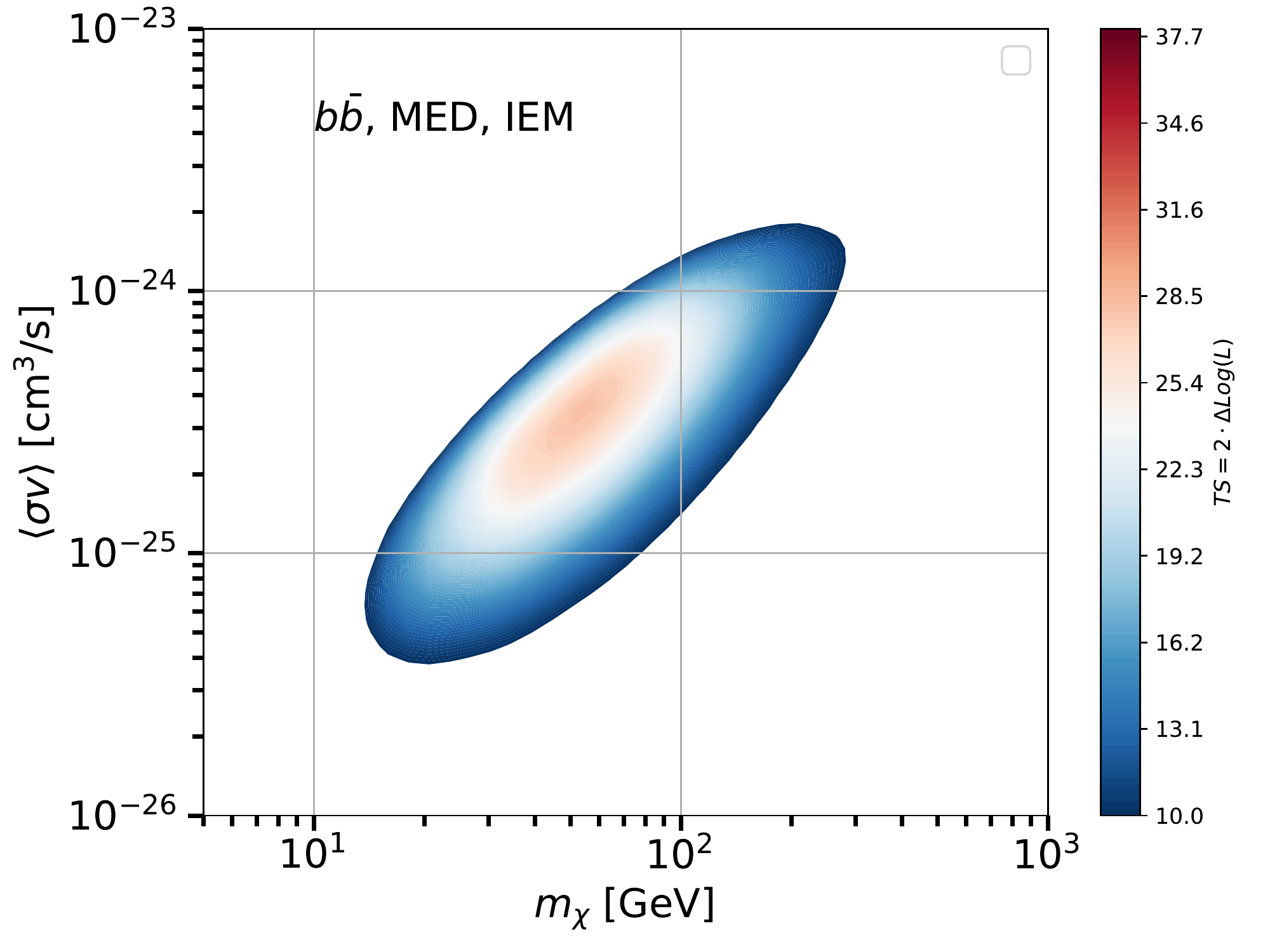}
\includegraphics[width=0.49\textwidth]{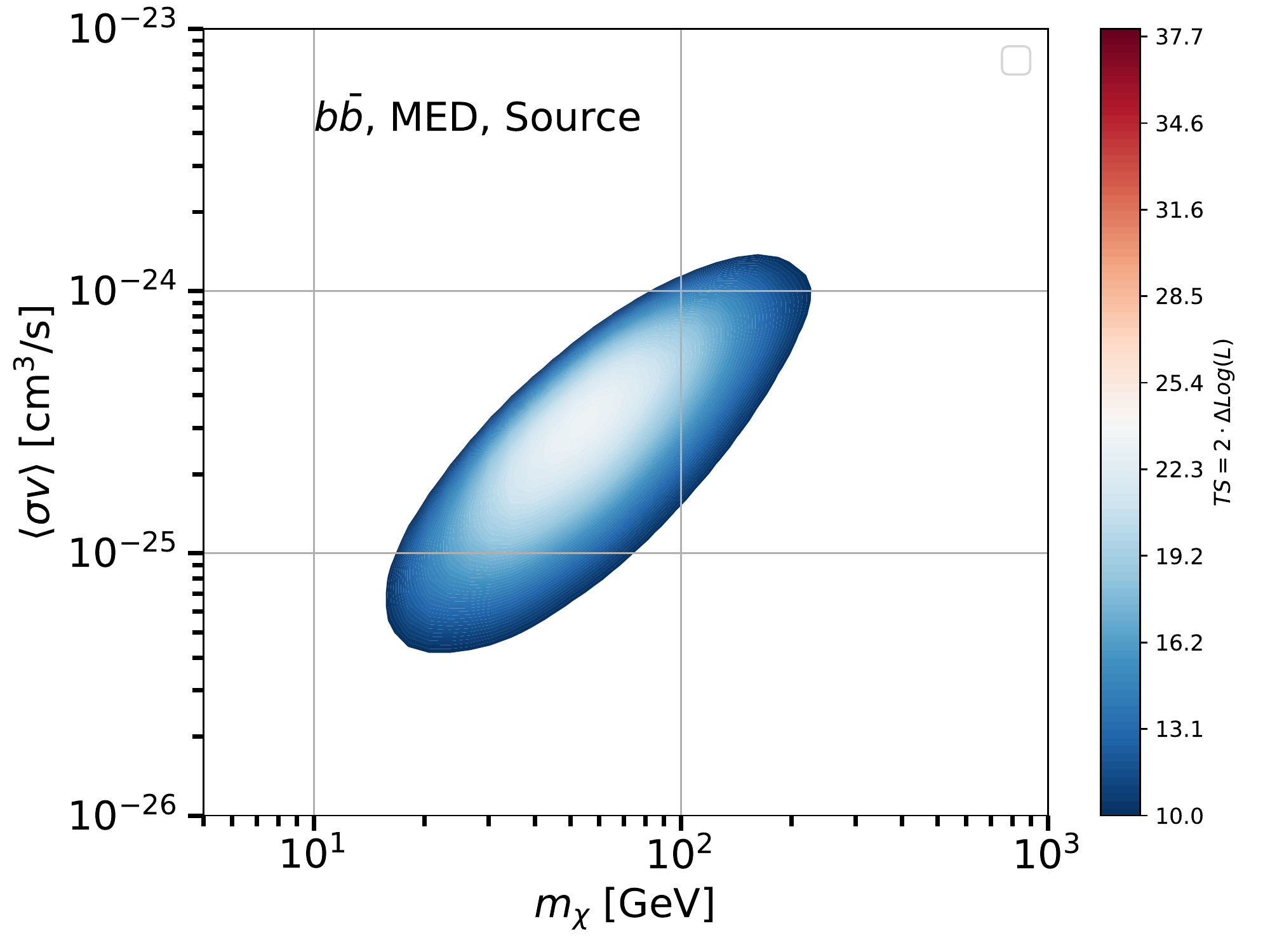}
\caption{Same as Fig.~\ref{fig:contourcluster} for some of the cases tested in the analysis to check the systematic on the results due to choice of the energy range ({\tt 100 MeV}, top left), ROI size ({\tt ROI}, top right), of the IEM ({\tt IEM}, bottom left) and data selection ({\tt Source}, bottom right).}
\label{fig:contourcases}
\end{figure*}

We report in Fig.~\ref{fig:TScases} the values of the $TS$ as a function of the DM mass obtained for the different tested cases for the MED DM model, that gives for the {\tt Baseline} setup the highest significance. We choose the $b\bar{b}$ annihilation cross-section.
This plot shows that lowering the energy range of the analysis to {\tt 100 MeV} increases the $TS$ from 27 of the {\tt Baseline} model to 33. 
However, a higher $TS$ does not guarantee that also the significance of the signal is higher. In fact, by including the low-energy data, the $TS$ distribution for the null hypothesis has a higher tail at large $TS$. This is due to the fact that most of the photons at energies smaller than 500 MeV are related to the IEM and isotropic background. Therefore, the contamination of background photons is larger for the case labeled as {\tt 100 MeV}.
A similar conclusion can be drawn for the cases {\tt ROI} and {\tt ROI 100 MeV}. For the latter, the peak of the $TS$ increases to 52 but the distribution of the null hypothesis signal is much larger for large $TS$. This keeps the level of significance of the signal similar to the {\tt Baseline} case.
Even assuming for the cases {\tt 100 MeV} and {\tt ROI 100 MeV} which are the ones that give highest $TS$ values, the same null hypothesis $TS$ distribution as the one we obtain for the {\tt Baseline} case, we would obtain a signal at $3.1\sigma$ and $3.5\sigma$ significance, respectively. The real signal significance for these two cases is surely below these values for the reasons explained above.

We also show the results obtained when we fix $\sigma_J$ for all clusters to 0.4. In this case the peak of the $TS$ increases to 60. As shown in Sec.~\ref{sec:null} and Fig.~\ref{fig:null}, the null hypothesis distribution of the $TS$ reaches much larger values of $TS$ and thus the resulting significance of the signal is similar to the {\tt  Baseline} case. Instead, the case with $\sigma_J=0$ has a much smaller peak of $TS$.

The {\tt Baseline} and the {\tt IEM} cases provide almost the same result, while the {\tt SOURCE} and {\tt ULTRACLEANVETO} data selections slightly worsen the value of the $TS$ peak.
Nevertheless, the best-fit mass is very similar among the tested cases and stands between $40-60$ GeV.

In Fig.~\ref{fig:contourcases} we show the contour plots of the $TS$ as a function of DM mass and annihilation cross section using the $b\bar{b}$ annihilation channel and the MED DM annihilation model.
Also from these plots it is clear that the {\tt 100 MeV} and $\sigma_J=0.4$ cases are the ones that give the highest $TS$.
We also show that the best-fit properties of the possible DM signal are very similar among the tested setups. In particular the annihilation cross section is around $2-5\times 10^{-25}$ cm$^3$/s.

\subsection{Constraints on a dark matter contribution}
\label{sec:UL}

We have demonstrated that the signal we find from the combined analysis of clusters is not significant and that if interpreted as a DM signal is not compatible with the results from the dSphs DM search.
Therefore, we are motivated to find upper limits for the annihilation cross section and a lower limit for the DM decay time.
We obtain the DM limits by proceeding in the following way.
For a fixed DM mass we find the value of $\langle \sigma v \rangle$ ($\tau$) for which the $\Delta \mathcal{L} = 2.71/2$ (see Eq.~(\ref{eq:TS})), which is associated with the one-sided $95\%$ CL upper limits.

\begin{figure*}
\includegraphics[width=0.49\textwidth]{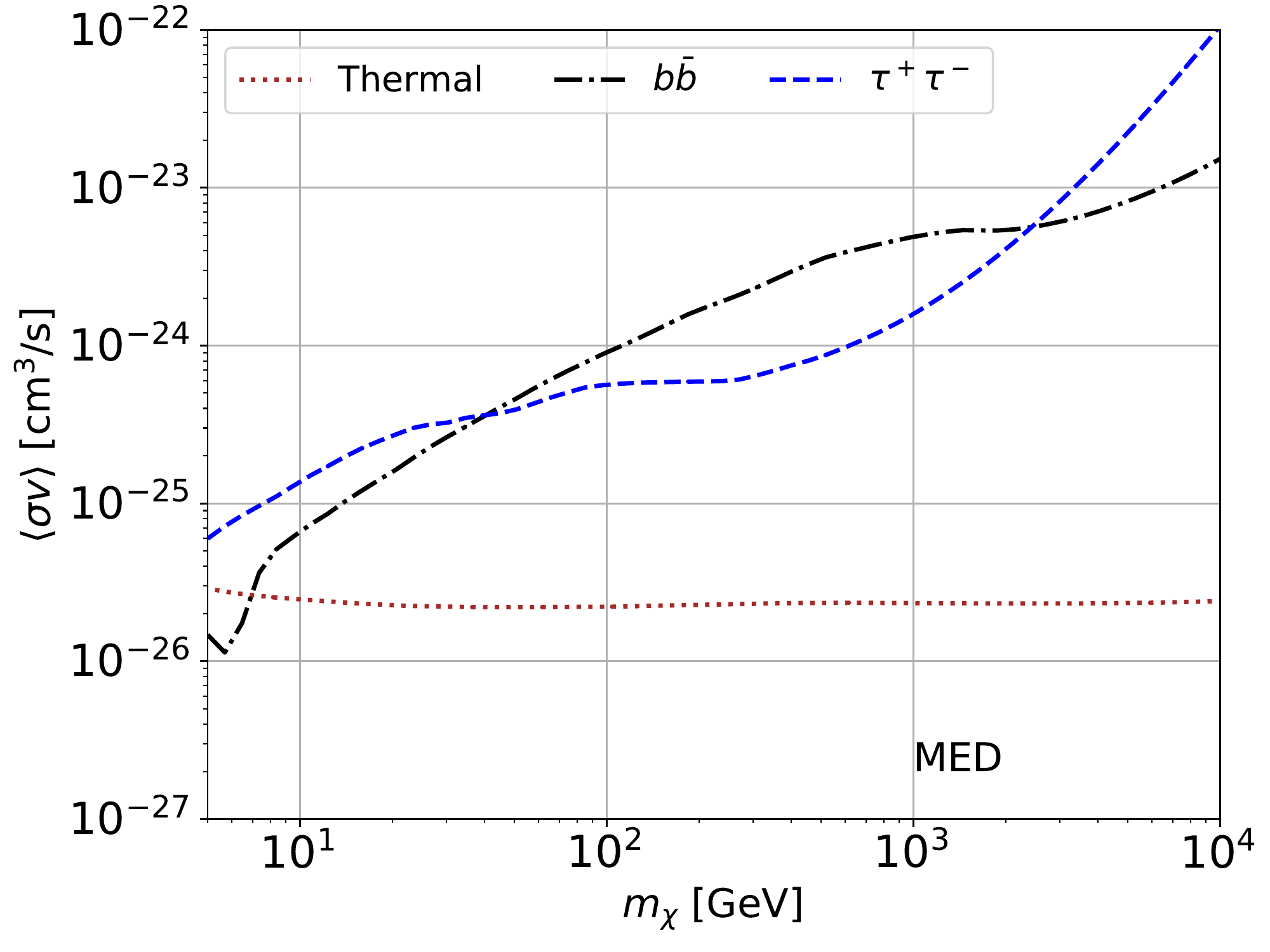}
\includegraphics[width=0.49\textwidth]{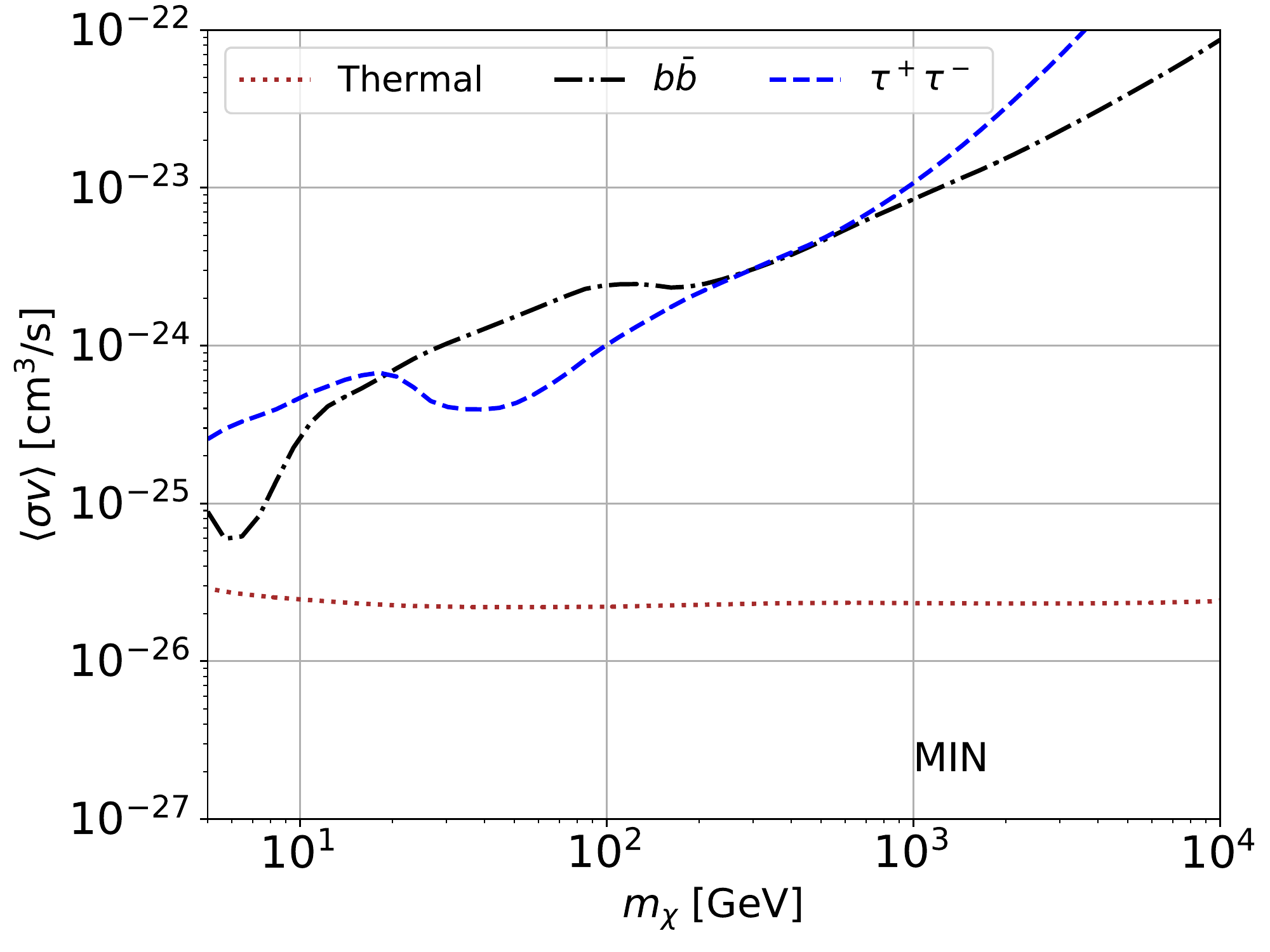}
\includegraphics[width=0.49\textwidth]{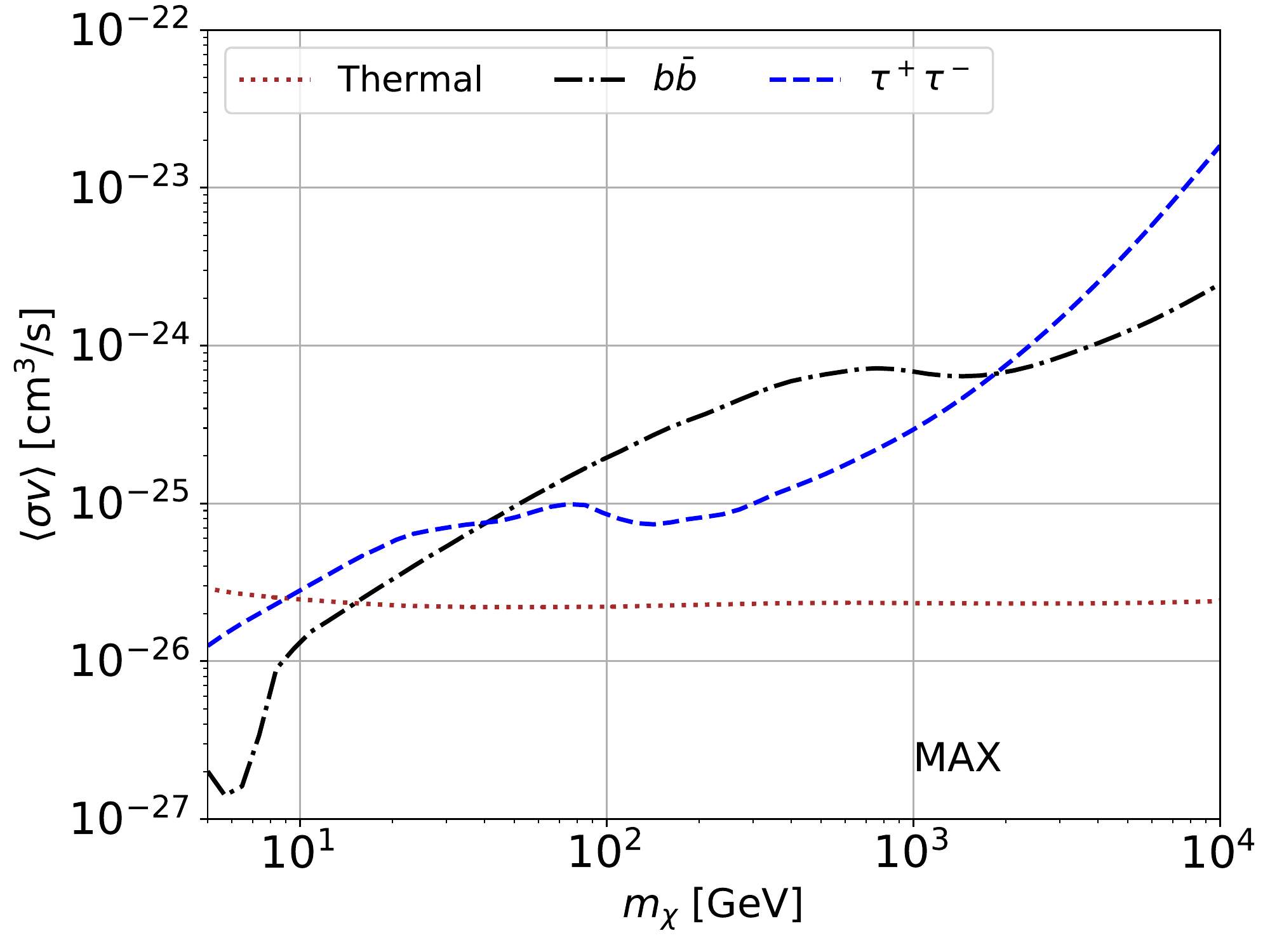}
\includegraphics[width=0.49\textwidth]{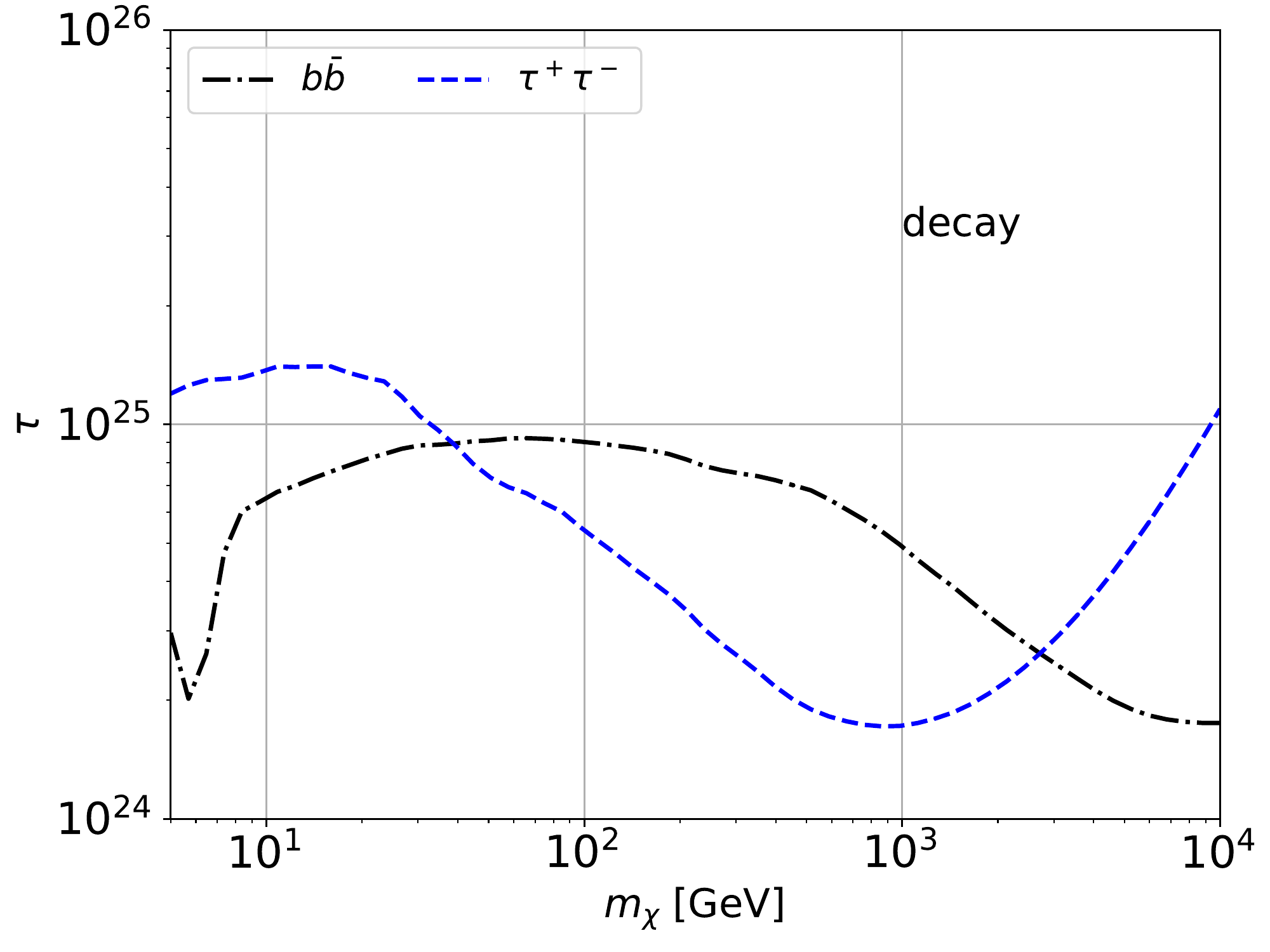}
\caption{Upper limits for $\langle \sigma v \rangle$ and the lower limit for $\tau$ found with our analysis. We show the results obtained for the MIN, MED and MAX models for annihilating DM and for the $b\bar{b}$ (blue dashed) and $\tau^+\tau^-$ (black dot dashed) annihilation channel. We also show the thermal relic cross section in the annihilation panels (brown dotted).}
\label{fig:ul}
\end{figure*}

We show in Fig.~\ref{fig:ul} the upper limits for $\langle \sigma v \rangle$ and lower limit for $\tau$ we obtain for the $b\bar{b}$ and $\tau^+\tau^-$ annihilation channels, and for the MIN, MED and MAX DM models.
The constraints on $\langle \sigma v \rangle$ and $\tau$ are weaker with respect to the ones obtained with dSphs (see, e.g., \citep{DiMauro:2021qcf}).
In particular, the upper limits for $\langle \sigma v \rangle$ are above the thermal relic cross section for almost all the masses.
The results for the MIN/MED/MAX models scale inversely proportional to the value of the corresponding $J$-factors.  

More precise, robust upper (lower) limits for $\langle \sigma v \rangle$ ($\tau$) could be obtained by first calculating and taking into account the astrophysical $\gamma$-ray production from CRs interacting against the ICM. We will work on this task in a following paper. We note that our current limits are, in this sense, conservative, as the inclusion of CR-induced $\gamma$-ray emission in the computation of DM limits would only make these stronger.

\section{Conclusions}
\label{sec: conclusions}
In this work, we have analyzed 12 years of {\it Fermi}-LAT data in the direction of clusters of galaxies selected by their masses, distances from the Earth and their X-ray fluxes.
We searched for a signal of $\gamma$ rays coming from annihilating or decaying DM particles in the hypothesis of WIMPs.
We first built physically-motivated spatial and spectral templates for the DM emission, based on different assumptions for the distribution of DM and halo substructures within the clusters. For annihilation, we considered three models labeled as MIN, MED and MAX corresponding, respectively, to a case with no substructure or with substructure modeled with a slope of the subhalo mass function of 1.9 and 2.0, and a minimum mass of $10^{-6}$ and $10^{-9}$ $M_{\odot}$, respectively. 

The clusters detected with the largest significances are A3526-Centaurus, A3667, NGC~4626, A2256, NGC~5846, A2064 and A1656-Coma. However, in all cases, the found $TS$ is at most 15, i.e., well below the $5\sigma$ significance.
We then performed a combined analysis of the data for all the clusters, finding a signal that is at the level of $TS=6$, 27 and 23 for the MIN, MED and MAX models. The best-fit values for the mass in case of $b\bar{b}$ annihilation channel are about 40-70 GeV and the annihilation cross section is about $2-4$ ($0.4-0.9$) $10^{-25}$ cm$^3$/s for the MED (MAX) model. The signal is thus in tension with the non detection of a flux of $\gamma$ rays from dSphs. This implies that the interpretation of the signal as $\gamma$ rays produced by DM annihilation is excluded by the analysis of dSphs. 

In case of decaying DM the best-fit mass is similar to the annihilation case while the decay time is around $5-8 \cdot 10^{24}$ s and $8-12 \cdot 10^{24}$ s for the $b\bar{b}$ and $\tau^+\tau^-$ decay channels. The peak of the $TS$ is about 29 for both decay channels. However, the lower limits obtained with the {\it Fermi}-LAT isotropic diffuse $\gamma$-ray background rule out the DM interpretation.

A more probable interpretation is that this hint of a signal comes from photons produced in the ICM by CRs interacting against gas and photons fields.

We performed the same analysis in random directions to find the actual $TS$ distribution of the null hypothesis, i.e., no DM signal, and derived the corrected significance of the signal. Considering all the cases tested, different analysis techniques and assumptions for the IEM and $\sigma_J$, the signal is at $2.5-3.0\sigma$ significance level. 
This signal is robust against different selections of the data, analysis techniques, IEM templates and assumptions for the errors on the geometrical factors. 
In the future, we will perform a dedicated analysis of the signal using the hypothesis that it originates from the interaction of CRs accelerated within the clusters and colliding with atoms and photon fields in the ICM.

We finally derived upper limits for $\langle \sigma v \rangle$ and lower limit for $\tau$ with the conservative assumption that all the photons produced by the clusters come from DM.
The upper limits for $\langle \sigma v \rangle$ are less stringent than the one obtained with the dSphs in the MIN and MAX cases while are at the level of the ones found with the dSphs with the MAX DM distribution.
Instead, the lower limit for $\tau$ is at the level of $10^{24}-10^{25}$ s.

\begin{acknowledgments}
The {\it Fermi} LAT Collaboration acknowledges generous ongoing support from a number of agencies and institutes that have supported both the development and the operation of the LAT as well as scientific data analysis. These include the National Aeronautics and Space Administration and the Department of Energy in the United States, the Commissariat\'a l'Energie Atomique and the Centre National de la Recherche Scientifique / Institut National de Physique Nucl\'eaire et de Physique des Particules in France, the Agenzia Spaziale Italiana and the Istituto Nazionale di Fisica Nucleare in Italy, the Ministry of Education, Culture, Sports, Science and Technology (MEXT), High Energy Accelerator Research Organization (KEK) and Japan Aerospace Exploration Agency (JAXA) in Japan, and the K. A. Wallenberg Foundation, the Swedish Research Council and the Swedish National Space Board in Sweden.
Additional support for science analysis during the operations phase is gratefully acknowledged from the Istituto Nazionale di Astrofisica in Italy and the Centre National d'Etudes Spatiales in France. This work performed in part under DOE Contract DE- AC02-76SF00515.

The authors thank R\'emi Adam and Stephan Zimmer for insightful discussions.
MDM research is supported by Fellini - Fellowship for Innovation at INFN, funded by the European Union’s Horizon 2020 research programme under the Marie Sklodowska-Curie Cofund Action, grant agreement no.~754496. JPR work is supported by grant SEV-2016-0597-17-2 funded by MCIN/AEI/10.13039/501100011033 and ``ESF Investing in your future''. MASC was also supported by the {\it Atracci\'on de Talento} contracts no. 2016-T1/TIC-1542 and 2020-5A/TIC-19725 granted by the Comunidad de Madrid in Spain. The work of JPR and MASC was additionally supported by the grants PGC2018-095161-B-I00, PID2021-125331NB-I00 and CEX2020-001007-S, all funded by MCIN/AEI/10.13039/501100011033 and by ``ERDF A way of making Europe''. 
NF acknowledges support from: {\sl Departments of Excellence} grant awarded by the Italian Ministry of Education, University and Research (MIUR); Research grant {\sl The Dark Universe: A Synergic Multimessenger Approach}, Grant No. 2017X7X85K funded by the Italian Ministry of Education, University and Research (MIUR). NF and MDM acknowledges support from the research grant {\sl TAsP (Theoretical Astroparticle Physics)} funded by Istituto Nazionale di Fisica Nucleare (INFN).

\end{acknowledgments}

\newpage
\clearpage

\begin{sidewaystable}
\centering
\begin{tabular}{|c||c|c|c|c|c|c|c||c|c|c|c|c|c||c|}
\hline 
Cluster & $d_{L}$ & $M_{200}$ & $c_{200}$ & $\rho_s$ & $r_s$ & $R_{200}$ & $\theta_{200}$ & $\log_{10}J_{MIN}$ & $\log_{10}J_{MED}$ & $B_{MED}$ & $ \log_{10}J_{MAX}$ & $B_{MAX}$ & $\log_{10}D$ & TS \tabularnewline
\hline 
\hline 
 & [Mpc] & [10$^{14}$ M$_{\odot}$] &  & [M$_{\odot}$/kpc$^3$] & [kpc] & [kpc] & [deg] & [GeV$^2$cm$^{-5}$] & [GeV$^2$cm$^{-5}$] &  & [GeV$^2$cm$^{-5}$] &  & [GeV cm$^{-2}$] & \\
\hline 
A478 & 387.29 & 6.08 & 5.06 & 303795 & 345.37 & 1747.71 & 0.30 & 16.05 & 17.00 & 9.03 & 17.77 & 52.90 & 17.74 & 0.00 \tabularnewline
\hline 
A399 & 320.39 & 4.03 & 5.14 & 314222 & 296.58 & 1523.16 & 0.31 & 16.02 & 17.00 & 9.54 & 17.76 & 54.90 & 17.72 & 5.69 \tabularnewline
\hline 
A2065 & 325.13 & 4.73 & 5.10 & 309802 & 314.87 & 1607.11 & 0.33 & 16.08 & 17.05 & 9.46 & 17.82 & 55.00 & 17.78 & 4.94 \tabularnewline
\hline 
A1736 & 203.92 & 1.45 & 5.40 & 352863 & 200.77 & 1084.70 & 0.33 & 15.96 & 16.98 & 10.50 & 17.71 & 56.70 & 17.65 & 4.89 \tabularnewline
\hline 
A1644 & 208.50 & 1.55 & 5.38 & 349910 & 205.81 & 1107.83 & 0.33 & 15.96 & 16.98 & 10.50 & 17.72 & 56.70 & 17.66 & 1.90 \tabularnewline
\hline 
A401 & 339.38 & 5.92 & 5.06 & 304380 & 342.03 & 1732.25 & 0.34 & 16.14 & 17.11 & 9.34 & 17.88 & 54.90 & 17.84 & 8.07 \tabularnewline
\hline 
A2029 & 348.92 & 6.59 & 5.05 & 302105 & 355.64 & 1795.26 & 0.34 & 16.16 & 17.13 & 9.21 & 17.90 & 54.40 & 17.86 & 0.26 \tabularnewline
\hline 
Hydra-A & 240.76 & 2.60 & 5.24 & 328469 & 251.56 & 1317.25 & 0.35 & 16.06 & 17.07 & 10.20 & 17.82 & 57.70 & 17.76 & 3.74 \tabularnewline
\hline 
ZwCl1215 & 339.38 & 6.54 & 5.05 & 302272 & 354.58 & 1790.34 & 0.35 & 16.18 & 17.15 & 9.32 & 17.92 & 55.00 & 17.88 & 0.00 \tabularnewline
\hline 
MKW3S & 199.34 & 1.66 & 5.36 & 346794 & 211.39 & 1133.45 & 0.36 & 16.02 & 17.05 & 10.60 & 17.78 & 57.60 & 17.72 & 0.00 \tabularnewline
\hline 
A133 & 254.68 & 3.35 & 5.18 & 319842 & 276.74 & 1432.35 & 0.36 & 16.12 & 17.12 & 10.10 & 17.88 & 57.70 & 17.83 & 2.46 \tabularnewline
\hline 
A3158 & 263.99 & 3.97 & 5.14 & 314620 & 295.06 & 1516.19 & 0.37 & 16.16 & 17.16 & 9.99 & 17.92 & 57.70 & 17.87 & 5.39
\tabularnewline
\hline 
\end{tabular}
\caption{\label{tab:clusters_all_1} Table containing the main parameters of the whole sample of galaxy clusters, ordered from smaller to bigger in size ($\theta_{200}$). The columns correspond to: (1) cluster name; (2) luminosity distance, computed assuming $\Lambda$CDM and redshift from \citep{Schellenberger:2017wdw}, except for NGC~5044, NGC~5846 and M49, which is extracted from \citep{2002ApJ...567..716R}, and for Virgo, extracted from \citep{2011JCAP...12..011S}; (3) virial mass from \citep{Schellenberger:2017wdw}, except for NGC~5044, NGC~5846 and M49, which is extracted from \citep{2002ApJ...567..716R}, and for Virgo, extracted from \citep{2011JCAP...12..011S}; (4) concentration computed using \citep{Sanchez-CondeEtAl2014}; (5) normalization of the NFW profiles, where $\rho_s = \rho_0/4$ (see Eq.~(\ref{eqn:NFW})); (6) scale radius for the NFW profile (see Eq.~(\ref{eqn:scale radius})); (7) virial radius calculated as Eq.~(\ref{eq:R200}); (8) angle subtended by $R_{200}$ (see Eq.~(\ref{eqn:theta})); (9) $J$-factors assuming the MIN benchmark model for substructures (check Table~\ref{tab:dm_models}); (10) $J$-factors assuming the MED benchmark model (check Tab.~\ref{tab:dm_models}); (11) boost factor with respect $J_{MIN}$ (column 8) as defined in Sec.~\ref{subsec:subhalo_modelling}; (12) $J$-factors assuming the MAX benchmark model (check Tab.~\ref{tab:dm_models}); (13) boost factor with respect $J_{MIN}$ (column 8) as defined in Sec.~\ref{subsec:subhalo_modelling}; (14) TS of detection (see Eq.~(\ref{eq:TS})).}
\end{sidewaystable}


\begin{sidewaystable}
\centering
\begin{tabular}{|c||c|c|c|c|c|c|c||c|c|c|c|c|c||c|}
\hline 
Cluster & $d_{L}$ & $M_{200}$ & $c_{200}$ & $\rho_s$ & $r_s$ & $R_{200}$ & $\theta_{200}$ & $\log_{10}J_{MIN}$ & $\log_{10}J_{MED}$ & $B_{MED}$ & $ \log_{10}J_{MAX}$ & $B_{MAX}$ & $\log_{10}D$ & TS \tabularnewline
\hline 
\hline 
 & [Mpc] & [10$^{14}$ M$_{\odot}$] &  & [M$_{\odot}$/kpc$^3$] & [kpc] & [kpc] & [deg] & [GeV$^2$cm$^{-5}$] & [GeV$^2$cm$^{-5}$] &  & [GeV$^2$cm$^{-5}$] &  & [GeV cm$^{-2}$] & \\
\hline 
A4059 & 203.92 & 2.19 & 5.28 & 334997 & 235.56 & 1244.13 & 0.38 & 16.12 & 17.14 & 10.50 & 17.89 & 58.90 & 17.83 & 0.06 \tabularnewline
\hline 
A1795 & 278.01 & 5.17 & 5.09 & 307558 & 325.36 & 1655.37 & 0.38 & 16.23 & 17.22 & 9.81 & 17.99 & 57.50 & 17.94 & 0.42 \tabularnewline
\hline 
A2657 & 176.55 & 1.69 & 5.36 & 345942 & 212.97 & 1140.70 & 0.40 & 16.13 & 17.16 & 10.80 & 17.90 & 58.90 & 17.84 & 4.53 \tabularnewline
\hline 
A2147 & 153.91 & 1.17 & 5.47 & 363492 & 184.45 & 1009.48 & 0.40 & 16.09 & 17.13 & 11.00 & 17.86 & 58.70 & 17.79 & 5.72 \tabularnewline
\hline 
A3376 & 199.34 & 2.58 & 5.24 & 328779 & 250.74 & 1313.53 & 0.41 & 16.20 & 17.23 & 10.60 & 17.98 & 59.90 & 17.92 & 0.84 \tabularnewline
\hline 
A3562 & 222.29 & 3.53 & 5.16 & 318132 & 282.44 & 1458.40 & 0.41 & 16.24 & 17.26 & 10.40 & 18.02 & 59.70 & 17.96 & 0.03 \tabularnewline
\hline 
A85 & 250.04 & 5.09 & 5.09 & 307918 & 323.62 & 1647.33 & 0.42 & 16.30 & 17.31 & 10.10 & 18.07 & 59.00 & 18.03 & 0.31 \tabularnewline
\hline 
A3391 & 236.13 & 4.51 & 5.11 & 311034 & 309.49 & 1582.37 & 0.43 & 16.29 & 17.30 & 10.30 & 18.07 & 59.90 & 18.02 & 0.11 \tabularnewline
\hline 
A3667 & 250.04 & 5.30 & 5.08 & 306940 & 328.42 & 1669.45 & 0.43 & 16.31 & 17.32 & 10.10 & 18.09 & 59.50 & 18.04 & 13.31 \tabularnewline
\hline 
A2052 & 153.91 & 1.63 & 5.37 & 347614 & 209.89 & 1126.58 & 0.45 & 16.22 & 17.26 & 11.00 & 18.00 & 60.10 & 17.93 & 0.03 \tabularnewline
\hline 
2A0335 & 153.91 & 1.66 & 5.36 & 346659 & 211.64 & 1134.59 & 0.45 & 16.23 & 17.27 & 11.00 & 18.01 & 60.20 & 17.95 & 5.44 \tabularnewline
\hline 
A2589 & 185.64 & 2.99 & 5.20 & 323540 & 265.28 & 1379.98 & 0.46 & 16.31 & 17.34 & 10.70 & 18.10 & 61.20 & 18.04 & 0.13 \tabularnewline
\hline 
EXO0422 & 172.01 & 2.49 & 5.25 & 330093 & 247.36 & 1298.09 & 0.47 & 16.30 & 17.33 & 10.80 & 18.09 & 61.30 & 18.02 & 0.18 \tabularnewline
\hline 
\end{tabular}
\caption{\label{tab:clusters_all_2} Continuation of Table~\ref{tab:clusters_all_1}, containing the main parameters of the whole sample of galaxy clusters, ordered from smaller to bigger in size ($\theta_{200}$).}
\end{sidewaystable}

\newpage
\clearpage

\begin{sidewaystable}
\centering
\begin{tabular}{|c||c|c|c|c|c|c|c||c|c|c|c|c|c||c|}
\hline 
Cluster & $d_{L}$ & $M_{200}$ & $c_{200}$ & $\rho_s$ & $r_s$ & $R_{200}$ & $\theta_{200}$ & $\log_{10}J_{MIN}$ & $\log_{10}J_{MED}$ & $B_{MED}$ & $ \log_{10}J_{MAX}$ & $B_{MAX}$ & $\log_{10}D$ & TS \tabularnewline
\hline 
\hline 
 & [Mpc] & [10$^{14}$ M$_{\odot}$] &  & [M$_{\odot}$/kpc$^3$] & [kpc] & [kpc] & [deg] & [GeV$^2$cm$^{-5}$] & [GeV$^2$cm$^{-5}$] &  & [GeV$^2$cm$^{-5}$] &  & [GeV cm$^{-2}$] & \\
\hline
A576 & 167.47 & 2.37 & 5.26 & 331959 & 242.73 & 1276.91 & 0.47 & 16.31 & 17.34 & 10.90 & 18.09 & 61.30 & 18.03 & 0.99 \tabularnewline
\hline 
A2063 & 153.91 & 1.97 & 5.31 & 339288 & 226.15 & 1201.08 & 0.48 & 16.29 & 17.34 & 11.00 & 18.08 & 61.00 & 18.01 & 9.44 \tabularnewline
\hline 
A3558 & 213.09 & 4.89 & 5.10 & 308961 & 318.70 & 1624.70 & 0.48 & 16.41 & 17.42 & 10.30 & 18.19 & 60.90 & 18.14 & 0.35 \tabularnewline
\hline 
A2142 & 411.48 & 28.03 & 4.97 & 291172 & 585.57 & 2908.46 & 0.48 & 16.66 & 17.57 & 8.15 & 18.38 & 51.70 & 18.36 & 0.00 \tabularnewline
\hline 
A119 & 194.77 & 3.96 & 5.14 & 314731 & 294.64 & 1514.28 & 0.49 & 16.33 & 17.39 & 11.20 & 18.15 & 65.60 & 18.09 & 8.49 \tabularnewline
\hline 
A2634 & 135.92 & 1.55 & 5.38 & 349762 & 206.07 & 1109.02 & 0.50 & 16.30 & 17.35 & 11.20 & 18.09 & 60.90 & 18.02 & 4.31 \tabularnewline
\hline 
A2256 & 268.66 & 10.17 & 4.99 & 294929 & 415.33 & 2074.55 & 0.50 & 16.53 & 17.52 & 9.65 & 18.31 & 59.10 & 18.26 & 9.91 \tabularnewline
\hline 
A496 & 144.90 & 2.56 & 5.24 & 329080 & 249.96 & 1309.96 & 0.55 & 16.45 & 17.49 & 11.10 & 18.25 & 63.50 & 18.18 & 0.00 \tabularnewline
\hline 
A3266 & 263.99 & 13.44 & 4.97 & 292052 & 457.72 & 2276.43 & 0.55 & 16.67 & 17.65 & 9.57 & 18.44 & 59.60 & 18.40 & 8.19 \tabularnewline
\hline 
A1367 & 95.81 & 0.88 & 5.57 & 379136 & 164.49 & 916.83 & 0.57 & 16.36 & 17.42 & 11.50 & 18.14 & 60.80 & 18.06 & 0.99 \tabularnewline
\hline 
A4038 & 122.49 & 2.23 & 5.28 & 334336 & 237.08 & 1251.09 & 0.62 & 16.53 & 17.58 & 11.30 & 18.33 & 64.00 & 18.26 & 0.71 \tabularnewline
\hline 
A754 & 236.13 & 25.00 & 4.96 & 290649 & 564.09 & 2799.56 & 0.75 & 17.14 & 18.05 & 8.23 & 18.86 & 52.70 & 18.82 & 0.28  \tabularnewline
\hline
\end{tabular}
\caption{\label{tab:clusters_all_3} Continuation of Table~\ref{tab:clusters_all_2} containing the main parameters of the whole sample of galaxy clusters, ordered from smaller to bigger in size ($\theta_{200}$).}
\end{sidewaystable}


\begin{sidewaystable}
\centering
\begin{tabular}{|c||c|c|c|c|c|c|c||c|c|c|c|c|c||c|}
\hline 
Cluster & $d_{L}$ & $M_{200}$ & $c_{200}$ & $\rho_s$ & $r_s$ & $R_{200}$ & $\theta_{200}$ & $\log_{10}J_{MIN}$ & $\log_{10}J_{MED}$ & $B_{MED}$ & $ \log_{10}J_{MAX}$ & $B_{MAX}$ & $\log_{10}D$ & TS \tabularnewline
\hline 
\hline 
 & [Mpc] & [10$^{14}$ M$_{\odot}$] &  & [M$_{\odot}$/kpc$^3$] & [kpc] & [kpc] & [deg] & [GeV$^2$cm$^{-5}$] & [GeV$^2$cm$^{-5}$] &  & [GeV$^2$cm$^{-5}$] &  & [GeV cm$^{-2}$] & \\
\hline
A2199 & 131.44 & 5.07 & 5.09 & 308030 & 323.08 & 1644.84 & 0.76 & 16.80 & 17.85 & 11.10 & 18.62 & 66.00 & 18.56 & 1.86 \tabularnewline
\hline 
A3571 & 162.95 & 10.90 & 4.99 & 294084 & 425.60 & 2123.16 & 0.80 & 16.95 & 17.97 & 10.50 & 18.77 & 65.20 & 18.71 & 0.00 \tabularnewline
\hline 
NGC~5044 & 38.81 & 0.41 & 5.88 & 428317 & 121.16 & 711.87 & 1.07 & 16.82 & 17.90 & 11.90 & 18.60 & 60.50 & 18.51 & 0.00 \tabularnewline
\hline 
NGC~5813 & 27.55 & 0.27 & 6.06 & 460583 & 102.21 & 619.60 & 1.31 & 16.96 & 18.03 & 11.80 & 18.72 & 58.30 & 18.62 & 4.10 \tabularnewline
\hline 
A1656-Coma & 100.24 & 13.16 & 4.97 & 292223 & 454.37 & 2260.40 & 1.35 & 17.42 & 18.46 & 11.00 & 19.26 & 69.60 & 19.20 & 9.93 \tabularnewline
\hline 
NGC~5846 & 26.25 & 0.38 & 5.91 & 434293 & 117.22 & 692.90 & 1.53 & 17.13 & 18.20 & 11.90 & 18.91 & 60.40 & 18.81 & 10.81 \tabularnewline
\hline 
A1060-Hydra & 47.51 & 2.97 & 5.20 & 323860 & 264.34 & 1375.66 & 1.70 & 17.43 & 18.51 & 12.00 & 19.27 & 70.00 & 19.19 & 5.41 \tabularnewline
\hline 
A3526-Centaurus & 43.16 & 2.27 & 5.27 & 333726 & 238.51 & 1257.60 & 1.70 & 17.41 & 18.49 & 12.10 & 19.25 & 69.20 & 19.16 & 15.62 \tabularnewline
\hline 
NGC~1399-Fornax & 21.50 & 0.51 & 5.79 & 413641 & 131.82 & 762.97 & 2.05 & 17.41 & 18.50 & 12.20 & 19.21 & 62.60 & 19.11 & 4.01 \tabularnewline
\hline 
M49 & 18.91 & 0.46 & 5.82 & 419644 & 127.27 & 741.24 & 2.26 & 17.49 & 18.57 & 12.10 & 19.28 & 62.00 & 19.18 & 0.00 \tabularnewline
\hline 
NGC~4636 & 17.18 & 0.53 & 5.77 & 409991 & 134.72 & 776.79 & 2.61 & 17.63 & 18.71 & 12.20 & 19.43 & 63.00 & 19.33 & 13.09 \tabularnewline
\hline 
VIRGO & 15.46 & 5.60 & 5.07 & 305646 & 335.10 & 1700.27 & 6.32 & 18.65 & 19.74 & 12.30 & 20.52 & 74.80 & 20.44 & 1.05 \tabularnewline
\hline
\end{tabular}
\caption{\label{tab:clusters_all_4} Continuation of Table~\ref{tab:clusters_all_3} containing the main parameters of the whole sample of galaxy clusters, ordered from smaller to bigger in size ($\theta_{200}$).}
\end{sidewaystable}




\newpage
\clearpage
\appendix
\section{Study of the $J$- and $D$-factors vs. distance relation}\label{app:JD_parametrization}

Inspired by previous works on dwarf spheroidal galaxies (dSphs) \citep{Fermi-LAT:2016uux, 2019MNRAS.482.3480P}, we investigated the relation between the cluster $J$- and $D$-factors and their distances so as to provide a parametrization that could be used as a first-order approximation for the computation of the $J$- and $D$-factors, just knowing the cluster distance. This may be particularly useful for different cluster samples and/or upcoming cluster catalogs. 

\begin{figure*}
\includegraphics[width=0.49\textwidth]{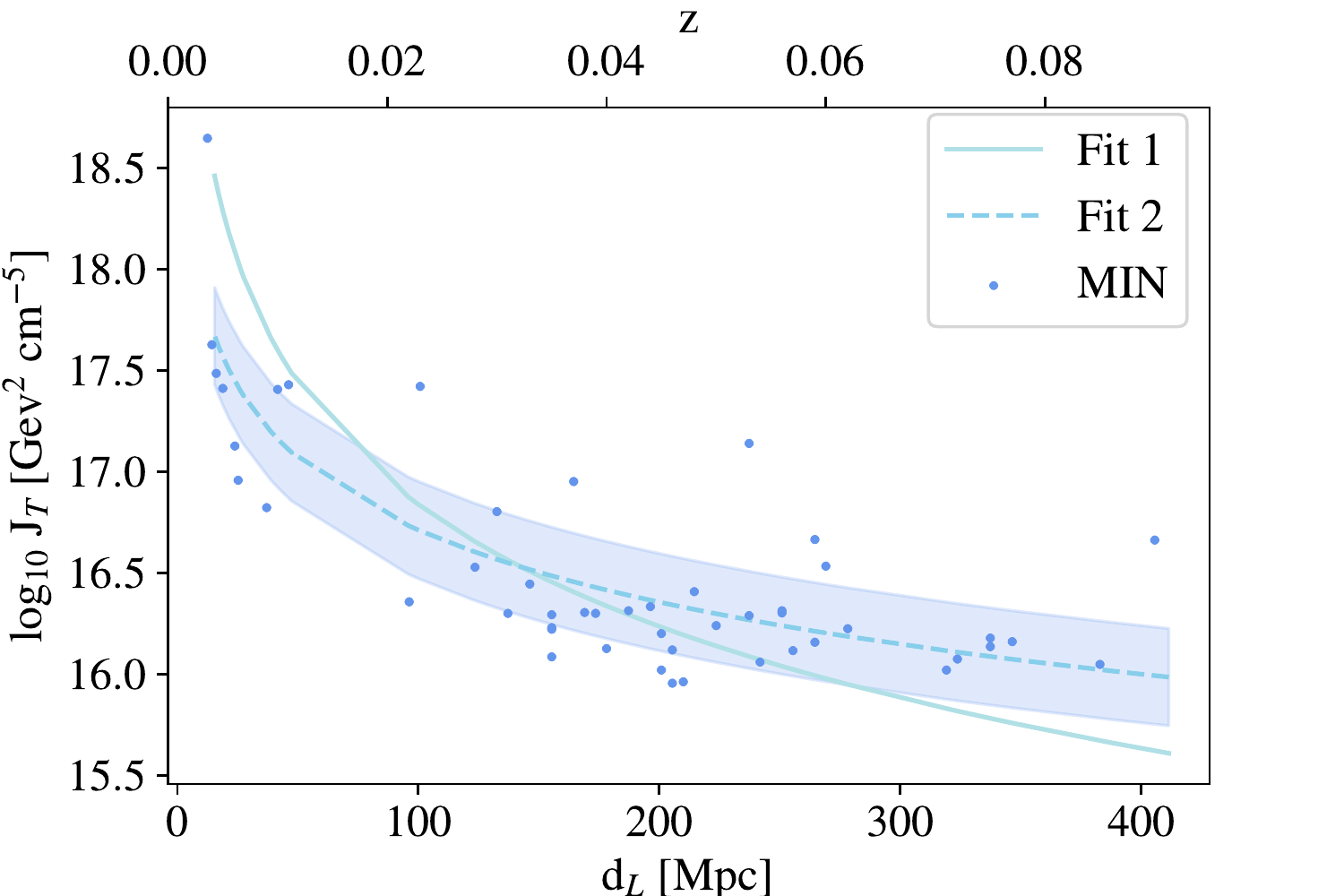}
\includegraphics[width=0.49\textwidth]{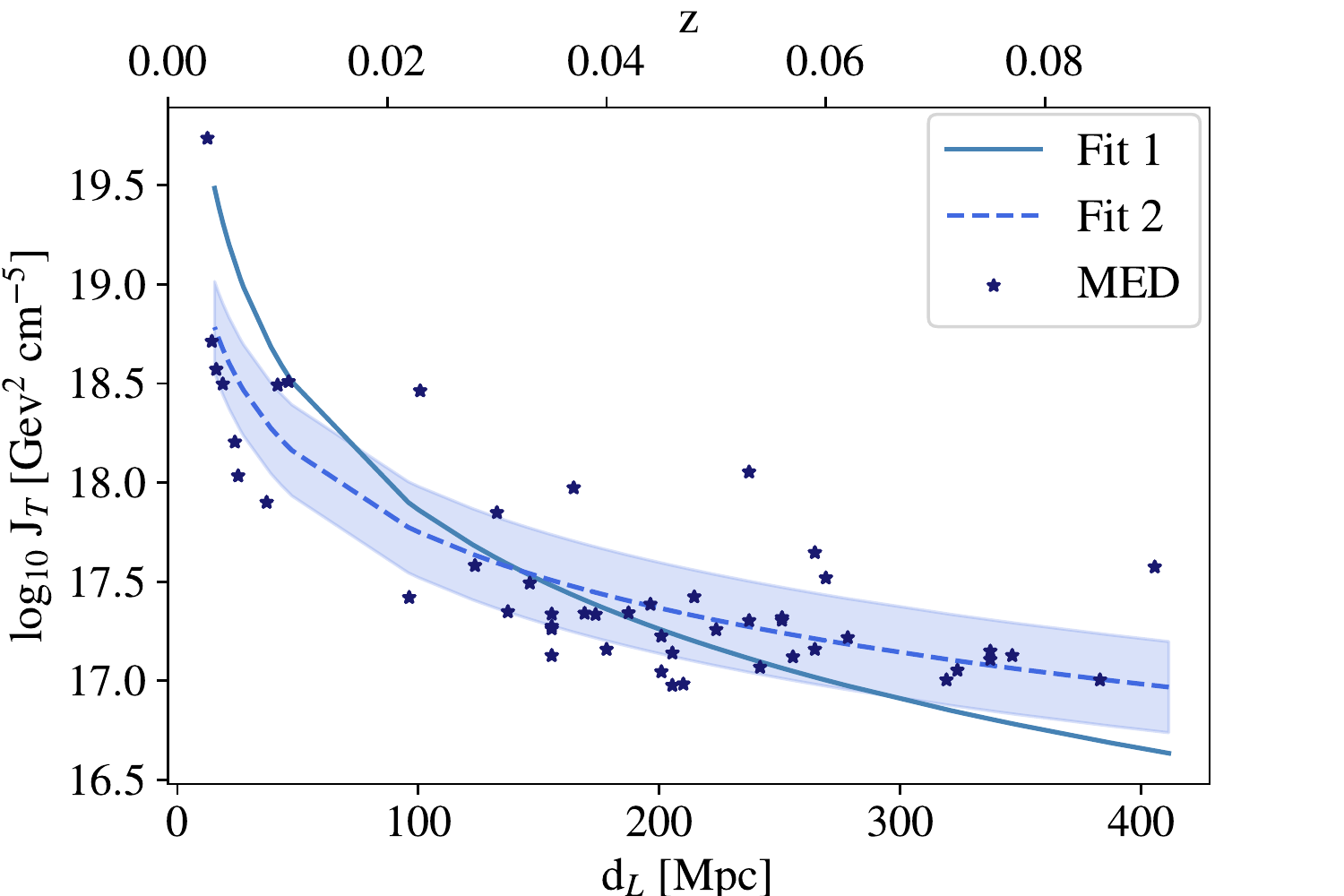}
\includegraphics[width=0.49\textwidth]{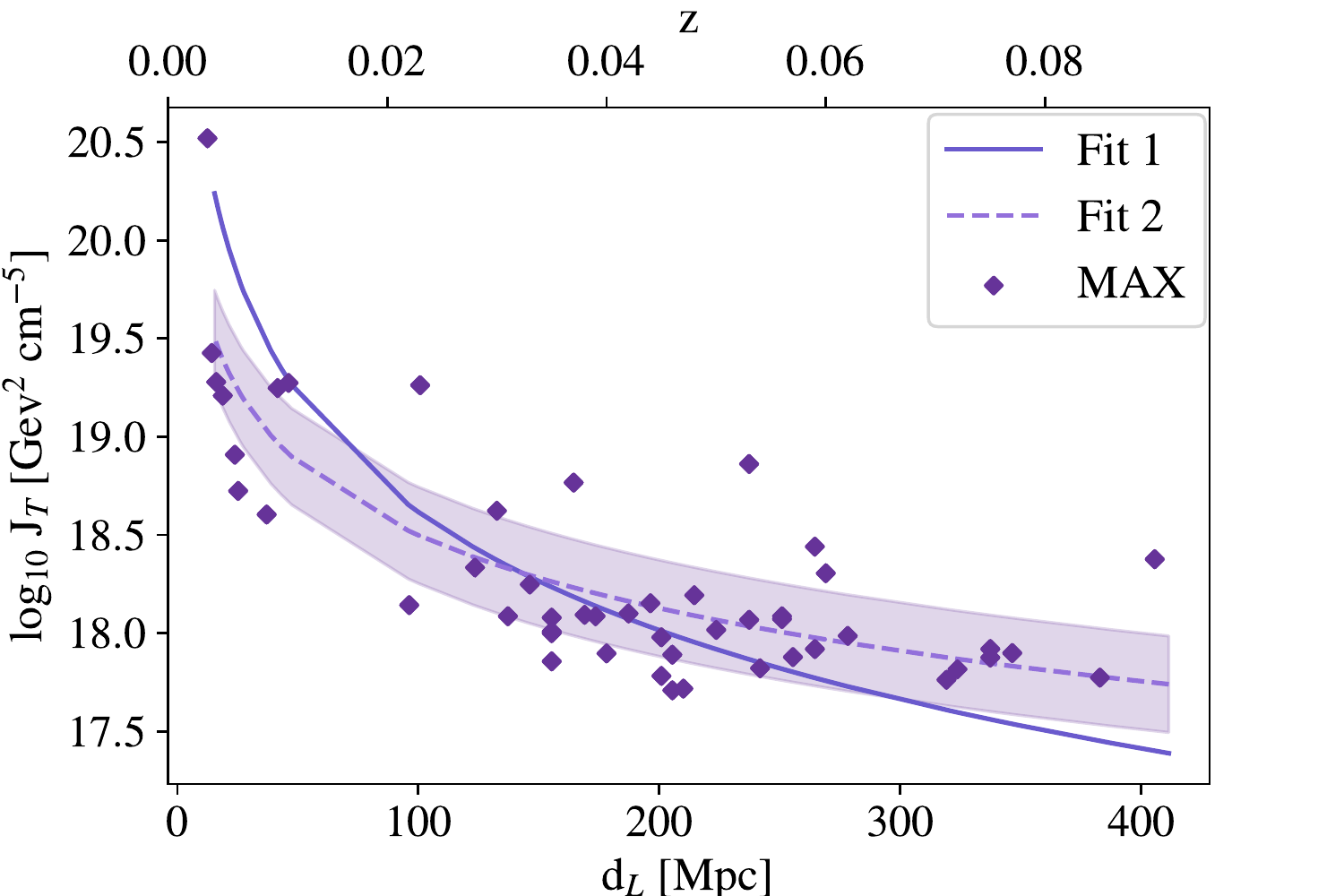}
\includegraphics[width=0.49\textwidth]{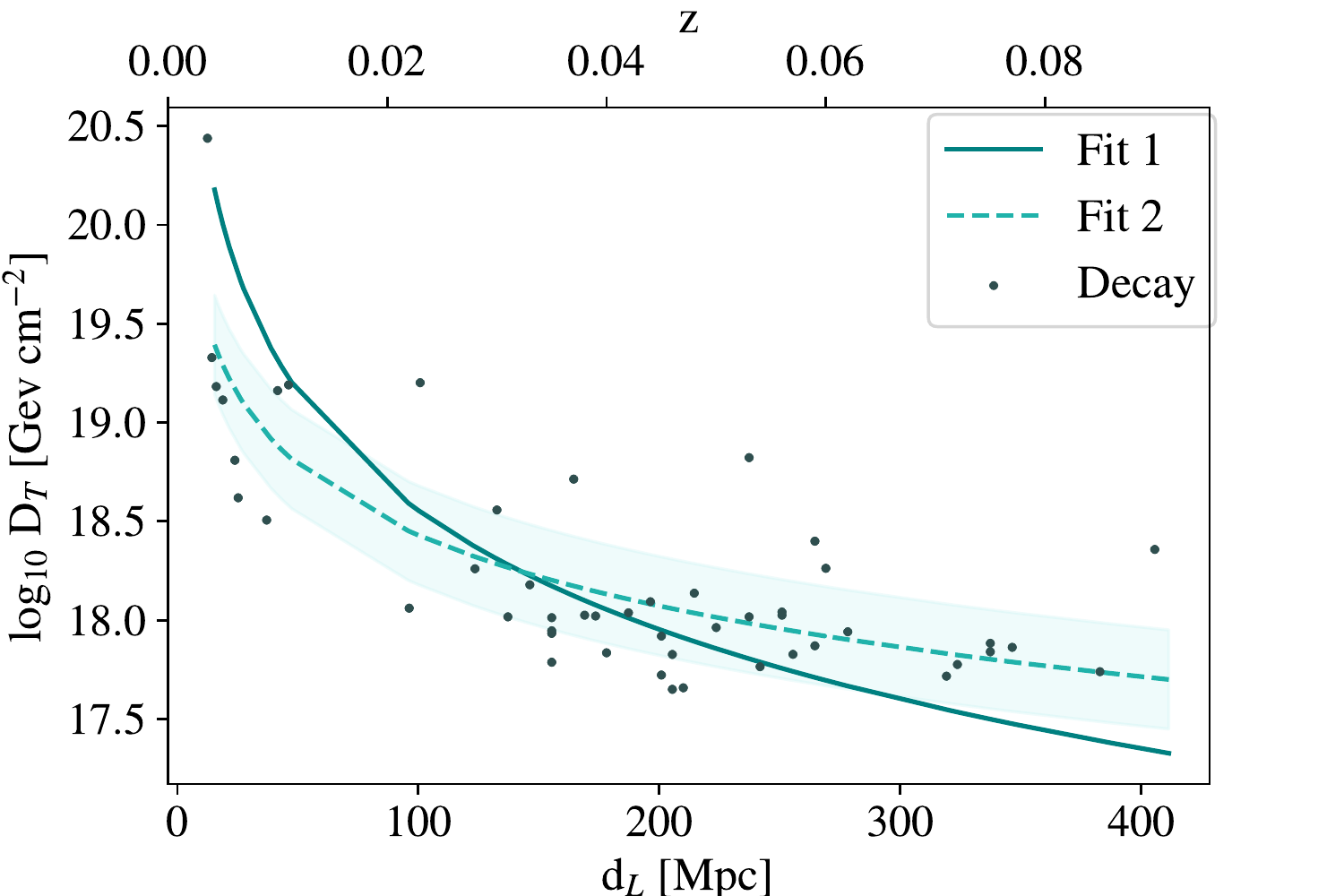}
\caption{Fits to the obtained $J$- and $D$-factors for the four benchmark models with respect to cluster's distance. Two fits are shown, both following Eqs.~(\ref{eq:J_fit})-(\ref{eq:D_fit}). For ``Fit 1'' (solid lines) we adopt a fixed value of $x=2$, while in ``Fit 2'' (dashed lines) the $x$ parameter is left free. We also show as shaded bands the mean spread of the $J$- and $D$-factors for ``Fit 2''.
}
\label{fig:sample_dm_parametrization}
\end{figure*}

According to the definition of the $J$- and $D$-factors shown in Eqs.(\ref{eq:j-factor})-(\ref{eq:d-factor}), the main dependencies are:
\begin{equation}\label{eq:J_dependencies}
J_{T} \propto \frac{M_{200}c_{200}^3}{d_{L}^2},
\end{equation}
\begin{equation}\label{eq:D_dependencies}
D_{T} \propto \frac{M_{200}}{d_{L}^2}.
\end{equation}
From these, we can check how much the $J$- and $D$-factors follow this expected dependence with the distance. Starting from these relations, we propose to fit the $J$- and $D$-factors to the following simplified scaling relations:
\begin{equation}\label{eq:J_fit}
\log_{10} J_{T} = -x\log_{10}\frac{d_{L}}{100 \rm{Mpc}} + \log_{10}J_0,
\end{equation}
\begin{equation}\label{eq:D_fit}
\log_{10} D_{T} = -x\log_{10}\frac{d_{L}}{100 \rm{Mpc}} + \log_{10}D_0,
\end{equation}
where $J_0$ and $D_0$ are nominal values that together with the exponent of the dependence with the distance, $x$, will be the parameters to fit. We perform a first fit where we fix $x=2$, thus $J_0$ and $D_0$ being the only free parameters; and a second fit where we set free also $x$, in order to quantify deviations from the expected inverse-of-the-distance-squared behavior. We perform the fits for the four benchmark models (three for annihilation -- MIN, MED, MAX -- and one for decay) and show the corresponding results in Fig.~ \ref{fig:sample_dm_parametrization}. Tab.~\ref{tab:dm_fits} summarizes the obtained best-fit parameter values.

\begin{table}
\begin{tabular}{|c||c||c|c|}
\hline
Model & Fit 1 ($x=2$) & \multicolumn{2}{c|}{Fit 2} \\ 
\hline
 & $\log_{10}J_0$ [GeV$^2$cm$^{-5}$] & x & $\log_{10}J_0$ [GeV$^2$cm$^{-5}$] \\
\hline
MIN & 16.839 & 1.18 & 16.713 \\
\hline
MED & 17.863 & 1.27 & 17.751 \\
\hline
MAX & 18.618 & 1.24 & 18.499 \\
\hline
\hline
 & $\log_{10}D_0$ [GeV cm$^{-2}$] & x & $\log_{10}D_0$ [GeV cm$^{-2}$] \\
\hline
Decay & 18.556 & 1.19 & 18.430 \\
\hline
\end{tabular}
\caption{\label{tab:dm_fits}Best-fit parameter values obtained for our sample of galaxy clusters and the four considered benchmark models when fits to Eqs.(\ref{eq:J_fit})-(\ref{eq:D_fit}) are performed, i.e. J/D factors vs. distance. ``Fit 1'' corresponds to the case in which we fix $x=2$, while ``Fit 2'' is the case in which $x$ is left free in the fit.}
\end{table}

From Fig.~\ref{fig:sample_dm_parametrization} we can see important departures from the expected simplistic $J_T, D_T \propto 1/d_L^2$ relation. Thus, these results imply that the dependence with the mass is much more significant in the case of clusters than in the case of dSphs \citep{Fermi-LAT:2016uux} and should not be neglected. 
We leave for a future work the inclusion of this mass dependence in the fit. From ``Fit 2'', an exponent of $x \sim 1.2$ seem to better fit the data in all cases\footnote{It should be further investigated if this departure from $x=2$ may be caused by neglecting the mass dependence.}. The uncertainty bands represent the mean discrepancy between the actual J/D-values with respect the corresponding ones but from the ``Fit 2''. In all of the panels we can see that most of the clusters lie within this uncertainty band, whose value is $\delta = 0.23 - 0.25$ dex depending on the fit. This discrepancy matches perfectly with the estimated $\sigma_J$ uncertainty from the scatter of the concentration-mass relation. We also note that the best-fit values found for $J_0$ and $D_0$ are close to the mean $J_T$ and $D_T$ of all the clusters for each benchmark model. 


\bibliography{paper}

\end{document}